\documentclass [12pt,a4paper,     ]{book} 

\usepackage{graphics}
\usepackage{times}

\DeclareFontFamily{OT1}{times}{}
\DeclareFontShape {OT1}{times}{m }{n }{ <-> ptmr }{}
\DeclareFontShape {OT1}{times}{bx}{n }{ <-> ptmb }{}
\DeclareFontShape {OT1}{times}{m }{it}{ <-> ptmri}{}
\DeclareFontShape {OT1}{times}{bx}{it}{ <-> ptmbi}{}
\begin{document}

\renewcommand{\chaptername}{Paper}  

\title{{\Huge {\bf The origin of Iraq's nuclear weapons program:}}\\
          ~\\ {\bf Technical reality and Western hypocrisy}}

\author{by Suren Erkman, Andre Gsponer,\\
           Jean-Pierre Hurni, and Stephan Klement\\
                         ~\\ 
    \emph{Independent Scientific Research Institute}\\ 
    \emph{Box 30, CH-1211 Geneva-12, Switzerland}}

\date{ISRI-05-09.11 ~~ 20 October 2008 }

\maketitle

\frontmatter

\chapter{Abstract}
\pagestyle{myheadings}
\markboth{ }{ }
\label{abstract}

\begin{quote}

This report is based on a series of papers written between 1980 and 2005 on the origin of Iraq's nuclear weapons program, which was known to one of the authors in the late 1970s already, as well as to a number of other physicists, who independently tried without success to inform their governments and the public.

It is concluded that at no point did the Western governments effectively try to stop Iraq's nuclear weapons program, which suggests that its existence was useful as a foreign policy tool, as is confirmed by its use as a major justification to wage two wars on Iraq.

\end{quote}

\newpage

\begin{quote}

Dedicated to {\bf Roy Adrian Preiswerk} (1934--1982).

Professor at the University of Geneva Graduate Institute of International Studies (IUHEI), creator and first Director of the Institute of Development Studies (IUED), founding member and first President (1980--1982) of the Directing Committee of the Geneva International Peace Research Institute (GIPRI).

Author of:

\emph{Ziele und Mittel schweizerischer Friedenspolitik auf internationale Ebene}, in: Urs Altermatt, Roy Preiswerk, und Hans Ruh, Formen schweizerischer Friedenspolitik (Iustitia et Pax und GIPRI, Fribourg, 1982) 7--66;

translated in French as:

\emph{Les objectifs et les moyens d'une politique internationale suisse pour la paix}, trad.\ Isabelle Schulte-Tenckhoff avec la collaboration de Gilbert Rist, in: Gilbert Rist, \'editeur, \`A Contre-Courants : l'Enjeu des Relations Interculturelles (\'Editions d'en bas, Lausanne, 1984) 61--146.

\end{quote}


\tableofcontents

\listoffigures

\chapter{Preface: From 1979 to 2005} 
\pagestyle{myheadings}
\markboth{Preface}{Preface}
\label{preface}

The existence of this report is due to a simple event:  The visit to CERN (the European center for nuclear research) in Spring 1979 of a few Iraqi engineers who showed a considerable interest in the construction details of a unique large magnet, which at the time was the key component of the NA10 experiment at CERN, in which Andre Gsponer was working as a physicist.

\begin{figure}
\begin{center}
\resizebox{12cm}{!}{ \includegraphics{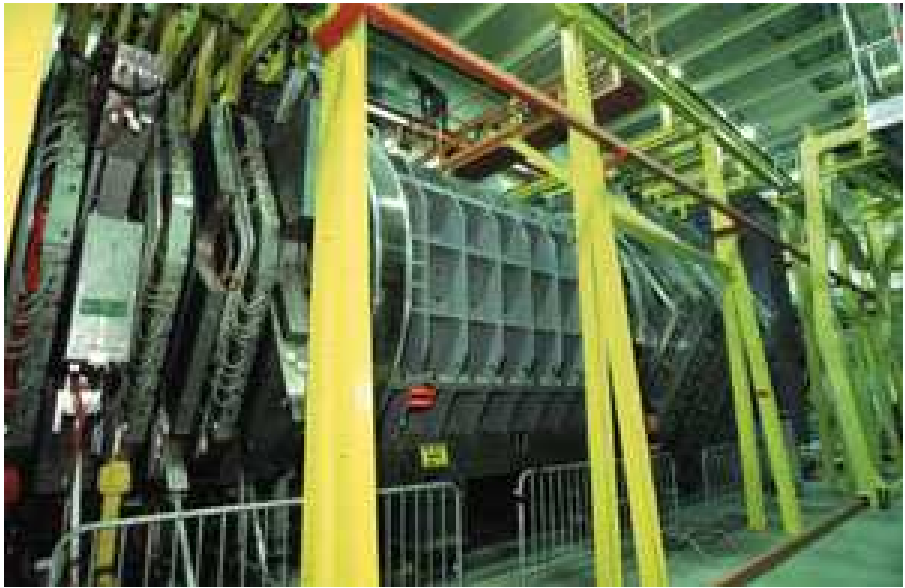}}
\end{center}
\caption[NA10 magnet at CERN]{\emph{The magnet of the NA10 experiment at CERN is one of which Iraq took inspiration to design its electromagnetic  uranium-enrichment facilities. The equipment on the photograph is about 20 m long and more than 6 m high. The magnet is at the center, between the two yellow posts.}}
\label{fig:NA10-magnet}
\end{figure}

Because of his unexpected discovery one year earlier of the relevance of particle accelerator technology such as developed at CERN to classified particle beam weapons research, which led him to study a lot of technical papers on the military implications of particle accelerator technology, Gsponer immediately understood that the only plausible explanation for Iraq's interest in large magnets was to use them as ``calutrons,'' i.e., as electromagnetic isotope separators, to enrich uranium in view of making an atomic bomb --- just like the American did in 1945 to produce the U-235 of the Hiroshima bomb.

Following this shocking realization of the crucial (but untold) importance of particle accelerator technology to the proliferation and further development of nuclear weapons, Gsponer left CERN in 1980 to work on technology assessment with the hope of contributing to nuclear disarmament by publishing his conclusions on the feasibility of calutrons, particle beam weapons, antimatter weapons, etc.

This report consists of five papers and a postface, presented in the order in which they were published or released.  They are authored by one or several people whose names are mentioned on the title page of this report, but whose responsibility extends only over those papers which bare their name.

Paper \ref{Chap.1} by Suren Erkman is a newspaper article published in \emph{Le Journal de Gen\`eve} of 22-23 April 1995.  It briefly explains the relation of CERN to the origin of Iraq's calutron program and announces the forthcoming publication of a comprehensive technical report on this subject, i.e., Paper \ref{Chap.2}.  Erkman's article is in French and it is expected that an English translation will be available for a future edition of this report.

Paper \ref{Chap.2} by Andre Gsponer and Jean-Pierre Hurni, released on 19 October 1995, is a technical assessment of high-current electromagnetic isotope separation (EMIS) technology for uranium and plutonium enrichment (i.e., calutrons), and a detailed account of the circumstances and significance of the 1979 discovery at CERN of Iraq's definite interest in calutron technology.  Numerous attempts made since that time to forewarn of the nuclear weapons proliferation impact of particle accelerator technology are recalled.

Paper \ref{Chap.3} by Andre Gsponer, Jean-Pierre Hurni, and Stephan Klement (a physicist and lawyer specialized in nuclear law), released on 12 February 1997, is a development of \ref{Sec.2.5.3} of Paper \ref{Chap.2}.  Its purpose is to show that the legal framework created in order to provide a legal basis for the operation of UNSCOM (the United Nations Security Council Special Commission) in Iraq also constitutes a unique legal precedent because it defines for the first time what is actually meant by ``proliferation prone nuclear activities,'' and what a truly ``nuclear-free zone'' (i.e., exempt of any nuclear facility, reactor, or weapon) should be.

Paper \ref{Chap.4} by Andre Gsponer, first posted on internet on 31 July 2001, i.e., ten years after the first Gulf War and only a few weeks before the 11th September 2001 events, is a discussion of the fate of Papers \ref{Chap.1}, \ref{Chap.2}, and \ref{Chap.3}.  It is shown that whereas these papers had only little visible impact, the correctness and pertinence of their content was amply confirmed by a number of additional facts, including the interactions of the authors with the IAEA, and similar attempts by other physicists who like Gsponer tried to inform their governments and the public.

Paper \ref{Chap.5} by Andre Gsponer, first published here, is a continuation of Paper~\ref{Chap.4}.  It relates, in particular, how in April 2003 (as soon as he could after the beginning of the US-led invasion of Iraq) Jafar Dhia Jafar, the former Head of Iraq's nuclear weapons program, got in contact was Gsponer in order to try to come to Switzerland.  Jafar believed that Switzerland was neutral in the war against his country, and that Geneva was the best place in the world to give evidence against the false claim that Iraq had resumed its  nuclear weapons program --- which was used as one of the main arguments in the case for war on Iraq.  However, after considerable efforts by Gsponer, it was found that Jafar could not come to Geneva because Switzerland was not neutral, as a result of decisions pleasing to the United States made by the Swiss government prior to the invasion of Iraq.

The final paper in this report is a postface by Andre Gsponer, in which he uses his first hand experience of the Swiss and Iraqi nuclear weapons programs, to express a very pessimistic judgment on the true nature of Western democracy, as well his strong disillusions about the real impact of public-spirited initiatives such as his 25 years of work at GIPRI and ISRI to produce high quality technical information on nuclear weapons non-proliferation and disarmament.

%
\mainmatter

\chapter{Atomic Bomb: Iraq Went through CERN}
\pagestyle{myheadings}
\markboth{Journal de Gen\`eve}{22--23 avril 1995}
\label{Chap.1}


by {\bf Suren Erkman}, published in the weekend edition of \emph{Le Journal de Gen\`eve}, Saturday 22 - Sunday 23 April 1995, pages 1 and 5.

\begin{figure}
\begin{center}
\resizebox{4cm}{!}{ \includegraphics{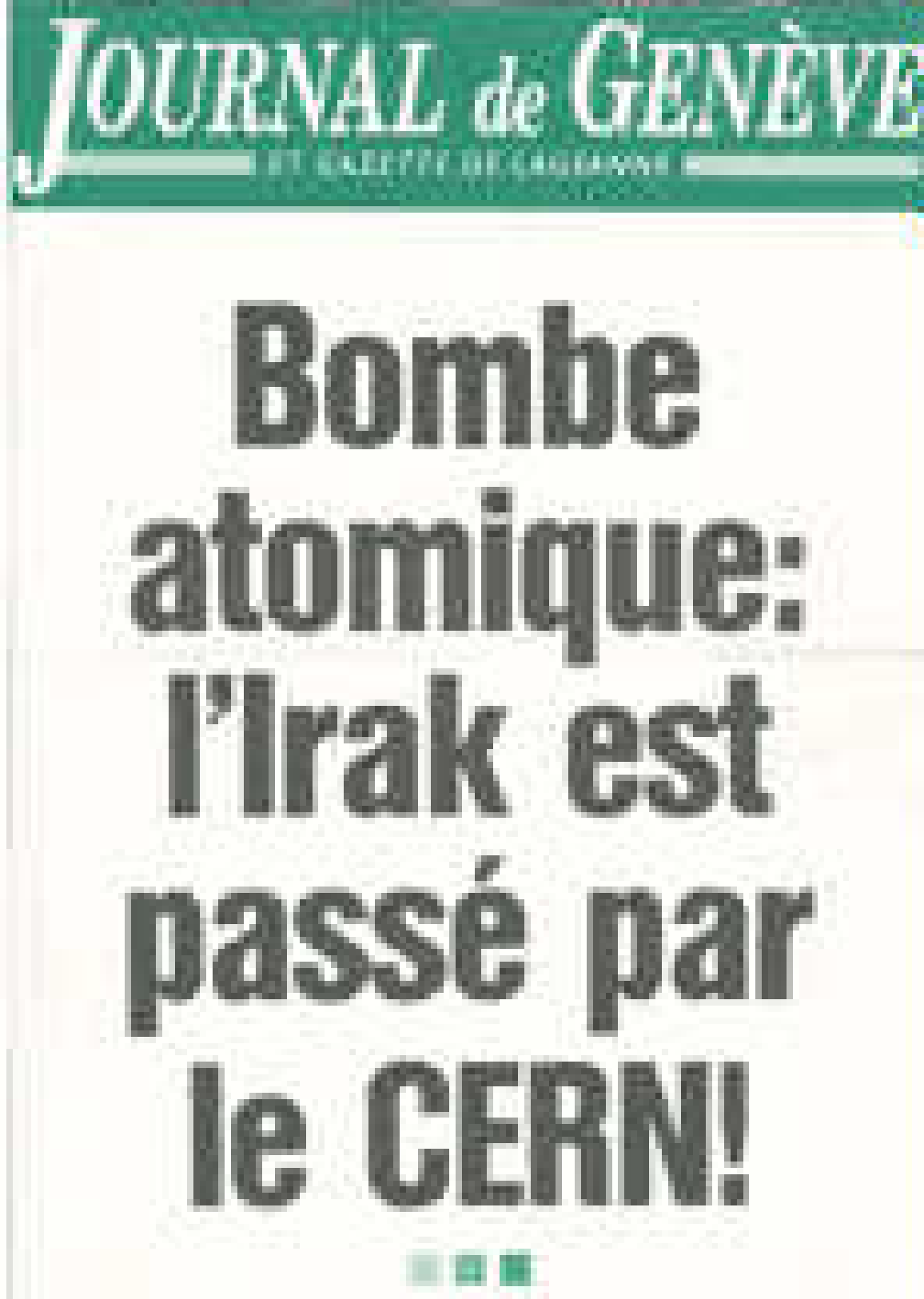}}
\end{center}
\caption[Poster announcing paper on Iraq's calutrons.]{\emph{Poster announcing the paper on Iraq's calutrons.}}
\label{fig:Poster}
\end{figure}


\section{NUCLEAR WEAPONS ---  A Researcher Blew the Whistle!}
\label{Sec.1.1}

{\it In the 1970s, the Iraqis visited CERN several times to gather information as part of their atomic bomb program.}

In the 1970s, the Iraqis visited CERN several times to gather information as part of their atomic bomb program. 

In the 1970s, the head of the Iraqi atomic bomb program, Jafar Dhia Jafar, made several trips to CERN in Geneva while working at the Imperial College London. On at least one occasion in 1979, he dispatched one of his engineers to CERN to gather technical information about a particular type of magnet. Among several possible ways to obtain enriched uranium, the Iraqis had decided to use and perfect calutrons, a technique that involves using powerful magnets to separate the atoms of uranium 235 (of military interest) from the atoms of uranium 238 (poorly fissile) that make up natural uranium. For this reason they were specifically interested in the types of magnets developed for one of CERN's experiments. In the end, the Iraqis chose a design other than CERN's for the magnets in their calutrons, but the visits reveal that they were already exploring this uranium enrichment technique some fifteen years ago. 
A researcher from Geneva, who at the time tried in vain to alert the scientific and disarmament communities, is now going public about the visits—at the moment when the Treaty on the Non-Proliferation of Nuclear Weapons is being renegotiated in New York. (Editor's Note: See articles by Suren Erkman on page 5.)

\section{Geneva: Cradle of the Iraqi Bomb}
\label{Sec.1.2}

{\it  Iraq gathered technical information in Geneva with an atomic bomb in mind. The head of Iraq's military nuclear program himself made several trips to CERN, the European Laboratory for Particle Physics.}

One day in 1979, an Iraqi engineer arrived at CERN, the European Laboratory for Particle Physics in Geneva. Although he was not involved in one of CERN's research projects, his visit was not out of the ordinary, the European laboratory being a place for sharing scientific information that is open to all regardless of nationality. He approached one of the physicists working on the NA10 experiment and asked questions about details of the construction of the highly specialized magnet that is the key component of the equipment in the experiment. The physicist, who was in charge of the magnet, mentioned the visit to one of his colleagues, Andre Gsponer, a young researcher who was concerned by the role of scientific research in the arms race.

\subsection{Jafar Dhia Jafar's Connections}
\label{Sec.1.2.1}

After a quick analysis, Gsponer concluded that, in all likelihood, the Iraqis were obviously interested in producing enriched uranium using a technology known as ``electromagnetic isotope separation.'' In principle, the technique involves separating, with the help of a powerful magnet, the atoms of uranium 235 (of military interest) from the atoms of uranium 238 (poorly fissile) that make up natural uranium. The technique has been around for many years; the United States used it between June 1944 and July 1945 to make the nuclear charge for the bomb dropped on Hiroshima. The machine, developed at the University of California, is known as a ``calutron,'' a contraction of ``California University Cyclotron.''

The details of the technique, which in some aspects is rudimentary, were made public between 1946 and 1956. Of the several possible methods for obtaining enriched uranium, the Iraqis had decided to use calutrons, but after modernizing and perfecting them. This is why they were interested specifically in magnets like those developed for the NA10 experiment at CERN. 

The Iraqis did not come to Geneva by accident. CERN has always been considered as the world's center of excellence in large magnet technology. Above all, the head of Iraq's bomb program, Jafar Dhia Jafar, had made several visits to CERN during the 1970s while he was working at the Imperial College London. Between 1967 and 1976, he published 12 scientific articles on research carried out at the synchrotrons of Birmingham and  CERN. 

Naturally, Jafar used personal contacts made during his long stays in Europe to send his researchers to collect crucial information from the best sources. This is how the Iraqi engineer was explicitly sent by Jafar to question the researcher responsible for the NA10 magnet. However, says the latter, who wishes to remain anonymous, the visit did not have any particular significance because ``the information the engineer was given is available in public scientific literature. To my knowledge, he didn't get access to the original engineering plans or to specific processes developed by CERN to make the magnet.''

\begin{figure}
\begin{center}
\resizebox{10cm}{!}{ \includegraphics{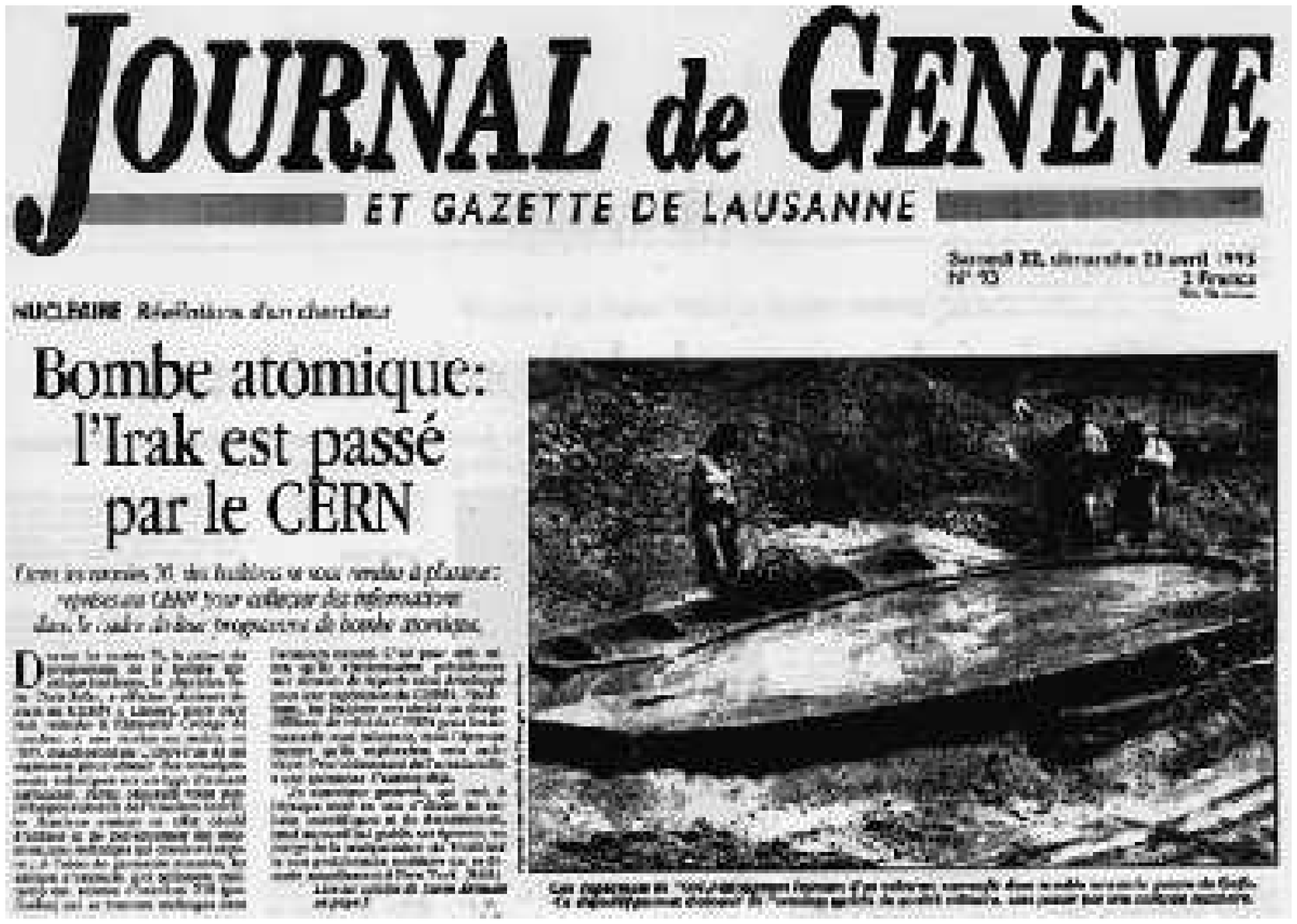}}
\end{center}
\caption[``Atomic bomb: Iraq went through CERN.'']{\emph{Journal de Gen\`eve, page 1: ``Atomic bomb: Iraq went through CERN.''}}
\label{fig:Page 1}
\end{figure}

\subsection{Proliferant Technology}
\label{Sec.1.2.2}

When the Gulf War broke out, Jafar Dhia Jafar, who now lives under high-level protection in Baghdad, officially took over the positions of Deputy Minister of Industry and Military Industrialization, Vice President of the Iraqi Atomic Energy Commission, and Director of Reactor Physics at the Center for Nuclear Research in Tuwaitha. (A compilation of information about Jafar Dhia Jafar, including mention of his time at CERN, can be found in William E. Burrows and Robert Windrem's Critical Mass: The Dangerous Race for Superweapons in a Fragmenting World, Simon \& Schuster, New York, 1994. Unfortunately, the book has many technical errors.) 

For his part, Andre Gsponer took the incident very seriously. He left CERN and in 1980 become the first director of GIPRI, the Geneva International Peace Research Institute, that he founded in 1979. His aim was to conduct a rigorous assessment of the possible military consequences of research carried out in the fields of nuclear and particle physics. In 1980, during the second review conference of the Treaty on the Non-Proliferation of Nuclear Weapons (NPT) in Geneva, he authored a document in which he stateed that the technology used for electromagnetic separation, which had seen many technical advances, had potentially become a very proliferant technology, and he specifically mentioned Iraq. 

After having tried in vain to awake the interest of the disarmament community about the issue, and after having been warned by experts about the dangers of disclosing such information for him and his former colleagues at CERN, Gsponer decided to no longer refer publicly to Iraq's keen interest in calutron technology. He left Switzerland in 1987 and returned to his first love: theoretical research in fundamental physics.

\subsection{The Calutrons Are Confirmed}
\label{Sec.1.2.3}

However, in 1991, after the Gulf War, Gsponer was startled to see the images coming out of Iraq --- more precisely, the pictures of calutrons used in the Iraqi bomb program! Andre Gsponer gathered all available information about calutrons in scientific literature and in reports from UN missions to Iraq. He is now finishing a technical study of calutrons in which he describes in detail how Iraq intended to manufacture the enriched uranium needed for a bomb (Andre Gsponer and Jean-Pierre Hurni, Calutrons: 1945---1995, ISRI, Case postale 30, 1211 Genève-12). Events had proved him right: Iraq, like other countries, had indeed taken the path of calutrons, confirming the proliferation potential for this technology. ``If one had taken the matter seriously at the time, the course of events may have been different,'' notes Gsponer.
 
One major question remains, nonetheless. Western intelligence claimed to have been surprised to have uncovered a significant calutron program in Iraq. ``I can't believe that I was the only one, along with a few others, to know the Iraqis were extremely interested in calutrons in the late 1970s,'' wonders Gsponer. ``Western intelligence agencies must have known. Why wait so long to take action then, and why pretend to discover the program after the Gulf War?''

\begin{figure}
\begin{center}
\resizebox{8cm}{!}{ \includegraphics{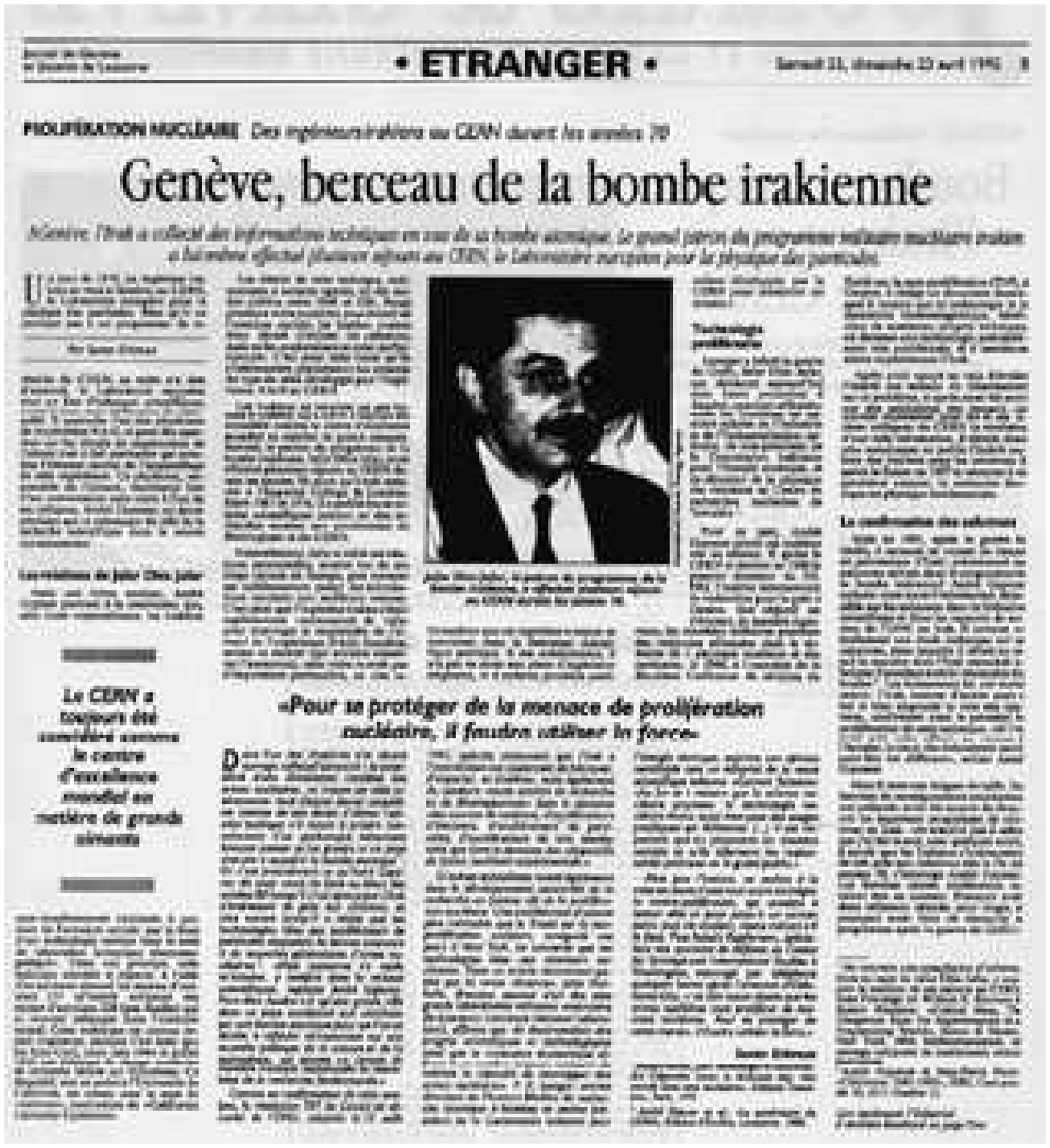}}
\end{center}
\caption[``Geneva, cradle of the Iraqi bomb.'']{\emph{Journal de Gen\`eve, page 5: ``Geneva, cradle of the Iraqi bomb.''}}
\label{fig:Page 5}
\end{figure}

\section{``To protect ourselves from the threat of nuclear proliferation,
           we will have to use force''}
\label{Sec.1.3}

One chapter in a recent collective work on the possibility of completely eliminating nuclear weapons offers an interesting idea: every citizen should consider it his or her responsibility to alert public opinion if he or she learns of any incident that suggests a group or country is seeking to acquire the atomic bomb (Joseph Rotblat, Jack Steinberger, and Bhalchandra Udgaonkar in Un monde sans armes nucléaires [A world without nuclear weapons], Editions Transition, Paris, 1995). Yet, that is exactly what Andre Gsponer said he tried to do in the early 1980s when he saw that Iraq was very interested in calutrons, and even more to the point, when he realized that the technologies related to particle accelerators could give rise to new generations of nuclear weapons (Andre Gsponer et al., La quadrature du CERN [The quadrature of CERN], Editions d'En-Bas, Lausanne, 1984]. ``But no one wanted to listen to me, including the scientific community,'' laments Gsponer. ``Perhaps it'll take a major city in a Western country going up in smoke after an atomic bomb for us to decide to seriously think about a new policy for science and technology that takes into account the consequences of fundamental research in a truly responsible way.'' 

Seemingly confirming this analysis, the UN Security Council adopted on 15 August 1991 Resolution 707, which clearly specifies that Iraq is prohibited not only from constructing, importing, and using, but also from conducting ``any activity such as research and development'' in the field of ``neutron sources, electron accelerators, particle accelerators, heavy ion accelerators'' and in the field of ``nuclear fusion experimental devices.'' 

Other experts also see the uncontrolled development of research as a key factor in nuclear proliferation. A proliferation that is all the more inevitable as the Treaty on the Non-Proliferation of Nuclear Weapons now being renegotiated in New York covers only technology related to nuclear reactors. In an article recently published by Science, John Nuckolls, associate director of one of the largest American military laboratories, Lawrence Livermore National Laboratory, states, ``the dissemination of scientific and technological advances as well as economic growth offer an ever-growing number of nations the capacity to develop nuclear weapons.'' P. K. Iyengar, former director of the Bhabha Atomic Research Center in Bombay and former chairman of India's Atomic Energy Commission, expresses a similar opinion in an editorial in the Indian journal Current Science: ``As nuclear science advances, nuclear technology evolves as much for peaceful purposes as for military ones....It is important for physicists to recognize this and to inform political leaders and the general public.'' 

However, for the moment, we are witnessing an entirely different strategy being put in place: counterproliferation, which involves allowing a country to go to a certain point and then stopping it, with military force if necessary. For Robert Kupferman, an expert in defense issues at the Center for Strategic and International Studies in Washington, D. C., interviewed by telephone a few hours after the Oklahoma City bombing, ``there's no doubt that nuclear weapons are going to proliferate at a terrifying rate. To protect ourselves from this threat, we will have to use force.'' 


\chapter{Iraq's calutrons : Electromagnetic isotope separation,
         beam technology, and nuclear weapon proliferation}
\pagestyle{myheadings}
\markboth{Report}{ISRI-95-03}
\label{Chap.2}

by {\bf Andre Gsponer} and {\bf Jean-Pierre Hurni}\\ ~\\ Report ISRI-95-03 ~~ --- ~~ 19 October 1995\\

\section{Abstract}
\label{Sec.2.1}

The past and present status of high-current electromagnetic isotope separation (EMIS) technology for uranium and plutonium enrichment (i.e., calutrons) is reviewed in the five nuclear weapons states and in four critical states: Japan, India, Israel and Iraq.

The circumstances and significance of the 1979 discovery at CERN, the European center for nuclear research in Geneva, of Iraq's definite interest in calutron technology, is discussed in detail, together with the problem of publishing independent opinions on the nuclear proliferation implications of particle accelerator and fusion technologies.

The conclusion stresses the potential of ``old'' beam technologies such as calutrons, e.g., for the transformation of reactor-grade into weapons-grade plutonium, and of particle accelerators for the efficient production of plutonium or tritium.  UN Security Council Resolutions 687 and 707, obliging Iraq to all proliferating nuclear activities, are shown to provide a legal precedent for the unambiguous definition of strictly peaceful nuclear activities.  The ``failure'' of Western intelligence in detecting Iraq's gigantic calutron program is questioned, and the relation of this ``failure'' to the justification of past and possible future coercive counter-proliferation actions is investigated.

\section{Introduction}
\label{Sec.2.2}

	Shortly after the Gulf War, under the terms of UN Security Council resolution 687, several International Atomic Energy Agency (IAEA) teams inspected known or suspected nuclear sites in Iraq. Between June and September 1991, substantial enrichment activities were discovered, including two industrial-scale facilities using the electromagnetic isotope separation (EMIS) method, and a program to produce enriched uranium with ultracentrifuges. From that time on, the fact that Iraq did successfully put the EMIS method into practice, and the fact that Iraq had a complex, comprehensive nuclear weapons program, have been presented as big surprises and as major failures of Western intelligence.  Why therefore, in the months before the Gulf War, did President Bush and his administration give such prominence to Iraq's nuclear bomb ambitions? ``As I report to you, air attacks are under way against military targets in Iraq. We are determined to knock out Saddam Hussein's nuclear bomb potential,'' the President said, before ticking off other objectives of the assault, just two hours after U.S. warplanes began attacking Iraq on January 16, 1991.

	There were many indications of Iraq's nuclear ambitions even before an Israeli air raid in 1981 destroyed the Iraqi Tammouz 1 (Osiraq) reactor just before its completion. For instance, the first author of this report (A. Gsponer) learned in 1979 that Iraq was already interested in the construction of an industrial-scale facility using the EMIS method \cite{AA.1}. This important discovery was one of the reasons why he decided to quit high energy physics and to start working full time on disarmament. With limited success, he tried to inform the arms-control/disarmament community of the military impact of particle accelerator technology, and in particular of their implications on both vertical and horizontal proliferation of nuclear weapons \cite{AA.1,AA.2,AA.3}.

	The purpose of this report is to review some of the historical and technical aspects of EMIS, to summarize what is publically known of Iraq's attempt to use this technology in its nuclear weapons program, and to draw the main disarmament conclusions.

	In the first part, it will be seen that EMIS technology is an integral part of the nuclear programs of all nuclear weapons states (USA, Russia, England, France and China) and that it has been developed to various degrees in many countries including India, Israel and Japan.

	In the second part, it will be seen that Iraq's EMIS design was not a simple copy of the rather crude one used by the USA during World War Two, but an improved design which incorporated many of the refinements made since 1945. In this part details of a biographical and historical nature will be given. For convenience, these will be referred to in the third person.
	 
	In the final part, it will be stressed that not only EMIS technology, but also a whole range of technologies, comprising old as well as new ones, are going to make nuclear proliferation an increasingly likely possibility. UN Security Council Resolution 687 and 707, which impose a comprehensive moratorium on Iraq in order to avoid the resumption of its nuclear program, and which recognize the proliferation potential of these technologies, are analysed in the perspective of their contribution to the unambiguous definition of strictly peaceful nuclear activities. Finally, the problem of the ``failure'' of intelligence to detect the massive Iraqi nuclear weapons program is addressed in the light of the discovery of Iraq's interest in calutrons as early as 1979.

	The authors would like to thank Ms Heather Serdar, graduate student at The Graduate Institute of International Studies, University of Geneva, for sharing some of her documentation on the United Nations and Iraq with us. The authors thanks are also due to Ms Louise Dance for a considerable amount of work to ensure that the English language of the report became acceptable.

\begin{figure}
\begin{center}
\resizebox{8cm}{!}{ \includegraphics{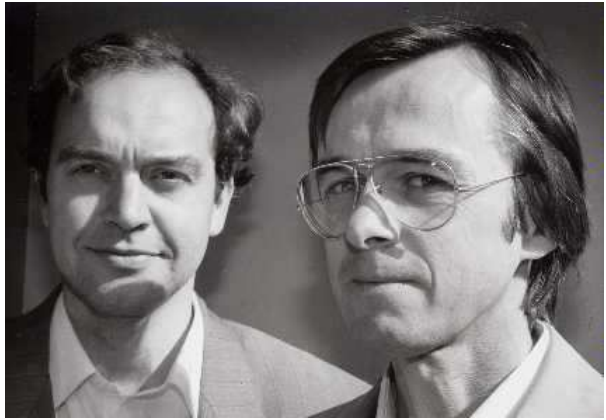}}
\end{center}
\caption[J.-P. Hurni and A.A. Gsponer, circa 1995]{\emph{Jean-Pierre Hurni and Andre A. Gsponer, circa 1995.}}
\label{fig:JPH+AAG}
\end{figure}

\section{Principle and state of the art in calutron technology}
\label{Sec.2.3}

\subsection{Enrichment technologies in perspective}
\label{Sec.2.3.1}

	There are two general methods for producing fissile materials for military or civilian purposes: breeding and enrichment. For a small or developing nation with limited military nuclear ambitions, in the absence of any technical or political obstacle, the cheapest and fastest option is breeding.

	With breeding, a source of neutrons is used to convert a non-fissile material such as U-238 or Th-232 into fissile Pu-239 or U-233. The neutron source is generally a nuclear reactor or, possibly, a more complicated device such as a particle accelerator or (in the future) a fusion reactor \cite{AA.3}.

	In practice, a small fission reactor with a power of a few tens of MW(thermal) is sufficient to breed enough Pu-239 to make one atomic bomb per year. Such a reactor was built in Israel in 1960 and it is now widely accepted that it has been used to produce enough fissile material for several nuclear weapons. A reactor of similar size was used to produce the plutonium for the bomb India exploded in 1974. In 1981, concerned that Iraq could be developing nuclear weapons, an Israeli air raid destroyed the Iraqi 70 MW(th) Tammouz 1 (Osiraq) reactor.

	In enrichment, the concentration of fissile U-235, which is only 0.72 per cent in natural uranium, is increased beyond the normal isotopic concentration. Weapons-grade uranium should be more than 80 per cent pure in U-235. A number of enrichment technologies are available; electromagnetic isotope separation (EMIS), gaseous diffusion, ultracentrifugation, laser isotope separation, plasma isotope separation, etc. Of these technologies, gaseous diffusion is the most mature and the method on which the present industrial production is based. The physical principle of gaseous diffusion is such that plants using this method are necessarily very large and expensive. Since the existing diffusion plants are aging, two major alternatives for large scale commercial production (plasma and laser isotope separation) have been extensively studied. In the USA, it is the atomic vapor laser isotope separation (AVLIS) process that has been selected as the uranium enrichment method of the future \cite{AA.4}. And the plasma separation process (PSP) has been made available for other applications \cite{AA.5}. Both of these processes are technologically highly sophisticated and do not constitute a near term threat for nuclear proliferation in developing countries.

	From the horizontal proliferation point of view, the ultracentrifuge method is now possibly the most attractive technique for building a relatively small enrichment plant --- one which would turn out enough fissile uranium for one or two nuclear weapons a year. This method was used by South Africa and Pakistan to produce the fissile material for their nuclear weapons. Ultracentrifugation requires comparatively little energy and leads to small plants that can be easily concealed. Technologically however, ultracentrifugation is somewhat more sophisticated than electromagnetic enrichment, the technique which was historically the first to be used on an industrial scale and produced the uranium for the Hiroshima bomb.

	Compared with the other enrichment methods, the main advantage of EMIS is that it uses only well known classical technologies (ion sources, vacuum, magnets, etc). Its main disadvantage is that it is not a continuous process. It involves a complicated and labour intensive series of physical and chemical tasks which pose considerable problems during plant operation. As shown by Iraq however, many of these problems can be alleviated by the use of microcomputer control systems.

\subsection{Basic principle and main characteristics of EMIS}
\label{Sec.2.3.2}

	The electromagnetic isotope separation method is based on the principle that ions of the same energy, but of different masses, describe trajectories with different curvatures in a magnetic field.

	At the heart of an EMIS system is an electromagnetic separator which comprises three main parts: a source in which the mixture of isotopes is ionized and the resulting ions merged into a beam which is accelerated to some energy; an analysing magnet providing the field in which the accelerated beam is separated into as many beams as there are isotopes in the original mixture; and a receiver in which different pockets collect the ions from the separated ion beams. The source and the receiver are located in a vacuum tank situated between the pole faces of the electromagnet. Some residual gas is left in the vacuum tank in order to pinch and stabilize the ion beams. The associated chemical operations consist of preparing the feed-material (usually uranium tetrachloride), extracting the enriched material from the receivers (usually made of graphite) and cleaning the vacuum tank for recovering the material lost in the separator by scattering, sputtering and stray beams.

	The most important parameters which characterize an electromagnetic separator are the ion \emph{beam current}  and the \emph{mass separation power}. These two characteristics are antagonistic: a very high isotopic purity can only be achieved at the expense of a low current, which implies a low productivity. In practice, the dividing line between laboratory separators for high precision electromagnetic separation and industrial separators for high productivity is of the order of 1 mA (one milliampere). Here the term ``calutron'' refers to a production electromagnetic separator with a current of 1 mA or more.

	In industrial scale separation, for various technical reasons, there is an upper limit of about 100 mA to the calutron beam current, which in the case of uranium, leads to a maximum enrichment of only 10 to 20\%. There are two consequences. First, since the production of 50 kg of U-235 per year corresponds to a total beam current of over 100 A, there must be at least 1000 separators working in parallel to share the load. Second, in order to achieve a final enrichment of more than 90\%, there must be a second enrichment stage which will require a somewhat smaller number (about 300) lower current but higher resolution separators. Hence, to produce enough U-235 for one atomic bomb a year a very large number of calutrons is required.

\begin{figure}
\begin{center}
\resizebox{8cm}{!}{ \includegraphics{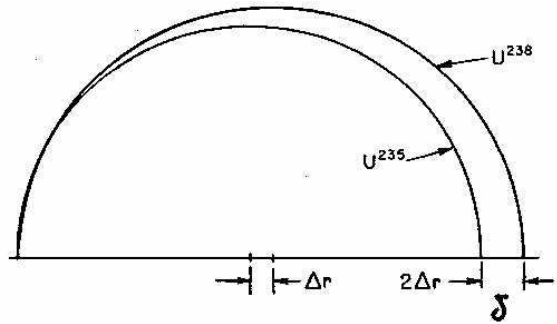}}
\end{center}
\caption[Trajectories of U-235 and U-238 ions.]{\emph{Trajectories of U-235 and U-238 ions in a 180$^{\rm{o}}$ calutron.}}
\label{fig:Fig.1a}
\end{figure}

	The most straight forward design of a calutron makes use of a uniform magnetic field, i.e., a field that is constant through space. In such a field the trajectory of an ion is a circle whose radius is a function of its mass. The basic arrangement consists of placing the source and the receiver in the field in such a way that after half a turn, both of the ion beams are caught by appropriately placed collector pockets (Fig.~\ref{fig:Fig.1a}), one for U-238, another for U-235. This design, often referred to as the \emph{180$^{\rm{o}}$ method}, was brought up from laboratory to industrial scale by E.O. Lawrence during World War Two \cite{AA.6} and now serves as a reference design for comparing other calutron designs.

\begin{figure}
\begin{center}
\resizebox{8cm}{!}{ \includegraphics{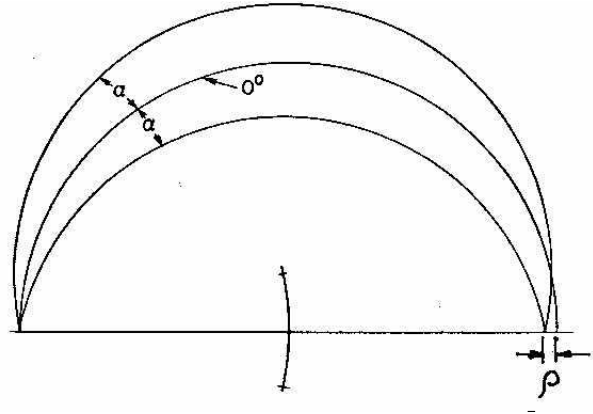}}
\end{center}
\caption[Trajectories of ions of the same mass.]{\emph{Trajectories of ions of the same mass leaving the source at different angles in a 180$^{\rm{o}}$ calutron.}}
\label{fig:Fig.1b}
\end{figure}

	The mass separation power is a function of two parameters: the \emph{dispersion} and the \emph{resolution}.

	The dispersion is the spacing $\delta$  at the receiver between two beams of different masses. In the 180$^{\rm{o}}$ method, the dispersion is given by the formula
\begin{equation}
                  \delta   = R  \frac{m}{M}
\end{equation}
where $M$ is the average beam mass and $m$ the mass difference. Hence, in the separation of natural uranium (i.e., $M = 238$  and $m = 238 - 235 = 3$ ) with a calutron of mean beam radius $R  = 100$ cm,  the mass dispersion is of  $100 \times 3 / 238 = 1.26$ cm.

	The resolution $\rho$ is given by the width of the beams at the receiver. This is a function of many parameters such as the width of the source, image aberrations (error in focusing), scattering on the residual gas, stability of the magnetic field, etc. For high intensity separators, the main contribution to the resolution is from the aberrations due to the angular spread of the initial ion beam. In particular, in the 180$^{\rm{o}}$ method, if ions of the same mass leave the source at different angles, their trajectories do not meet at the same point on the receiver (Fig.~\ref{fig:Fig.1b}). In the small angle approximation, the resolution is then
\begin{equation}
                   \rho   = R  \alpha^2
\end{equation}
where  $\alpha$  is half the opening angle of the initial beam.

	For isotope separation to be possible, the width of the beams at the receiver must be smaller than their separation, i.e.,  $\rho < \delta$  . Hence, from the above two expressions, it is seen that uranium isotope separation with the 180$^{\rm{o}}$ method is only possible with beams of an initial divergence less than 6.4$^{\rm{o}}$ (i.e., $\sqrt{3/238}$ radians), a rather small divergence for a high intensity beam.

	To improve the resolution, and thus to enable separation of high intensity beams, various methods were studied during World War Two. The method finally used in Lawrence's calutrons was to make the magnetic field slightly non-uniform (by introducing specially shaped iron shims in between the pole faces of the magnet) in order to bring the beams to a better focus at the collectors \cite{AA.6}. After the War, in the period 1945-1955, many other methods were tested to improve the performance of electromagnetic isotope separation, a problem that is directly connected to the improvement of several closely related techniques such as mass spectroscopy and particle acceleration. In general, these methods tried to make use of non-uniform magnetic fields in order to improve the dispersion and/or the resolution, and thus to increase the separation power which may be characterized by the ratio  $\delta / \rho$ .

	An important example of an improved calutron design was invented in 1946 by the Swedish physicists Nils Svartholm and Kai Siegbahn \cite{AA.7} who were studying the general problem of momentum spectroscopy in a non-uniform field. They found that if the field decreases with radius going outward from the center of the orbits, there is focusing in the direction parallel with the magnetic field as well as normal to the field --- the focusing is stigmatic. (If the field is uniform, there is focusing only in the radial direction.) In particular, in a magnet of rotational symmetry, if the field falls off in inverse proportion to the square root of the radius, maximum separation and focusing is obtained when the ion's trajectory through the field makes an angle of 255$^{\rm{o}}$ between the source and the receiver. The Svartholm-Siegbahn method is therefore often called the \emph{255$^{\rm{o}}$ method}. In this method, the dispersion and resolution are given by 
\begin{equation}
                    \delta   = 2 R  \frac{m}{M}
\end{equation}
and
\begin{equation}
                     \rho   = \frac{4}{3} R  \alpha^2
\end{equation}
respectively. Thus, compared with the 180$^{\rm{o}}$ method, while the dispersion is better by a factor of 2, the resolution is worse by a factor 4/3. The separation power is therefore only 1.5 times better. There is however, a further advantage in the Svartholm-Siegbahn method: because of the double-focusing effect of the non-uniform magnetic field, it is possible to use beams of higher intensity.

	This discussion of aspects of the 255$^{\rm{o}}$ method is typical of the kind of improvements possible over the standard 180$^{\rm{o}}$ method: no break through is possible and only factor of two improvements are feasible. Nevertheless, by combining several such factors, substantial progress has been made since World War Two.

\subsection{Sources of information on EMIS and calutrons}
\label{Sec.2.3.3}

	In comparison with other enrichment technologies, and more generally with the secrecy which surrounds the construction of nuclear weapons, essentially all the information concerning EMIS has been declassified since World War Two. In particular, the technical details of the American calutron program have been declassified in two steps. First, a series of fundamental research reports appeared in division I (Electromagnetic Separation Project) of the U.S. National Nuclear Energy Series between 1949 and 1952. Second, a collection of specialized reports were declassified and started to appear towards the end of 1955 (Technical Information Service, Oak Ridge, Reports number TID-5210 to TID-5219). At the same time, many patents relating to the construction details of crucial calutron components were filed in the United States. This was the era of the optimistic ``Atoms for Peace'' program and it was believed that, apart from their scientific applications and potential to produce small amounts of separated isotopes for industrial and medical use, no country would ever turn to EMIS to produce the relatively large amounts of enriched material needed for atomic weapons.

	The main practical reason for declassifying information on calutrons is that electromagnetic isotope separation involves no scientific or technological principle which could be effectively protected by a patent or kept secret. The principles of EMIS are common to several neighboring techniques which include mass spectroscopy, momentum spectrometry, electron microscopy and circular accelerator technology. EMIS is also a very important tool for fundamental research in nuclear physics (where it is essential for separating the various isotopes of a natural element in order to study their properties). All major components of an EMIS system (ion sources, magnets, vacuum system, high voltage power supplies etc) are widely used in all research laboratories which use low or high energy particle accelerators to study nuclear reactions or the interactions of elementary particles.

	As a result, after the declassification of the information from the Manhattan project in the late 1940s and mid 1950s, most of the progress in EMIS technology is now reported in open scientific literature. Important sources of information are the proceedings of the twelve ``EMIS conferences'' which have taken place in Europe, the United States, Israel and Japan between 1955 and 1992. The proceedings of these ``EMIS conferences'' are generally published in the journal \emph{Nuclear instruments and methods}, the editor of which is Kai Siegbahn, the co-inventor of the 255$^{\rm{o}}$ method.

	Because of the intrinsic simplicity of EMIS technology, there has been no major break-through since the Manhatten project and those improvements made between 1945 and 1955. For this reason, most of the contributions at recent EMIS conferences are concerned with perfecting the many low-current separators used world-wide in fundamental or applied research. An increasing number of papers deal with the so-called \emph{on-line} EMIS facilities, which are complex research instruments enabling the separation of very short-lived isotopes, and comparatively fewer papers with the standard \emph{off-line} facilities. Exceptions to this are a few papers presented at the last two conferences. These gave information on several old Russian and Chinese high-current off-line EMIS facilities on which previously there was little information in open scientific literature.

\subsection{EMIS in the United States}
\label{Sec.2.3.4}

	At the beginning of the Manhattan project, the method that would ultimately become practical for producing the fissile material for an atomic bomb was unknown. It was clear however, that any method that would process material in bulk would certainly be more efficient than any method, such as electromagnetic enrichment, that relies on passing small amounts of material through some kind of an analyser. For this reason, reactor production of plutonium, or gaseous diffusion of uranium, were expected to be the best methods. By mid-1942 no reactor worked and it was not clear whether an industrial scale gaseous diffusion plant would ever work. On November 5, 1942, General Groves decided that the design of the prototype electromagnetic separator, built by E.O. Lawrence at the University of California (hence the name ``calutron''), would be frozen and that a plant called Y-12, with a capacity of about 100 grams per day would be created in Oak Ridge \cite{AA.8}.

	In Lawrence's original design, called ``alpha,'' the evacuated tank containing the source and collector assembly was placed between the circular pole pieces of a large magnet originally intended for a cyclotron. This 184-inch magnet was completed in 1942 as a ``mechanism of warfare,'' with an A-1-a priority for steel \cite{AA.8}. In Oak Ridge there were many calutrons to be put into operation. A simple rectangular dipole magnet was designed and the calutrons were assembled into ovals comprising 96 magnets alternating with 96 calutron tanks. This had the advantage of forming a closed magnetic loop and minimizing magnetic losses and steel consumption. In the end there were 9 such ``racetracks,'' making 864 alpha calutrons in total.

	In the alpha calutrons an attempt was made to get maximum possible output per separator, and one method of doing this was to use equipment of a fairly large size. The models adopted for production had a source-collector distance of 244 cm. The lateral width of the ion beam is limited by the dimension of the tank in the direction of the magnetic field, which was of the order of 3500 gauss. The tanks had an inside dimension of 61 cm and the lateral width of the ion beams was about 50 cm. The radial width of the beam varied from about 1 cm near the receiver and acceleration system to approximately 60 cm in the 90$^{\rm{o}}$ position \cite{AA.6}.

	The output of the alpha calutrons was enriched only to about 15\%. To produce weapons-grade uranium, $8 \times 36 = 288$ improved calutrons were built to provide a second enrichment stage called ``beta.'' The beta tank equipment, including sources and collectors, was made with linear dimensions just half the corresponding alpha dimensions. Beta calutrons worked with a lower current and emphasized recovery, not only of the further enriched output but also of the already enriched feed. Between January and June 1945, using feed from the alpha calutrons and the small output of the gaseous diffusion plant, the Y-12 production was of 6 kg weapons-grade uranium per month \cite[p.494]{AA.9}. Neglecting losses, such an output means that the average alpha and beta calutron currents were of the order of 150 and 20 mA, respectively.

	By mid-1945, the gaseous diffusion process had demonstrated that there was a cheaper way of obtaining U-235 and soon after the cessation of hostilities, the electromagnetic plant was declared obsolete and the shutdown of the facilities was initiated. Only two of the nine buildings housing calutrons were retained intact --- the pilot plant with two alpha and two beta separators and a production building containing 72 beta separators \cite{AA.10}. From that time on, these remaining calutrons were used for the production of enriched stable isotopes (embracing more then 250 different nuclidic species), and selected radioactive isotopes, for use in military, scientific, industrial and medical applications.

	In order to increase the versatility of the Oak Ridge facility many improvements have been made over the years. In particular, to make it more suitable for multi-element enrichment, the original magnetic configuration (which linked the 72 beta calutrons into two sets of 36 separators in a common magnetic field) was modified. Installing 5 magnetic shunts resulted in the subdivision of the track into seven independent groups of calutrons. Moreover, six beta calutrons were modified into 255$^{\rm{o}}$ inhomogeneous magnetic-field separators to provide a factor of two enhancement in dispersion. In one calutron, the source and collector have been made external to the analysing magnet, providing a separation power roughly ten times that of the standard 180$^{\rm{o}}$ calutron, with a correspondingly lower ion throughput \cite{AA.11,AA.12}. 

	Apart from the large calutrons in Oak Ridge, many electromagnetic separators have been built in various universities and research laboratories. In 1981, Los Alamos National Laboratory and Lawrence Livermore National Laboratory decided to upgrade the quality of their isotope separation facilities to achieve better resolution and dispersion \cite{AA.13}. The project was named LLORIS, ``Los Alamos, Livermore, Orsay Isotope Separator,'' the design was the result of a collaboration with the Laboratoire Ren\'e Bernas of Orsay, France. A special feature of this 0.5 mA ion beam current separator is the use of a magnet with an adjustable quadrupole component. A total of three units have been constructed, two for Los Alamos and one for Livermore.

	For future large scale enrichment requirements, the plasma separation process (PSP) is expected to provide the best option \cite{AA.12}. ``In comparison with the calutron, the PSP has a lower enrichment factor and is capable of enriching only one isotope per pass. However, it is a very high throughput machine that could augment the present enrichment program by making available large quantities of material at medium enrichments and by providing pre-enriched feed-material for the calutron'' \cite{AA.11}. The main component of a PSP system is a large superconducting magnet \cite{AA.14}.

\subsection{EMIS in the Soviet Union}
\label{Sec.2.3.5}

	The first documented history of the Soviet atomic bomb has only recently been published \cite{AA.15}. The details of the enrichment program, and more particularly those concerning the diffusion and electromagnetic processes, are based on a document \cite{AA.16} by Igor Golovin, who worked closely with Igor Kurchatov in the 1950s and later wrote his biography.

	Like the United States, the Soviet Union worked from the beginning on all possible enrichment methods. Work on electromagnetic enrichment started at the Kurchatov Institute of Atomic Energy in 1943 when the institute was founded \cite{AA.17}, and was directed by L.A. Artsimovitch, I.N. Golovin and G.Ia. Shchepkin. Early in 1946, sites were selected for the gaseous diffusion plant, Sverdlovsk-44, central Urals, and the electromagnetic plant, Sverdlovsk-45, northern Urals. The first tests of the electromagnetic process were made in 1946 with the help of an electromagnet from Germany.

	Things did not go smoothly for either the electromagnetic or the diffusion process. Artsimovich was unable to obtain ion sources with the required current. The problems with the diffusion process were even more severe. Construction of the production plant, which had a planned output of one kilogram U-235 per day, was completed in 1948. But, in 1949, the year of the first Soviet plutonium atomic bomb, the degree of uranium enrichment obtained was only 40 per cent. This 40 per cent enriched uranium was brought to Sverdlovsk-45, and after a month of round-the-clock work, Artsimovich and his group, using their experimental apparatus, managed to produce 400 grams of uranium enriched to 92--98 per cent \cite[p.191]{AA.15}, \cite[p.20]{AA.16} (Neglecting losses and down-times, this corresponds to a total effective ion beam current of about 150 mA).

	With the help of German scientists, the problems at the diffusion plant were solved at the end of 1950. At about the same time the priority for electromagnetic separation was reduced. It was decided not to build a large-scale electromagnetic plant, and the small plant at Sverdlovsk-45, which had already been completed, was no longer treated as a top-priority project.

	As a reaction to Eisenhower's ``Atoms for Peace'' proposal in 1953, a large amount of information was published on the Soviet achievement in the nuclear domain. In particular, at the 1958 Atoms for Peace Conference in Geneva, a paper on electromagnetic isotope separation gave many details on more than a dozen large electromagnetic separators of various types, then operational in the Soviet Union \cite{AA.18}. Of special interest were the largest separators which had a two-story, four-tank design within a single 400 tons magnet. Such characteristics suggest that these calutrons were the prototypes for the Sverdlovsk-45 plant. Photographs of these calutrons can be seen in the paper \cite[Fig.2]{AA.18} and in a booklet distributed at the 1958 Atoms for Peace exhibit \cite[p.53]{AA.19}.

	In 1957, L.A. Artsimovitch and others published a particularly elegant example of a high-resolution calutron using the 255$^{\rm{o}}$ method \cite{AA.20}. (In fact, for some technical reasons, the focusing angle was 225$^{\rm{o}}$ instead of the optimum 255$^{\rm{o}}$ value.) This design used a magnetic field of rotational symmetry and provided excellent single-pass enrichment for heavy elements such as uranium or plutonium with ion beam currents of the order of 10--15 mA.

	In 1969, an electromagnetic separator, S-2, especially designed for high efficiency separation of isotopes of the heavy radioactive elements which have a small relative mass difference, was built in Arzamas-16 \cite{AA.21}, the Soviet equivalent of the Los Alamos laboratory. The magnet is C-shaped and the pole tips have a slope creating a field which decreases in inverse proportion to the radius of the beam trajectory. The ion source provides a beam of up to 10 mA. Although Arzamas-16 was originally a Soviet military laboratory, isotopically pure samples in the form of layers, targets, solutions or other forms may now be obtained elsewhere on a contractual basis \cite{AA.22}.

	The future of enrichment in the USSR was discussed by A. Tikhomirov at the EMIS-12 conference \cite{AA.17}. The centrifugal method, which was considerably developed in the USSR and led to a practical plant in 1959 at Sverdlovsk-44, was presented as a good option for the large scale enrichment of medium weight isotopes (such as germanium for semiconductor applications). However, the plasma separation process (which in contrast to the centrifugal one does not require volatile compounds) is seen as an important option for the future. In particular, it can be compared with the electromagnetic method in universality, and the centrifugal method for the productivity. A similar conclusion had earlier been reached by J.G. Tracy in an assessment of the future of enrichment from an American perspective \cite{AA.12}.

\subsection{EMIS in the United Kingdom}
\label{Sec.2.3.6}

	A comprehensive review of the state of the art in electromagnetic enrichment in the 1950s, giving details on the British and French early calutron efforts (as well as some details on similar efforts in other European countries, South Africa and Japan), was published in 1958 \cite{AA.23}. The British project was started in 1945 by some of the British scientists who had been working for the Manhattan project. They designed units similar to those they had been working on in the U.S.A. The first separator built at Harwell (south of Oxford), completed early in 1950, was a 24-inch beam-radius separator similar to the American beta calutron.

	The main purpose of this large capacity 180$^{\rm{o}}$ machine was to produce material in sufficient quantity rather than to obtain very high enrichment. Due to the success of the Capenhurst diffusion plant, which came into operation between 1954 and 1957, there was no need to further develop the high production calutron technology. Both scientific and military applications however, demanded smaller quantities of highly enriched materials, for which a beam current of the order of 1 mA is sufficient.

	A disadvantage of the 180$^{\rm{o}}$ type of machine is the restriction placed on the source and receiver design because of the cramped space and the magnetic fields in the region of these units. In order to obtain high enrichment factors, a solution is to consider a ``sector machine'' in which the source and collector are external to the magnet. This led to the construction in Harwell of HERMES, ``Heavy Elements and Radioactive Material Electromagnetic Separator,'' in which the ion beam trajectory made a 90$^{\rm{o}}$ angle at a 48-inch radius in the magnetic field \cite[150--165]{AA.23}.

	Typical applications of HERMES comprise the separation of plutonium isotopes \cite[p.165]{AA.23}. At Oak Ridge, U.S.A., a beta calutron had been modified for the same purpose while a smaller version had been constructed for use as a second stage machine \cite[p.165]{AA.23}. In the U.S.S.R., the 225$^{\rm{o}}$ machine of Artsimovich had been specially designed for the separation of radioactive heavy elements such as plutonium \cite{AA.20}.

\subsection{EMIS in France}
\label{Sec.2.3.7}

	In 1940, Alfred O. Nier at the University of Minnesota in collaboration with Booth, Dunning and Gross at Columbia University, were the first to use the electromagnetic method to separate the uranium isotopes in order to investigate their fission properties \cite{AA.24}. Ren\'e H. Bernas, who studied physics at the University of Minnestota and was to become the leader of electromagnetic separation in France, developed a high current version of the type of separators ordinarily used in Nier's laboratory \cite{AA.25}. In France in 1952, Bernas built a 60$^{\rm{o}}$ sector machine with an ion beam radius of 50 cm \cite[p.82--95]{AA.23}. This separator was installed in a laboratory of the Commissariat \`a l'Energie Atomique (CEA), Saclay. In September 1955, out of 290 hours of collection time, 200 were dedicated to uranium with a production of 1.4 gram of U-235 \cite[p.93]{AA.23}, corresponding to an effective beam current of 0.1 mA. The success of the French machine was a positive factor in the British decision to build HERMES.

	Contrary to the United States, England and the Soviet Union, there has been no attempt in France to pursue simultaneously the uranium and plutonium routes to atomic weapons. Indeed, in 1952, France chose plutonium as the priority for its own nuclear weapons program. Therefore, the construction of the first French uranium enrichment plant, using the gaseous diffusion process, started in 1960 only. The Pierrelatte enrichment plant became operational in 1964 and was completed in 1967. It is interesting that at about the same time some effort was made to build large size calutrons. In effect, a 255$^{\rm{o}}$ double focusing calutron operating with a maximum ion beam current of about 150 mA, was built in Saclay between 1962 and 1965 to produce isotopes in commercial quantities \cite{AA.26}. In the construction of this device, several field configurations were tested. Similarly to the Oak Ridge calutrons \cite{AA.8,AA.10}, this 24-inch separator was based on an ordinary rectangular magnet (which normally creates a uniform field). The required non-uniform field was obtained by introducing suitably shaped iron ``shims'' between the pole pieces \cite{AA.27}. While experiments with linear shims (corresponding to the 180$^{\rm{o}}$ method) enabled the use of ion beam currents with maximum intensities of the order of 100 mA, circular shims (corresponding to the 255$^{\rm{o}}$ method) enabled the use of currents of the order of 200 mA, clearly demonstrating the superiority of 255$^{\rm{o}}$ method. An important aspect of the publications describing this work is that they provide a good summary of the mid-sixties' state of the art in industrial-scale EMIS technology \cite{AA.26,AA.27}. In particular, they show the usefulness of computer programs to study complicated non-uniform field systems and the simultaneous focusing of two separate ion beams. 

	In France, as in the other countries which have mastered the technique of gaseous diffusion, calutrons have not become a means for large scale production of enriched uranium. Over the years however, several large separators were built in order to suit various other needs. For example, at the ``Laboratoire Ren\'e Bernas,'' in Orsay, two separators, SIDONIE and PARIS, were built to prepare extremely enriched isotopes \cite{AA.28}. PARSIFAL, a separator providing safe handling of radioactive materials, was built at the military laboratory of Bruy\`ere le Chatel to separate specific isotopes from a strong radioactive background \cite{AA.29}. Typical applications are the purification of very small quantities of isotopes, with half-lives greater than 12 days, that are produced in special monitoring targets exposed to the neutron flux of a reactor or nuclear explosion. To increase the availability of PARSIFAL, a copy of the source block was built in 1990, so that secondary operations such as source-outgassing could be performed at the same time as actual separation using the other source \cite{AA.30}.

	Expertise gained by French scientists in the design and construction of high resolution electromagnetic separators led to a collaboration with the EMIS specialists at Los Alamos in an effort to upgrade the isotope separation facilities at the Los Alamos and Livermore nuclear weapons laboratories \cite{AA.13}. This is a typical example of the collaboration/competition relationship characterizing the arms race between France and the United States since the early 1980s, and even more so since the break up of the former Soviet Union. Where each state in relation to the other takes the place of the Soviet Union as the challenger in the development of the most sophisticated military technology.

	In order to replace its aging gaseous diffusion facilities, France has perfected the AVLIS process (SILVA in French) and expects to build an industrial scale laser enrichment facility by 2010.

\subsection{EMIS in China}
\label{Sec.2.3.8}

	Details on Chinese calutrons began to appear in open scientific literature in the proceedings of the EMIS-10 conference \cite{AA.31}. The main characteristics of the four calutrons built at the China Institute of Atomic Energy, Beijing, are given in these articles, and a summary of the elements processed from 1965 to 1986 was given at EMIS-11 conference \cite{AA.32}.

	A laboratory-scale 90$^{\rm{o}}$ separator (F-1) was installed in 1962. Since then two 180$^{\rm{o}}$ production separators (F-2 and F-3) have been installed in 1965 and 1968 respectively, and a 255$^{\rm{o}}$ double-focusing separator (F-4) was constructed in 1980. A comparison of the characteristics of F-2 and F-3 with those given in 1958 for the Soviet calutrons, show that they are in fact identical to the two large single-tank calutrons (with 220 and 280 tons of magnet weight) previously built in the USSR \cite{AA.18}. The summary of the elements processed shows that the first elements to be isotopically separated were lithium and uranium, clearly indicating that the applications were connected with the Chinese atomic weapons program. These separators are still in use. A computer-aided inspection system was developed as a first step in designing a computer control system to improve the quality of the products and reduce operator supervision \cite{AA.33}.

	Since Soviet aid to China in the 1950s and 1960s included help in the construction of a gaseous diffusion plant, the Chinese calutrons have not played a direct role in the large-scale production of enriched uranium. Recently, the Chinese know-how in calutron technology has become an element of concern for the proliferation of nuclear weapons. Indeed, the agreement that China signed with Iran in 1990 includes the shipment of several million dollars worth of calutron equipment \cite{AA.34}.

\subsection{EMIS in India}
\label{Sec.2.3.9}

	A laboratory-scale 255$^{\rm{o}}$ isotope separator, designed after the spectrometer originally built by Svartholm and Siegbahn \cite{AA.7}, was completed in 1958 at the Saha Institute of Nuclear Physics, Calcutta \cite{AA.35}. The construction of this EMIS device took place at the time of the construction of the research reactor supplied by Canada. This reactor began operating in 1960 and was used to produce the plutonium for the 1974 Indian nuclear explosion.

	This separator was modified in the mid-1980s in order to be used for off-line separation of short-lived isotopes produced using the variable-energy cyclotron of the Bhabha Atomic Research Center \cite{AA.36}.

\subsection{EMIS in Israel}
\label{Sec.2.3.10}

	Israeli activities in the domain of EMIS were described at the EMIS-8 conference of 1973 and further explained at the EMIS-9 conference which took place at Kiryat Anavim, Israel, May 10-13, 1976. The two basic instruments developed in Israel for this purpose, the SOLIS and MEIRA separators, are operating at the Soreq Nuclear Research Center.

	The Soreq on-line isotope separator (SOLIS) is connected to a fission source placed at an external beam port of the Soreq Research Reactor. It is a research instrument dedicated to the study of short lived isotopes produced in the fission of U-235 \cite{AA.37}.

	The separator MEIRA is a high output electromagnetic isotope separator whose typical source oven charges are one hundred grams and beam currents are of the order of 50 mA. ``The aim of the MEIRA development was not to arrive at the highest quality mass analysis, but to invest the minimum development, in both equipment and modifications, in order to arrive reliably at the required analysis capability \cite{AA.38}.'' This capability is typically the production of high purity tellurium-124, as is required for the production of iodine-123 for radiopharmaceutical applications \cite{AA.39}. The first objective of MEIRA however, has been the ``systematic experimental investigations (leading to) an understanding of the basic phenomena in the separation process \cite{AA.39}.'' The depth of this understanding can be measured by the fact that tellurium-124 at a purity in excess of 99\% has been obtained in a single pass with MEIRA \cite{AA.39}. Following this success, an automatic system, allowing for unattended operation after the initial adjustment of the separation conditions, was developed \cite{AA.40}. The nuclear proliferation significance of the MEIRA separator is that it has the capability of efficient separation of plutonium isotopes, and shows that Israel has mastered the technological challenge of building high productivity calutrons for uranium enrichment or plutonium purification.

	The key man behind the development of high current EMIS technology in Israel is I. Chavet who received his PhD in France while working on the Orsay separator built by R. Bernas \cite{AA.40}. The Orsay separator (i.e., a sector-type separator derived from the one Bernas built in Saclay in 1953) is the model on which MEIRA was built.

\subsection{EMIS in Japan}
\label{Sec.2.3.11}

	Until the EMIS-11 conference in 1986, at which it was agreed that the following EMIS conference would be held in Japan, there was little information published on Japanese calutron activities in non-Japanese publications. The participation in EMIS conferences by Japanese scientists was minimal. At EMIS-11, as an introduction to a review paper on the on-line separator facilities in Japan, M. Fujioka presented a brief history of off-line isotope separators in Japan \cite{AA.41}. As early as 1941 an isotope separator was used for mass separation and identification of a radioactive isotope, sodium-24. This study was made as a test experiment for a larger research project concerning fission products. Construction of a larger isotope separator of 180$^{\rm{o}}$ deflection for such a purpose was started but interrupted due to World War Two. At that time there were five existing cyclotrons in Japan. They were seized by the Americans on November 20, 1945, and totally destroyed \cite{AA.42}. According to a leading Japanese physicist, a student during the War, Japan's atomic bomb effort focussed on enrichment. Two approaches were pursued, the electromagnetic separation of uranium-235 and separation by a thermal diffusion process \cite{AA.43}. The thermal process was abandoned before the end of the war and, according to the Japanese physicist, ``... if we spent 100 times more in research effort, we could have developed the bomb in one year.''

	The construction of off-line separators (mainly for stable isotopes) started again ten years after the War. Two small separators were built in 1955. Then followed the construction of four separators in succession. All four are of the Bernas or Saclay type, and were built in 1956, 1958, 1959 and 1961. The last separator, which is the only isotope separator constructed for off-line use that is still working today, has been used up to the present mainly for implantation experiments using stable as well as radioactive materials \cite{AA.41}.

	EMIS-12 was held at Sendai, Japan, 2-6 September 1991. Out of 149 registered attendees, 91 were from Japan and 58 were from elsewhere. There were several contributions regarding on-line separation and related technical problems, a subject on which Japanese scientists have developed considerable expertise \cite{AA.41}. There were also four contributions concerning off-line isotope separation dealing with ion cyclotron resonance and plasma separation processes \cite{AA.44}. These later contributions are an indication that, besides laser isotope separation, Japan might be interested in new enrichment technologies which have the capability of separating in industrial quantities, all kinds of isotopes, including uranium or plutonium. In effect, contrary to laser isotope separation, which may require different types of lasers for different isotopes, the plasma separation process is much more flexible because all parameters are continuously adjustable.

\subsection{EMIS in other countries and at CERN}
\label{Sec.2.3.12}

	In the previous sections, EMIS activities have been reviewed in the five nuclear weapon states and India, Israel and Japan. Similar activities, at one level or another, for either scientific or industrial purposes, are under way in many industrialized countries, especially in Europe. Since the previous coverage includes most of the countries (except Germany) which have significant nuclear activities, they will not be reviewed in detail.

	A special case has to be made for CERN, the European nuclear research center in Geneva, Switzerland. When CERN was created, in 1954, nuclear physicists and accelerator specialists from all over Europe came to Geneva and played an essential role in the highly successful development of the laboratory. These people were to become staff members as machine designers or researchers, or to become users of the CERN facilities while remaining attached to their home university. Let us just mention two people with direct previous experience in off-line and on-line isotope separation. C.J. Zilverschoon, who's PhD thesis (University of Amsterdam, 1954) was on the construction of the ``Amsterdam Separator,'' originally designed as a production machine and later used for more academic studies \cite[p.119]{AA.23}, was to join the CERN staff and become a leading accelerator scientist. And O. Kofoed-Hansen who, in 1950, with K.O. Nielsen, was to perform the first experiment to make use of the beams of short-lived radioactive nuclei produced by an on-line electromagnetic separator \cite{AA.45}. Kofoed-Hansen became a major advocate and user of the worlds largest on-line separator, the Isotope Separator On-Line (ISOLDE), built at CERN in 1967 \cite{AA.46}.

	While CERN's mission was clearly defined in 1952 as a laboratory for fundamental research in high energy nuclear physics, using the various large accelerators to be built in Geneva, it was also to become a centre of excellence in many specialized technologies with considerable industrial or military potential \cite{AA.47}. That this was clear from the beginning is indicated by the explicit mentioning of ``isotope separation and beryllium engineering,'' as examples of specialized technologies in which CERN had to collaborate with external institutions \cite{AA.48}. As a result, in time, CERN became a world leading institution, not just in pure science, but also in many advanced technologies of importance to various nuclear activities including electromagnetic separation of isotopes, together with a tradition of openness and international collaboration which was to make it an ideal place to acquire detailed information on such technologies.

\section{Iraq's calutron program}
\label{Sec.2.4}

\subsection{Spring 1979 at CERN}
\label{Sec.2.4.1}

\begin{figure}
\begin{center}
\resizebox{12cm}{!}{ \includegraphics{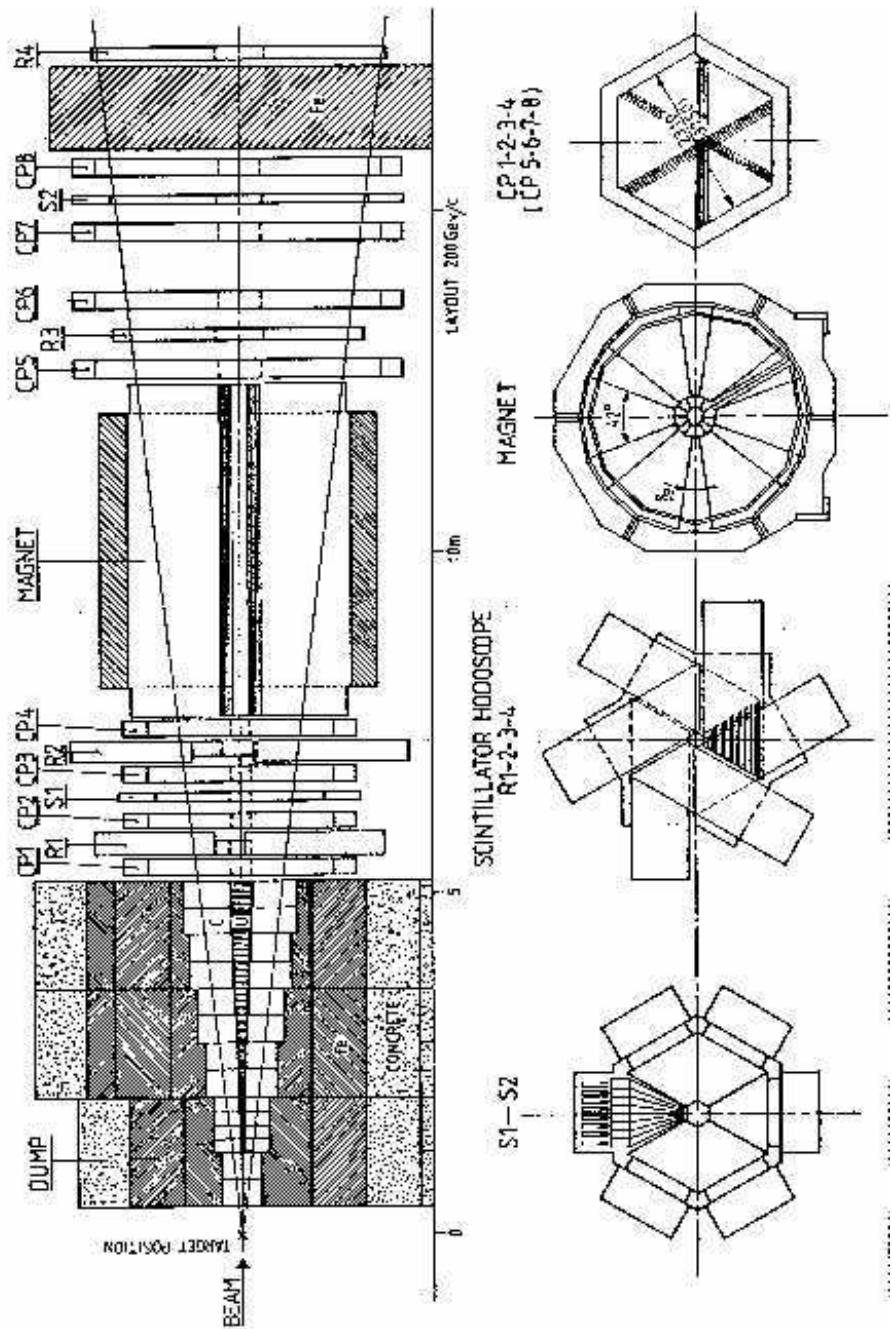}}
\end{center}
\caption[Spectrometer of experiment NA10.]{\emph{Spectrometer of experiment NA10. The outer diameter of the magnet is 410 cm.}}
\label{fig:Fig.2}
\end{figure}

	In 1979 Andre Gsponer was at CERN working on an experiment, NA10, designed to measure, with good statistics and good mass resolution, the production of high-mass muon pairs by pions. The apparatus (Fig.~\ref{fig:Fig.2}) consisted of a beam dump followed by a high-resolution spectrometer whose central piece was a large toroidal magnet (for a detailed description of the apparatus see reference \cite{AA.49}). In the experiment, Gsponer was responsible for the data acquisition system, while Klaus Freudenreich was the physicist in charge of the magnet. In Spring of 1979, Freudenreich informed Gsponer that he had recently been visited by an Iraqi engineer wanting to know everything about the magnet, including all sorts of construction details. As a justification for his interest, he claimed that he was motivated by the possibility of using such a magnet for storing electromagnetic energy. But, for such an application, a superconducting magnet would have been necessary, and the NA10 magnet was a conventional one. This contradiction prompted the discussion between Freudenreich and Gsponer.

	While the NA10 magnet was non-superconducting, it was a very special one, and unique in the world at least because of its size. In particular, it had the property of maximizing the amount of magnetized air (through which high energy particles can pass with minimum disturbance), while minimizing the amount of steel. Since the magnet had an axial symmetry, its construction had required the resolution of a number of difficult engineering problems. It was well known to all physicists of the NA10 collaboration that Mario Morpurgo, the designer of the NA10 magnet and one of the world's greatest specialists of magnet technology, considered the NA10 magnet as one of his masterpieces.

	Gsponer knew that during World War Two, in the Oak Ridge Y-12 plant, there were hundreds of calutrons working in parallel. Each of them consisted of an evacuated separation tank placed in a magnetic field. In Y-12, the magnets providing the field and the calutron tanks were assembled into ovals comprising 96 calutrons: this had the advantage of forming a closed magnetic loop and of minimizing magnetic losses and steel consumption. If this optimization process is pushed to its limit, the result is a configuration which looks like a cut through an orange, with one slice made out of steel, and the next one empty to contain a calutron tank. This is exactly what the NA10 magnet looked like (see lower part of Fig.~\ref{fig:Fig.2}), with room for six calutron tanks and a clever design calculated to use as little steel as possible.

	The conclusion was that the Iraqi engineer was most probably interested in a magnet for electromagnetic isotope separation. Due to the nature of the technical questions asked by the engineer, it was quite possible that Iraq, at the time, was already comparing the engineering problems of various options for the construction of an industrial scale EMIS plant. According to Freudenreich however, its seems that the Iraqi engineer did not get access to the engineering drawings, nor to the specific processes developed by CERN for the construction of the NA10 magnet.

\subsection{Jafar Dhia Jafar and the origin of Iraq's calutron program}
\label{Sec.2.4.2}

\begin{figure}
\begin{center}
\resizebox{12cm}{!}{ \includegraphics{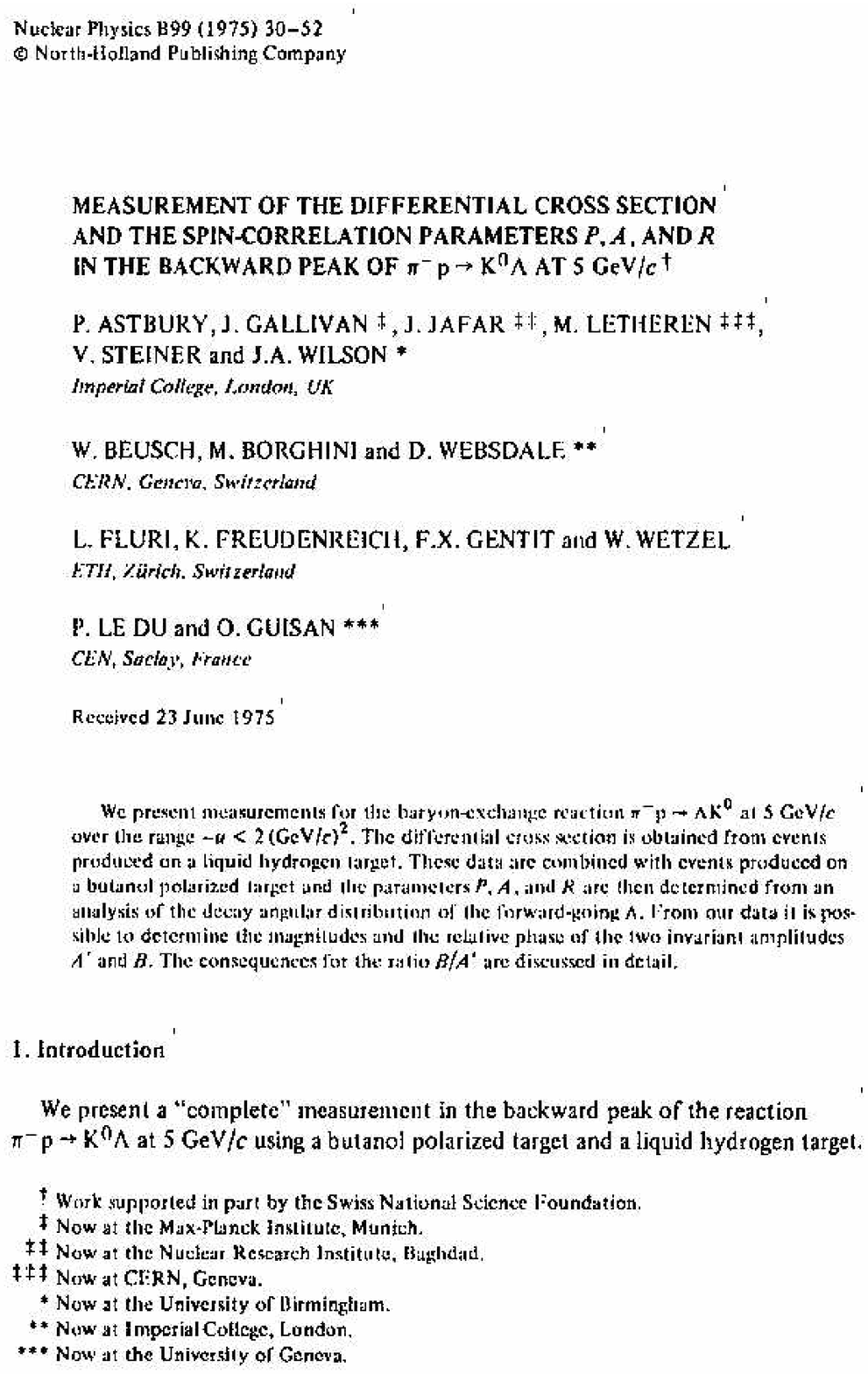}}
\end{center}
\caption[Front page of reference \cite{AA.49}.]{\emph{Front page of reference \cite{AA.49}.}}
\label{fig:Fig.3a}
\end{figure}

\begin{figure}
\begin{center}
\resizebox{12cm}{!}{ \includegraphics{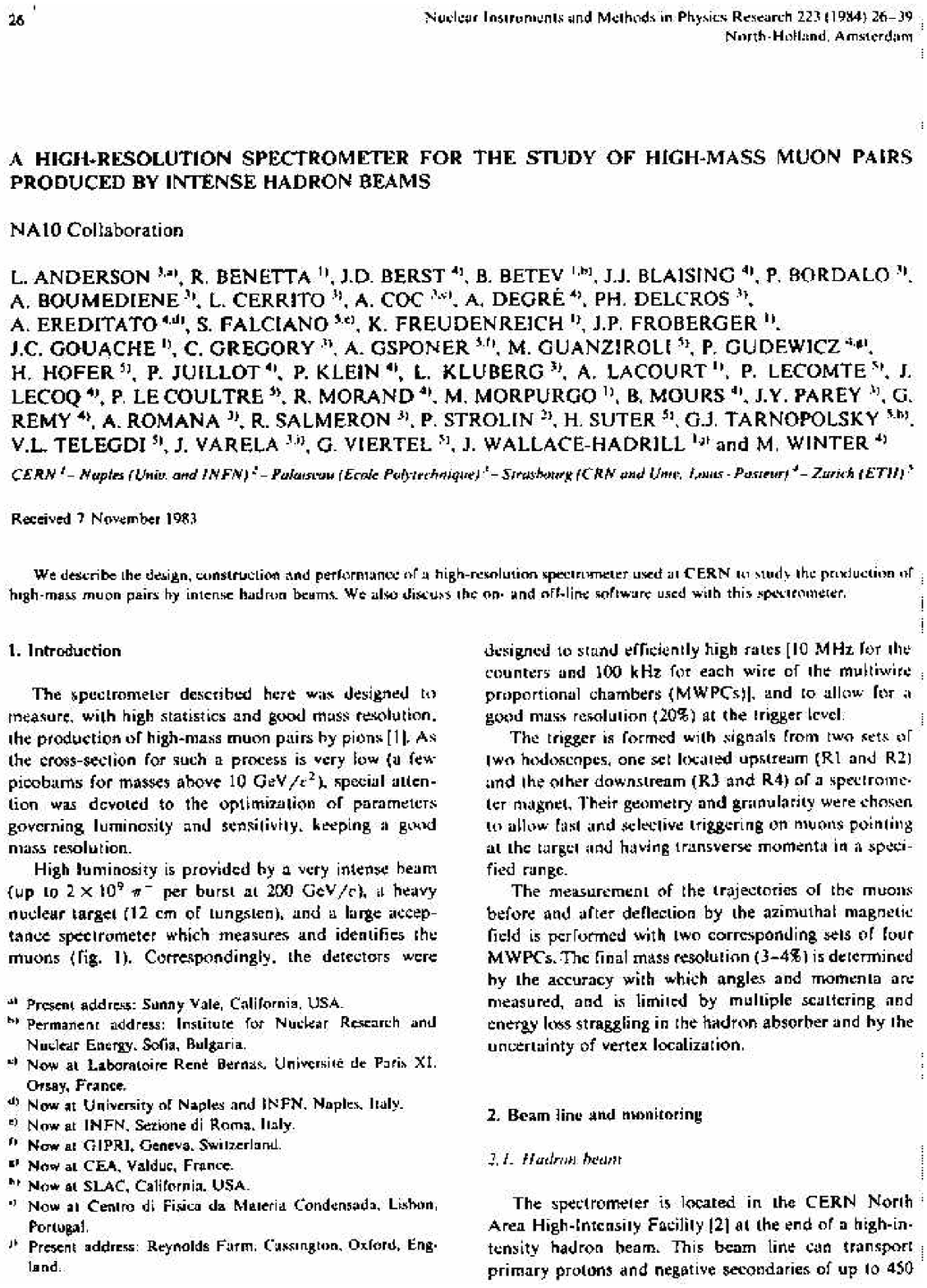}}
\end{center}
\caption[Front page of reference \cite{AA.50}.]{\emph{Front page of reference \cite{AA.50}.}}
\label{fig:Fig.3b}
\end{figure}

	When the Iraqi engineer came to CERN to gather technical information on the NA10 magnet, he introduced himself to Freudenreich as having been sent by Jafar Dhia Jafar, an Iraqi physicist who had worked with Freudenreich at CERN in the 1970s. Jafar who was to become the head of Iraq's atomic bomb program, was trained as a high energy physicist at the University of Birmingham and at Imperial College, London. Between 1967 and 1976 he published 12 papers on various high energy physics experiments, first at the Birmingham synchrotron and later at CERN. The results of the experiment on which Jafar had worked with Freudenreich were published in 1975, at which time Jafar was back in Iraq and working at the Nuclear Research Institute, Baghdad \cite{AA.50}. (The connections between Jafar and Freudenreich, and later those between Freudenreich and Gsponer, are illustrated in Figs.~\ref{fig:Fig.3a} and \ref{fig:Fig.3b}, in which the front pages of references \cite{AA.49} and \cite{AA.50} are reproduced.)

	The return of Jafar to Baghdad in 1975 coincided with the start of a government drive to recruit nuclear scientists \cite{AA.51}. By 1979, Jafar became Vice Chairman of the Iraq Atomic Energy Commission and was responsible for dealing with the French on Osirak, the French nuclear reactor under construction in Iraq. On April 7, 1979, two days before a pair of reactor cores were to be shipped to Iraq, seven Israeli agents broke into a warehouse in the port of La Seyne-sur-Mer, near Toulon, and blew them up. On June 13, 1980, an Egyptian chemist hired by Iraq to work on spent fuel reprocessing was killed in Saclay were he had been sent for training. On August 7, 1980, the office of the Italian firm that supplied plutonium reprocessing technology to the Iraqis, was bombed. Finally, on June 7, 1981, Israeli aircrafts dropped several bombs on Osirak, scoring enough hits to permanently knock out the reactor.

	The destruction of Osirak is certainly the first deliberate act of ``counter-proliferation,'' and the event that must have given priority to uranium enrichment over plutonium production in Iraq's nuclear weapons program. The assertion put forward by most analysts, i.e., that Iraq's calutron effort started \emph{after} the bombing of Osiraq, is not plausible. In common with all other nations with nuclear weapon ambitions, all possible options for either producing plutonium or enriching uranium must have been studied right from the beginning. As has already been stressed, the production of plutonium with a small reactor was certainly the easiest route. But, in the case of Iraq, the visit of an Iraqi engineer to CERN in 1979 indicates that a significant amount of theoretical work on calutrons had already taken place. Not only because the engineer was inquiring about the construction details of a large magnet, but also because a clear understanding of the special properties of this unique magnet was shown. In addition, the mere knowledge that such a magnet had been built at CERN, could only have been obtained after a thorough investigation of both calutron theory and the state of the art of magnet construction. In this context, the fact that Jafar personally knew the man in charge of the NA10 magnet, Freudenreich, appears as a pure coincidence, a favorable circumstance which he tried to exploit, and not as the starting point of some new investigation.

	Of course, since nothing is known by us about the connections between Jafar and Iraq's secret nuclear weapons program while he was in Europe, it is not possible to know what role Jafar played in the early days of Iraq's nuclear program. In particular, we do not know if Jafar had anything to do with this program when, after graduation at Birmingham and before joining Imperial College, he spent some time in 1968-1969 at the Nuclear Research Center, Baghdad \cite{AA.52}. Later, when Jafar worked with Freudenreich at CERN, it was well known in the experimental team that Jafar had been an officer in the Iraqi army, and jokes were made because another member of the team was a reserve officer in the Israeli army. Nothing can be inferrerd by this, military service being an obligation in both countries. It is interesting however, to point out the irony of this collaboration. Neither do we have confirmation of the possibility that Jafar had a look at the British calutrons while working at Harwell, or at the near-by Rutherford Laboratory, as it is suggested by Burrows and Windrem \cite[p.36]{AA.51}. What is clear, is that when Jafar returned to Baghdad in 1975, because of his background as a high energy physicist, it must have been more natural for him to work on calutrons than on the other parts of the program. Later, after the bombing of Osirak, when enrichment became the preferred option, Jafar was in a leading position at the Iraq Atomic Energy Commission. It must not have been too difficult for him to push the calutron method: compared to the centrifuge technique it required much less foreign expertise and it was likely that he had a well studied design ready to be tested in a pilot plant.

	In order to understand Iraq's calutron design, and to appreciate the significance of Jafar's interest in the NA10 magnet, it is important to examine first its potential as an analysing magnet for a calutron.

\subsection{The NA10 magnet as a calutron magnet}
\label{Sec.2.4.3}

	In 1977, a group of Swiss Federal Institute of Technology physicists (including Freudenreich and Gsponer) proposed an experiment at CERN to study the inclusive production of massive muon pairs with intense pion beams. They suggested in their letter of intent \cite{AA.53} the use of an axially symmetric spectrometer consisting of four magnetized iron toroids, with various detectors placed in between. Once the experiment was accepted, it was found during the preparation stage, that a much better technique than magnetized iron could be used for the magnetic analysis. The idea came from Mario Morpurgo, one of the original builders of CERN, who came to Geneva after his studies in Rome and immediately applied his intelligence to the design and construction of conventional magnets \cite{AA.54}.

\begin{figure}
\begin{center}
\resizebox{6cm}{!}{ \includegraphics{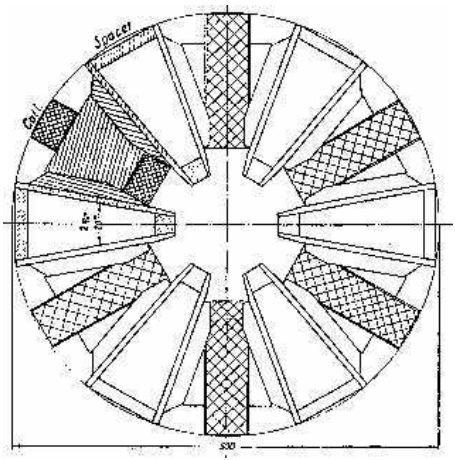}}
\end{center}
\caption[Section through the Ris{\o}  spectrometer.]{\emph{Section through the Ris{\o}  spectrometer. The outer diameter of the magnet is 50 cm.}}
\label{fig:Fig.4}
\end{figure}

	Morpurgo's idea was to build a large scale version of the six gap ``orange'' spectrometer built in 1955 at Ris{\o}, Denmark, by O.B. Nielsen and O. Kofoed-Hansen \cite{AA.55}: while the Ris{\o} magnet had a 0.25 meter radius and a length of about 0.5 meter, the NA10 magnet was to have a 2 meter outer radius and a 4.8 meter over-all length. A section through the Ris{\o} spectrometer can be seen in Fig.~\ref{fig:Fig.4} and a similar section through the NA10 magnet can be seen in Fig.~\ref{fig:Fig.2}. For the experimenters, the advantage of this magnet over the use of magnetized iron is that the magnetic analysis could be done by measuring the deflection of particles travelling through air instead of steel, thus with a subtantially better resolution. And for Morpurgo, the construction of such a magnet was a challenge, a special design which implied the use of several unusual techniques.

	Like the Ris{\o} magnet, the NA10 magnet has hexagonal symmetry \cite{AA.49}. An essentially azimuthal field $B = B(r)$ is excited between six laminated, wedge-shaped, iron pole pieces, each subtending 18$^{\rm{o}}$ in azimuth. The air-core part of the magnet, consisting of sectors between the iron wedges, thus subtends 70\% of the azimuthal acceptance. Over almost the entire air-core volume, the field has a $1/r$ dependence with a high degree of precision. The $1/r$-dependence of the field is useful for the magnetic analysis of particles of both low or high energy. That such a field, instead of a uniform field, could be used for an improved 180$^{\rm{o}}$ separator was first recognized by H.O.W. Richardson \cite{AA.56}. He also found that under suitable conditions both lateral and longitudinal focusing could occur. The idea was then further developed by O. Kofoed-Hansen, J. Lindhard and O.B. Nielsen, who in particular discussed the case where both source and focus are placed outside the magnetic field \cite{AA.57}. Finally, a magnet producing a field with a $1/r$ dependence is used in the S-2 separator at Arzamas-16, thus confirming that high productivity EMIS is possible with such a field \cite{AA.21}.

	The various methods proposed between 1940 and 1955 to increase the performance of magnetic spectrometers and electromagnetic separators, and the relative importance of the ideas developed by physicists of different countries with different kind of applications in mind, could be the subject of a study by an historian of sciences. Even a superficial survey of the literature however, would easily point to some of the most important contributions. For example, the general analysis of the double focusing problem by Svartholm and Siegbahn (which led to the 255$^{\rm{o}}$ method), or the ingenuity of the sector-type separators built by Ren\'e Bernas. In this context, the six gap ``orange'' spectrometer of Kofoed-Hansen, constructed at the famous Niels Bohr Institute, was also a clever and rather well known device. It is therefore not surprizing that Jafar must have taken considerable interest in investigating the possibility of using this concept for the separation of the uranium isotopes. In that case, the calutron tanks would have been placed in the air-cored segments, and the optimum size of the magnet would have been of precisely the same magnitude than that of the NA10 magnet.

	In practice, the construction of a NA10-like magnet is not a trivial thing. In particular, the windings have to be assembled into coils converging radially to the axis of the magnet. The problem is then one of assembly because the coils cannot simply be wound and later fitted together with the iron pieces: the coils have to be made by wrapping the windings around the steel once the full magnet carcass has been assembled. The solution was to make the coils by using a sophisticated high frequency welding technique to join copper bars. This implied the resolution of a number of ancillary problems, such as the electrical insulation of the coils, etc.

	In conclusion, even though a multitank calutron concept based on the NA10 magnet is attractive in theory, it leads to a number of engineering problems which would certainly have been considerable for Iraq. In fact, Jafar ultimately settled on another design, based on the 255$^{\rm{o}}$ concept, which in one respect has some similarity with the original NA10 proposal, i.e., the use of magnetized cylinders of steel \cite{AA.53}. Such magnets are much easier to build and were in fact used in another CERN experiment, NA4.

\subsection{Iraq's calutron design}
\label{Sec.2.4.4}

	Because of the rather limited amount of reliable technical information published until now, it is still not possible to give a truly accurate technical description of the Iraqi attempt to produce highly enriched uranium by means of calutrons, and to make a well documented independent assessment of this effort. The only non-classified first hand information on the subject are the reports to the UN Security Council on the fifteen IAEA on-site inspections under Security Council resolution 687 (1991) of Iraqi nuclear capabilities, carried out between May 1991 and November 1992. A number of photographs and articles by members of the UN inspection teams have also been released by the IAEA over the same period of time. In this section, using at best these documents, we will try to describe the scientific principles of the specific calutron design developed by Iraq. To keep our discussion at the level of the scientific principles, and due to the fragmentary nature of the available information, we will not address the many technical details that would be discussed in a comprehensive assessment.

	The central piece in any electromagnetic isotope separation system is the magnet which provides the mass analysing field. It is therefore fortunate that precise information is available on the magnet used by Iraq. Engineering drawings of the steel part of this magnet have been published \cite{AA.58,AA.59} and several good photographs are available \cite{AA.59,AA.60}.

	The distinctive feature of Iraq's calutron design is the use in an industrial-scale separation facility, of an axially symmetric magnetic field with the right kind of non-uniformity to produce double-focusing of the ion beam \cite{AA.7}. Such a field had been used for example, in the calutron built by Artsimovitch et al. in 1958 \cite{AA.20}, and was recognized by the French in 1965 to ``offer considerable advantage as regards better focusing and increase of the transmission yield, allowing separation of more intense beams with a  good enhancement factor and yield'' \cite{AA.26,AA.27}.

\begin{figure}
\begin{center}
\resizebox{10cm}{!}{ \includegraphics{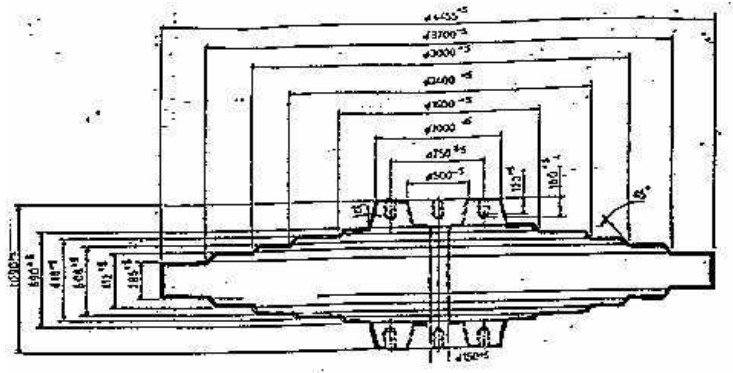}}
\end{center}
\caption[Pre-machined Iraq's alpha calutron iron core.]{\emph{Cross section of Iraq's calutron magnet cores: Pre-machined iron core for the 120 cm beam radius magnets.}}
\label{fig:Fig.5a}
\end{figure}

\begin{figure}
\begin{center}
\resizebox{10cm}{!}{ \includegraphics{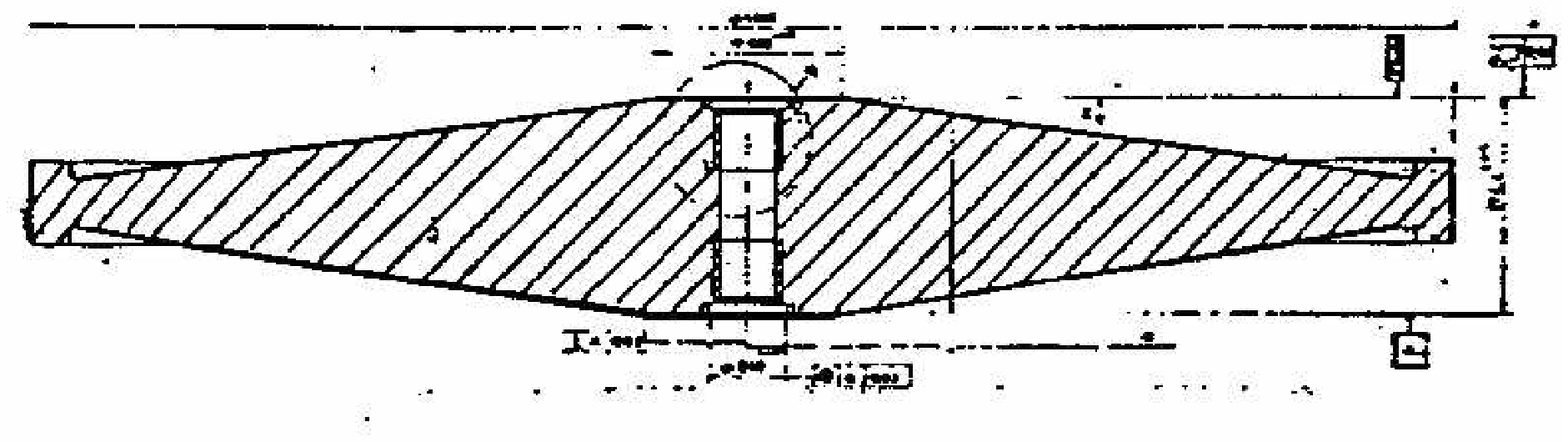}}
\end{center}
\caption[Final dimensions of Iraq's calutron iron core.]{\emph{Cross section of Iraq's calutron magnet cores: Final dimensions of 120 cm beam radius magnet cores.}}
\label{fig:Fig.5b}
\end{figure}

	In order to have a multi-calutron configuration in which the separation tanks are placed in between adjacent magnets to utilize the magnetic field more economically, a special modular design is necessary. In Iraq's case, the basis of this design is a rotational symmetric dipole magnet shaped as a truncated flat double cone with a slope of about 8$^{\rm{o}}$ (Figs.~\ref{fig:Fig.5a} and \ref{fig:Fig.5b}). For the first enrichment stage (i.e., the alpha-process) the outer radius of the magnet is about 220 cm and the maximum thickness about 70 cm. For the second enrichment stage (i.e., the beta-process) no drawings are available. It is known however, that the magnet was similar in shape but half the size.

\begin{figure}
\begin{center}
\resizebox{12cm}{!}{ \includegraphics{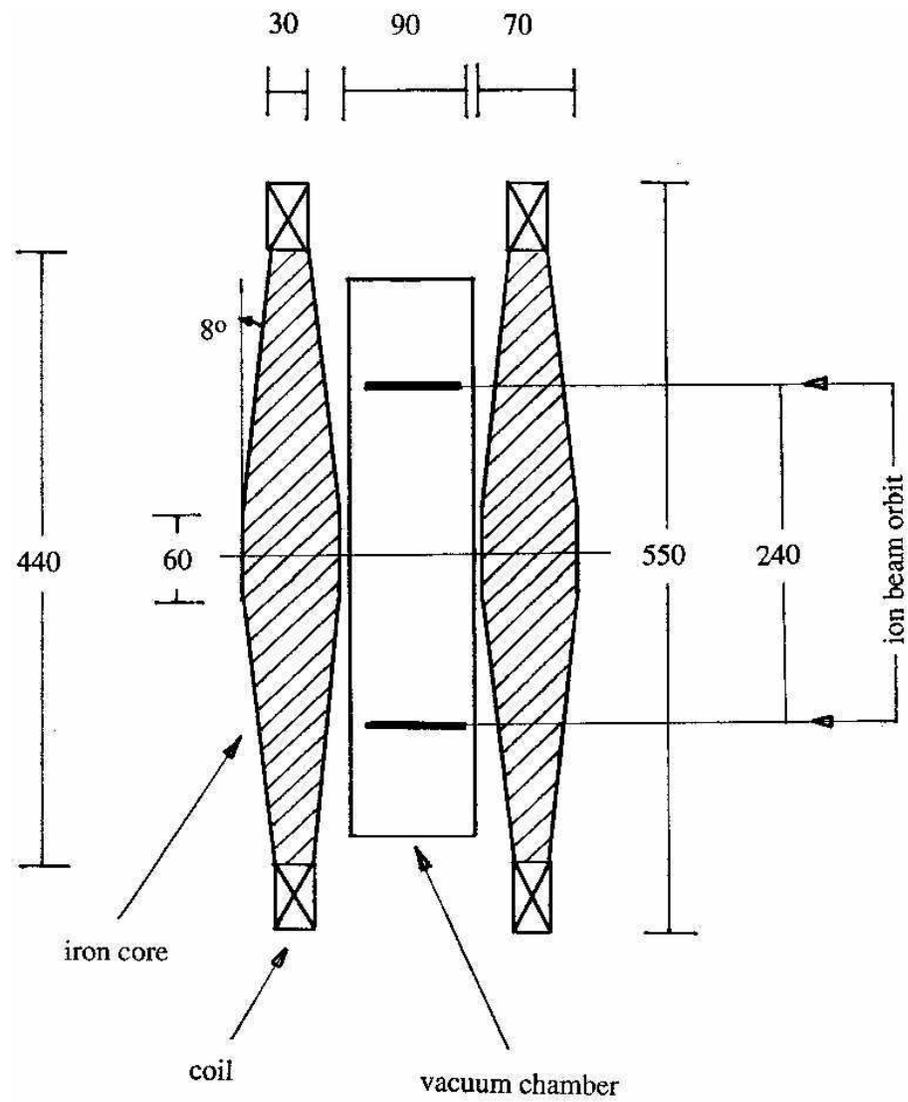}}
\end{center}
\caption[Schematic of Iraq's alpha calutron track.]{\emph{Schematic of Iraq's alpha calutron track.  Only two adjacent magnets and only one separation chamber are shown.}}
\label{fig:Fig.6}
\end{figure}

	The general arrangement of the magnets and interleaving separation chambers is known from the first three IAEA inspections \cite{AA.61,AA.62} and an Iraqi letter commenting on the third IAEA inspection \cite{AA.63}. Iraq's first industrial-scale EMIS facility was constructed in Tarmiya, 40 kilometers northwest of Baghdad. A second facility was planned at al Sharqat, 200 kilometers northwest of Baghdad. The two plants were to be identical with a total of 70 alpha-calutrons and 20 beta-calutrons in each. The 70 alpha separators were to be installed in two large (5 m by 60 m) parallel piers \cite[p.11]{AA.61} with 35 separators in each line. This gives an average space of 170 cm per separator. Removing 70 cm for the magnet width, a space of about 100 cm is left for the vacuum chamber. At the end of each line, a half magnet provides an end-pole. Interconnecting the end-poles of both lines with iron plates, the magnetic flux can be closed and the resulting configuration is similar to the ``race-tracks'' built in Oak Ridge during World War Two. A schematic of this arrangement is shown in Fig.~\ref{fig:Fig.6}. The same basic principles were to be used in the beta separators, with everything scaled down by a factor of two. While the alpha ion-beam mean radius was 120 cm, the beta ion-beam mean radius was 60 cm.

	From the shape of the dipole magnet, Fig.~\ref{fig:Fig.5b}, and the configuration depicted in Fig.~\ref{fig:Fig.6}, it is possible to discuss the main properties of the mass analysing field system used in Iraq's calutrons. Good focusing requires very accurate fabrication of the pole-pieces in order to obtain exactly the required field form. In Iraq's case, the problem of fabricating a number of large magnets with complicated non-linear pole faces has been avoided by using dipole magnets with simple conical pole-pieces. A sufficiently precise knowledge of the field form can then be gained from the simplified second order analysis of Snyder et al.\ \cite[p.854]{AA.64}. In this approximation, the field in the midplane between two pole faces is described by a field-index  $n$  which is a function of the slope of the pole-pieces, the beam radius and the spacing between the two pole faces. In particular, $n = 0$  for a uniform field, and $n = 1/2$  for the Svartholm-Siegbahn field of the 255$^{\rm{o}}$ method. With Iraq's calutron magnet and the configuration of Fig.~\ref{fig:Fig.6} in which there is room for a 90 cm separation chamber, the field index is about $n = 1/4$. Iraq's choice for the alpha-separator field form is therefore a compromise between those for the 180$^{\rm{o}}$ and 255$^{\rm{o}}$ methods.

	The compromise is a trade-off between productivity and quality. The larger the space between the magnets, the more room there is for the ion-beams in the separation chambers. For a given number of magnets, this increases the output, but at the expense of the enrichment of the product.

	On one hand, if the greatest possible enrichment is desired, the field form required is that of the 255$^{\rm{o}}$ method, which is obtained when the minimum spacing between the magnets is only 30 cm. This leaves just enough room for a 30 to 60 cm wide vacuum chamber. In such a chamber, it is difficult to use more than one or two beams. In effect, to generate an intense beam, the ion current is usually extracted from the source through a narrow slit, about 10 to 40 cm long. The source assembly is thus at least 20 to 50 cm wide, not much less than the width of the separation chamber itself. To have more than one beam, the ion-sources must be put below one another. This possibility, investigated by the French in the 1960s, is in practice limited to two concentric beams with two different radii \cite{AA.26}. The reason is that when beams from several independent sources overlap the system becomes unstable and the failure of one beam can cause all the beams to fail.

\begin{figure}
\begin{center}
\resizebox{8cm}{!}{ \includegraphics{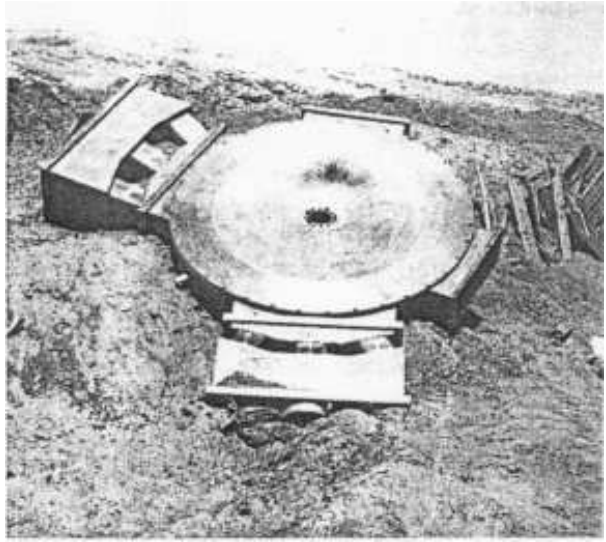}}
\end{center}
\caption[Alpha calutron chamber on its side.]{\emph{Photograph of Iraqi alpha calutron chamber on its side.}}
\label{fig:Fig.7a}
\end{figure}

\begin{figure}
\begin{center}
\resizebox{8cm}{!}{ \includegraphics{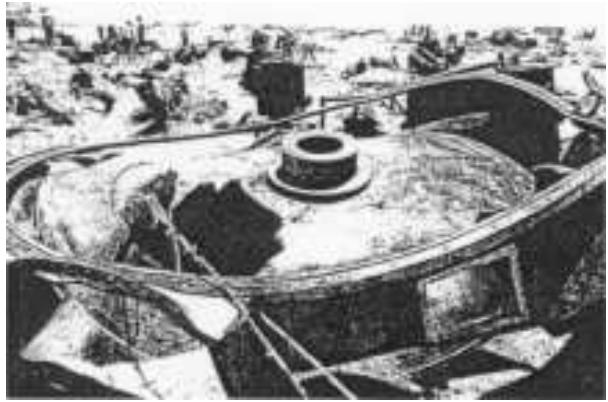}}
\end{center}
\caption[Crushed beta calutron chamber.]{\emph{Photograph of crushed Iraqi beta calutron chamber.}}
\label{fig:Fig.7b}
\end{figure}

	On the other hand, if a lower enrichment is acceptable an increase in productivity is possible by widening the space between the magnets. This allows the use of wider sources (i.e., with a longer extraction slit) or the use of several ion-sources positioned side by side. For example, with a spacing of 100 cm, there will be room for two or three double-sources, i.e., as many as six ion-sources in one 90 cm thick vacuum chamber. With this spacing the field index takes a lower value (1/4 instead of 1/2) and the benefits of the non-uniform field become less significant resulting in lower enrichment of the output. According to the theory \cite{AA.7}, the optimum focusing angle also decreases from 255$^{\rm{o}}$ down to about 210$^{\rm{o}}$. The widening of the gap between the magnets would in principle require a redesign of the vacuum chambers. Such a modification was eventually not necessary to achieve a sufficiently enriched product in the first-stage separation units. That this interpretation is plausible is supported by the photographs shown in Figs.~\ref{fig:Fig.7a} and \ref{fig:Fig.7b}.

	In Fig.~\ref{fig:Fig.7a}, an alpha calutron chamber is seen on its side. Assuming a diameter of 440 cm its thickness is about 90 cm. The upper side is on the left, with two protrusions on each side of a hook. The bottom side is on the right with two ports that may have been connected to the vacuum pumps. The angle between the two upper protrusions is close to 255$^{\rm{o}}$, and each of them has three axially directed rectangular channels that may have contained one double ion-source or one ion-collector assembly. That Iraq's calutron could have had two to six sources has been reported in at least one publication \cite[p.18]{AA.65}.

	In Fig.~\ref{fig:Fig.7b}, a beta calutron chamber can be seen. Assuming a diameter of 200 cm, its thickness is about 30 cm. Nothing definite can be said concerning the focusing angle, but it is most likely that the beta calutrons used the 255$^{\rm{o}}$ method to achieve the higher separation power required by the second enrichment stage.

	In conclusion, having used the engineering drawings of the magnet and the published photographs as the main input, our analysis is summarized as follows. The basis of Iraq's design is not the 180$^{\rm{o}}$ method used during World War Two but the 255$^{\rm{o}}$ method. For that purpose, a dipole magnet with a 8$^{\rm{o}}$ conical shape was built for the first enrichment stage. Using a 30 cm vacuum chamber, the 60 m long process bay of the Tarmiya plant could have accommodated a maximum of 120 magnets and 120 concentric double-beam separation units operating in the 255$^{\rm{o}}$ mode. Since the first-stage enrichment did not require the full separation power of the 255$^{\rm{o}}$ method, it must have been decided to put only 70 magnets in the process bay. The advantage is that substantially fewer magnets are required, freeing enough room for $70 \times 3 = 210$ double-beam units, and therefore increasing throughput by a factor of 210/120 . This conclusion is independent of the specific assumptions made here, i.e., the use of double ion-sources with non-overlapping beams \cite{AA.26}. It is not impossible that Iraq's calutrons had multiple-beam ion-sources in which several extraction slits, with lengths of the order of 90 cm, were placed just below on another to produce two or more overlapping beams of similar radii.

	As previously explained, a comprehensive assessment capable of answering whether or not the Iraqi effort was close to producing enriched uranium in significant amounts, would have required the detailed analysis of many other important aspects such as Iraq's ion-source technology. According to the IAEA inspection reports \cite[p.6]{AA.62}, and consistent with experience elsewhere in the world, Iraq found that the only real obstacle in the development of an efficient EMIS system is the design of a good ion-source. This is probably why the alpha calutrons under installation when the Gulf War began, had only four 150 mA ion-sources in each separation tank. Assuming an availability of 55\%, a straight-forward calculation shows that the maximum theoretical production of Tarmiya would have been about 14 kilograms of U-235 per year, in the form of 12 and 90\% enriched uranium, for the first and second enrichment stage respectively \cite{AA.63}.

\subsection{The difficulties of publishing}
\label{Sec.2.4.5}

	In 1980, two years after receiving a PhD in physics, Gsponer left elementary particle physics and stopped working at CERN. His aim was to establish a \emph{scientific} research institute in which himself and other researchers could apply their professional skills to the analysis of important disarmament problems. Together with Roy Preiswerk, then Director of the Institute of Development Studies of the University of Geneva, GIPRI, the \emph{Geneva International Peace Research Institute}, was created. An association whose most active members were physicists and social scientists, primarily from CERN and the University of Geneva, and several prominent local political figures.

	GIPRI's initial research program concentrated on the military applications of particle accelerators. The idea was to include the military impact of particle accelerator technology, of which calutrons are an example, in the arms-control/disarmament debate and to provoke a discussion in the scientific community on the subject.

	The third review conference of the Non-proliferation Treaty (NPT) was to be held in Geneva in August 1980. This presented a good opportunity to publish a paper and so the first GIPRI report was written for the occasion. The title was ``Particle accelerators and fusion technologies: implications on horizontal and vertical proliferation of nuclear weapons.'' The draft of this paper \cite{AA.1} had five chapters, the first one on enrichment. This chapter, after reviewing various methods, concluded with the following paragraph on EMIS:
\begin{quote}
	{\it ``Finally, the oldest enrichment method (which has the capability to achieve an almost complete separation of U-235 in a single step), i.e., \emph{electromagnetic isotope separation}, may become interesting in special circumstances thanks to technological advances. In effect, while this method is relatively expensive and economically unattractive, it uses only well known classical technologies. These technologies (ion sources, vacuum, large magnets) are routinely used in nuclear physics and are certainly within reach of many countries. The construction of an enrichment plant based on this method (of the kind the United States built during World War Two) could however, hardly be justified for civilian purposes and would be very difficult to conceal. Nevertheless, the appeal of this enrichment method is illustrated by the fact that Iraq recently showed strong interest in the technology of large magnets''} \cite[p.3]{AA.1}.
\end{quote}

	The other chapters dealt with the following subjects; accelerators for fissile material production, application of accelerators to nuclear weapons technology, thermonuclear fusion and hybrid reactors, applications of inertial confinement fusion to nuclear weapons technology and finally, implications of emerging nuclear technologies for developing countries.

	Obviously, at a time when the debate on the risk of nuclear weapons proliferation was centered on problems arising from nuclear power generation by means of fission reactors, an article with such a vast content was extremely ambitious. Despite efforts to keep the content as simple as possible, the average reader would have had some difficulty understanding the numerous technical concepts involved. In addition, Gsponer was a totally unknown young physicist and did not have the authority necessary to present new ideas to the rather closed and conservative arms-control/disarmament community. This lack of authority was particularly detrimental because the paper stressed a number of conclusions Gsponer was able to draw using scientific deduction alone. It also included a statement on Iraq's interest in EMIS technology for which there was no reference because Gsponer had accidentally discovered the fact himself.

	Consequently, having given the draft for comment to several scientists and researchers in the field of disarmament, Gsponer decided to shorten the paper. The first and last chapters dealing with enrichment and the implications for developing countries were removed. The paper was simplified by narrowing its focus on the nuclear proliferation aspects of accelerator and fusion technologies, and avoiding the danger in publishing something new and crucial on Iraq's nuclear ambitions. The years 1979-1981 were marked by a number of violent incidents in which Israeli secret services and armed forces attempted to halt Iraq's efforts to acquire nuclear reactor and reprocessing technology. It was a serious concern therefore, that mentioning Iraq's definite interest in calutron technology could endanger Gsponer and former CERN colleagues.

	The abridged paper was finalized with the help of Bhupendra Jasani, a physicist working at the \emph{Stockholm International Peace Research Institute}. It was sent to the heads of delegations of the 1980 NPT review conference \cite{AA.2}. Later it was submitted to \emph{Science}, \emph{Technology Review} and \emph{The Bulletin of the Atomic Scientists}. All three journals rejected the paper. Once translated and printed in German and Italian Gsponer did not continue to try to find an English publisher. 

	In the Spring of 1981, soon after Ruth Adams was elected Editor of \emph{The Bulletin of the Atomic Scientists}, she made a tour of Europe visiting several disarmament research institutes to encourage European researchers to submit articles to the \emph{Bulletin}. The idea was to open the journal to broader perpectives than the American ones which had dominated the journal since its foundation in 1945. In Geneva, Adams visited GIPRI where Gsponer told her about his research on the military use of particle accelerators. Interested by the subject, she invited him to submit a short paper on a recent public debate in which the particle accelerator issue was raised in the context of the construction of LEP, a very large particle accelerator to be built at CERN. The result of the peer review was negative. On the advice of Frank Barnaby, Director of SIPRI in Stockholm and an Editorial advisor of the \emph{Bulletin}, Gsponer submitted a new version of the paper in February 1982. In March, he received copies of the proofs of the paper scheduled as a commentary for the May 1982 issue of the \emph{Bulletin}. Notwithstanding this approval and that he had corrected and returned the proofs, his commentary was never printed. Despite several attempts, there was never any answer from the \emph{Bulletin} to any of his letters enquiring as to why the commentary had not been printed.

	Had Gsponer's commentary been censured? Had his paper addressed sensitive issues which were not supposed to be discussed in non-classified literature? The only paragraph that could have justified such an action is as follows:
\begin{quote}
    {\it ``Elementary particle physics, at the forefront of fundamental research, is mainly carried out through the use of increasingly powerful and larger high energy accelerators and storage rings. For applications of accelerators requiring lower energies but very high currents, the advances are equally impressive. These developments are leading towards numerous new applications, especially in the fields of nuclear energy and weaponry. Particle accelerators are becoming usable for efficient breeding of fission and fusion materials, for driving inertial confinement fusion devices, for studying the physics of thermonuclear weapons, and so on. The concept of high energy particle beam weapons may become feasible and free-electron lasers using electron accelerators and storage ring technologies may provide a new generation of highly efficient and powerful lasers with many military applications''} \cite{AA.66}.
\end{quote}

	Of course, it is not possible to know if the commentary was censured for classification reasons or not. The only thing that can be said for sure, is that this kind of problem is a recurring one in the United States \cite{AA.67,AA.68}.

	Another less dramatic explanation for the censorship of Gsponer's commentary is that nuclear physicists working in non-military laboratories, and especially those using particle accelerators, are extremely nervous when it comes to the question of the military impact of their work. The origin of this irascible response can be found in the passionate desire of many atomic scientists to believe that their work has many more positive consequences, such as nuclear energy or a better understanding of the fundamental laws of nature, than terrifying consequences such as nuclear weapons.

	In 1982 Gsponer left GIPRI to create ISRI, the \emph{Independent Scientific Research Institute}. It was a much less ambitious enterprise than GIPRI, concentrating on the independent assessment of nuclear technologies. For instance, in a paper presented at the Third International Conference on Emerging Nuclear Energy Systems, Gsponer and others stressed again the nuclear proliferation impact of new technologies \cite{AA.3}. In particular, in a section on enrichment, (with the electromagnetic, plasma and laser separation processes in mind) they warned that 
\begin{quote}
{\it ``new technologies could completely change the situation prevailing since 1945, namely, the fact that enrichment facilities are in general much larger, more complicated and expensive than simple production reactors''} \cite[p.172]{AA.3}.
\end{quote}

Similar concerns about the impact of particle accelerator technology on enrichment were repeatedly expressed in \emph{La Quadrature du CERN}, a book Gsponer had written with others for CERN's thirtieth anniversary \cite[p.20]{AA.47}, \cite[70--73]{AA.60}.

	In 1990 Gsponer was in Mauritius. Like most, he followed the Gulf War events on the radio and television. He resumed his work on theoretical physics once military action was over.

	Gsponer had heard the alarming statements that Iraq might have been working on an atomic bomb. But at no point did he think that they could have made much progress otherwise it would have been known for a long time. It came as a terrible shock to him to learn about Iraq's use of calutrons from a \emph{New Scientist} article published in July 1991 \cite{AA.69}.

	More than 12 years had passed since Gsponer had discovered Iraq's interest in calutrons. An event that he was trying to forget. As with his evaluation of particle beam weapons research \cite{AA.70,AA.71}, he thought more and more that his appreciation of this discovery was possibly exaggerated, and that he had been wrong to quit his work as a particle physicist. After all the negative comments about his work on the military implications of particle accelerator technology, and the personal attacks he had endured from former friends and colleagues, a tragic event --- the largest military coalition since World War Two --- finally confirmed that his worries were well founded.

	The time had finally come to make public how Gsponer had discovered in 1979, Iraq's work on EMIS technology. For obvious reasons, in all his publications, he had never explicitly mentioned Iraq. After the Gulf War however, there was no reason for further caution. Since he was still in Mauritius, he had first to return to Geneva to meet former CERN colleagues, and then to wait for a good opportunity.

	In 1995, Gsponer thought that the fifth review conference of the NPT would be an appropriate time to put straight the historical record on Iraq's calutron program. Having investigated the possibility of publishing a report on calutron technology to be sold or used as the basis of a book, he went to New York to inquire whether the \emph{New York Times}, with the help of some personal contacts in the science section, would be interested in the story. These attempts were unsuccessful. Meanwhile, a Geneva based science journalist picked up the story and wrote an article for the \emph{Journal de Gen\`eve}, a daily newspaper with a good international audience. After double-checking the facts, and interviewing Freudenreich to clarify the circumstances of the 1979 visit to CERN by an Iraqi engineer, the article was published as the main news story of the 22-23 April weekend edition of the journal \cite{AA.72}, one week after the NPT review conference in New York had started.

	Except for some emotional reactions from the CERN staff and management, the impact of the \emph{Journal de Gen\`eve} article was minimal, especially on the NPT negotiations in New York. Its main merit was therefore to publish an important historical fact, and to highlight the blindness of the scientific community with regard to the military implications of its activities.

	Fifteen years after Gsponer's report for the 1980 NPT review conference was written, the nuclear weapon states are coming close to an agreement on a Comprehensive Test Ban Treaty (CTBT). While this treaty would forbid all nuclear tests, either above or under ground, the nuclear weapon states, the US and France leading the way, are starting to build very large inertial confinement fusion facilities \cite{AA.73}. These facilities enable the detailed study of the ignition process of thermonuclear explosions. The irony is that the aim of Gsponer's papers of 1980-83 \cite{AA.1,AA.2,AA.3}, was precisely to show that such a treaty would be meaningless if accelerator and fusion technologies were systematically applied to the development of new nuclear weapons. With these technologies it is indeed possible to conceive and test new weapons in the laboratory without large scale explosions.

\section{Implications for the proliferation of nuclear weapons}
\label{Sec.2.5}

\subsection{Beam technologies and nuclear weapon proliferation}
\label{Sec.2.5.1}

	Enrichment of fissile materials by means  of calutrons is a classical example of the use of particle beam technology in the nuclear fuel cycle. Another example is breeding of special nuclear materials by means of particle accelerators. The current interest in these technologies is due to the fact that \emph{beam technologies} are becoming increasingly competitive substitutes for traditional nuclear technologies. Laser and particle beam technologies are also key components of emerging nuclear energy systems such as fusion, and of increasing importance for the qualitative development of nuclear weapons.

	In the early eighties when Gsponer and his GIPRI/ISRI collaborators were stressing the nuclear proliferation risk of ``old'' technologies such as EMIS, their opinion was at odds with the generally accepted one. For instance, in an excellent review on uranium enrichment and nuclear weapon proliferation \cite{AA.5}, despite the recognition that electromagnetic processes were ``again under serious consideration'' \cite[p.186]{AA.5} the general conclusion was to consider the calutron as ``no longer a viable process'' for the production of enriched uranium \cite[p.22]{AA.5}. Today, after Iraq's construction of a large EMIS plant, the question is not so much which assessment was right or wrong, but why Iraq's enormous effort remained ``undetected'' until the end of the Gulf War. This is particularly disturbing because all analysis have concluded that the construction of an industrial-scale EMIS plant would be very difficult to conceal, and because Iraq's motivation of using the EMIS process was precisely the same as the motivation of the United States during World War Two, to produce fissile material for a nuclear weapon by all means and at any cost.

	In this context, it should be stressed that besides EMIS, other proliferation-prone ``old'' technologies should be given serious consideration \cite{AA.1,AA.2,AA.3}. Of these technologies, the most important one is certainly that of particle accelerators. In fact, from 1941 (when plutonium was first produced) until the end of 1943, circular accelerators were the sole source of plutonium, and over this period slightly more than 2 milligrams of plutonium were produced \cite{AA.2}. It has been calculated that a 1000 MeV proton accelerator with a beam of 1 mA (i.e., a beam power of 1 MW) could produce enough spallation neutrons to breed about 10 kg of plutonium per year, enough for two to four atomic bombs \cite{AA.3}. The construction of such an accelerator, and of the associated reprocessing plant, is certainly possible for many countries, including some in the developing world. Compared with a calutron plant of similar fissile material output, the cost would be considerably less. A major obstacle to the acquisition of such a technology is that, contrary to nuclear reactors, complete accelerators with the requisite characteristics cannot be bought on the market. Similarly with what Iraq had to do in the case of its calutron plant, an accelerator-breeder would have to be built by assembling it from its components, something that requires a considerable indigenous research and development effort. In practice, the main difference would come from the fact that particle accelerator technology is in several respects more sophisticated than calutron or even centrifuge technology. With the current proliferation of advanced scientific, industrial and military technology, this kind of barrier is likely to become less and less effective in the future.

	Of special concern is the fact that particle accelerators have recently established themselves as very serious candidates for replacing aging nuclear reactors in many kinds of military and/or civilian applications. For a given total thermal power, a major technical advantage of spallation based systems is that productivity (i.e the number of neutrons or the amount of plutonium or tritium produced) is roughly five times greater than the productivity of fission based systems \cite{AA.2}. As a result, the problems of radioactivity, containment and cooling will be proportionally smaller by about the same factor.

	In the field of neutron physics research, the current problem in Europe \cite{AA.74} as well as in the USA \cite{AA.75}, is the replacement or upgrading of aging reactor or accelerator facilities. In Europe, there is a proposal to replace an old German research reactor fueled by highly-enriched uranium with a spallation neutron source (based on a 5 MW beam-power accelerator) rather than replacing it by a new research reactor of the same type \cite{AA.75}. Such a switch from a reactor to a different type of neutron source has a positive non-proliferation impact, mainly by suppressing the risk of diversion or theft of highly-enriched uranium. The negative impact is that it will contribute to the spread and development of accelerator-based neutron generation technology.

	The most significant development however, is the US proposal to replace its military production reactors with accelerator-based facilities \cite{AA.76}. The possibility that the US will build a giant tritium-producing accelerator (with a 100 MW beam power), dubbed the APT, Accelerator Production of Tritium, is quite high, since it could simultaneously satisfy the needs of both military and scientific communities. In the words of Burton Richter, the Director of the Stanford Linerar Accelerator Center: ``With small modification to the APT, the U.S. can have both the world's premier neutron source and a secure tritium supply'' \cite{AA.76}. If the US goes ahead with this proposal, the road will be open for other countries to follow. Of direct concern would not only be the declared nuclear powers, but also countries like India or Japan, which already have substantial knowledge and skills for building accelerators and existing or projected spallation neutron sources \cite{AA.2,AA.3}.

	Another recent development is the proposal to use accelerators to incinerate long lived radioactive waste \cite{AA.77}. Since the transmutation process is producing a large number of spallation neutrons, the same technology could be used to breed plutonium or tritium \cite{AA.77}.

	Finally, while the idea was not new, Carlo Rubbia, Nobel laureate and at the time Director of CERN, unveiled in 1993 a ``method to produce safe and clean nuclear energy by aiming accelerated protons at a thorium target'' \cite{AA.78}. The method claimed to pose ``no risk of military proliferation,'' and to be ``most advantageous for developing countries.'' Since then, experiments were performed at CERN \cite{AA.79} and research is underway to investigate the possibility of using the system as an incinerator of radioactive waste. While the method may have some potential technical advantages, it must be stressed that its nuclear weapons proliferation impact is far from negligible.

	Since accelerator based power systems may lead to economically attractive designs with an electric output of about one tenth of a normal nuclear power station, the problem of safeguarding a large number of dispersed accelerator power plants will be considerable. Similarly, in case of the clandestine use of accelerator technology to breed plutonium or tritium, the problem of detecting the illicit activity will be magnified by the five fold reduction of heat and radioactive effluents. The main proliferation problem however, is the fact that the construction of any kind of accelerator system for commercial scale nuclear power generation, nuclear fuel generation for civil or military use, transmutation, etc., will open another Pandora's box of problems that can only exacerbate the current nuclear proliferation situation.

\subsection{EMIS for plutonium purification}
\label{Sec.2.5.2}

	It is well known that the kind of plutonium bred in commercial power-generating nuclear reactors is not suitable for the design of reliable nuclear weapons. The nuclear weapons in the contemporary military arsenals are made of weapons-grade plutonium, i.e., plutonium that is more than 95\% pure in the isotope Pu-239. A state determined to make nuclear weapons using reactor-grade plutonium (i.e., containing between 10 to 30\% of the unwanted Pu-240 isotope) would certainly first try to purify it to convert it into weapons-grade plutonium. This would simplify considerably the design of the weapon by enabling the use of less sophisticated implosion technique to achieve criticality.

	In theory, the isotopic separation of plutonium is a much less demanding task than enrichment of uranium. For instance, as the initial Pu-239 content of reactor-grade plutonium is over 70\%, while the U-235 content of natural uranium is only 0.7\%, a plutonium enrichment plant will be about 100 times smaller than an uranium enrichment plant of the same fissile material output. For example, a straightforward calculation shows that using the electromagnetic separation method, a single calutron with a beam current of less than 100 mA is sufficient to produce 5 kilograms of weapons-grade plutonium per year.

	The technology of plutonium isotopic separation however, is no more covered by secrecy than uranium enrichment technology. An important reason for this is that nuclear reactor research, as well as nuclear weapons diagnostic \cite{AA.80}, requires plutonium isotope separation in order to measure Pu-240 relative to Pu-239 production because these isotopes have alpha-decay energies very close to one another. In practice however, there are problems.

	First, any method suitable for separating kilogram quantities of pure Pu-239 from reactor-grade plutonium has to be capable of operating with highly radioactive feed material. For such an application, gaseous diffusion or ultracentrifugation for example, are unsuitable because the whole apparatus (including key components such as the porous barriers or the centrifuge rotors) would become highly radioactive so that repair or maintenance becomes impossible. Isotopic separation of radioactive materials requires that the process takes place in a containment vessel which can be removed for decontamination or recovery of the feed material. This is possible with the laser, plasma or electromagnetic separation methods in which the ionized feed material is generally processed within a removable ``liner'' enclosed in the vacuum chamber.

	Second, the atomic weight difference between Pu-239 and Pu-240 is one, while it is three between U-238 and U-235. Thus, the enrichment of plutonium requires a three fold increase in separation power over enrichment of uranium. This means a substantial increase in difficulty \cite[p.348]{AA.10} so that a typical EMIS plutonium separator will look much more like the 255$^{\rm{o}}$ calutron of Artsimovitch \cite{AA.20} than a World War Two 180$^{\rm{o}}$ calutron.

	A good idea of what a calutron for plutonium separation might look like is given by the S-2 separator of Arzamas-16 \cite{AA.21}. The construction of a facility comprising a number of such calutrons would in some respects be more complicated, and perhaps more costly than the chemical reprocessing plant, which would first of all be required to extract the plutonium from irradiated power-reactor fuel elements. A government having completed the first step would most probably go ahead with the plutonium isotopic purification step, even though it has been proved that by using the appropriate technique, a crude nuclear explosive could be made of power-reactor plutonium \cite{AA.81}.

	A first country for which it can be seriously argued that plutonium enrichment is a potential nuclear proliferation threat is Iran. Since 1990, Iran has been receiving aid from China for the construction of a small calutron \cite{AA.34}. In 1995, Russia agreed to build a nuclear power reactor on the site of a German reactor which was left incomplete after the collapse of the Shah regime. In a decade or so therefore, and in the case of a breakdown of international safeguards, Iran could have direct access to indigenously bred reactor-grade plutonium, and the potential capacity to turn it into weapons-grade plutonium. 

	A second country of concern is Japan. Considering the return to Japan in 1993, of more than a ton of Japanese plutonium from a nuclear fuel reprocessing plant in France, Japan's inventory of separated reactor-grade plutonium could reach several tens of tons by 2005-2010. In Japan, the main technical justification for extracting plutonium from spent reactor fuel is its potential use in fast-breeder reactors. There is, however, ``the conspiracy theory that in the long term Tokyo aims to develop the capacity to build nuclear bombs at short notice should the international situation so demand'' \cite{AA.82}. In such a case, with all its technological might, Japan would certainly not satisfy itself with crude nuclear devices. It is more likely to purify its reactor-grade plutonium in order to build a credible nuclear arsenal. To do so, Japan would be able to chose from the full range of the most sophisticated enrichment technologies, including the laser and plasma isotope separation processes (see Sec.~\ref{Sec.2.3.11}). Japan's current interest in the development of advanced enrichment technology, and in particular of the plasma separation process (the modern method with the greatest potential in terms of universality and productivity), is of great concern from the point of view of nuclear weapon proliferation.

\subsection{UN resolutions 687 and 707 and their implications for a
            halt of all proliferation prone nuclear activities}
\label{Sec.2.5.3}

	In paragraph 13 of Security Council resolution 687, adopted on 3 April 1991, the IAEA was requested by the Security Council to carry out immediate on-site inspection of Iraq's nuclear capabilities and carry out a plan for the destruction, removal or rendering harmless of items prohibited to Iraq under paragraph 12 of the resolution 687. On 15 August 1991 the Security Council adopted a further resolution, number 707, obliging Iraq to ``\emph{halt all nuclear activities of any kind, except for the use of isotopes for medical, agricultural and industrial purposes} until the Security Council determines that Iraq is in full compliance with resolution 707 and with paragraphs 12 and 13 of resolution 687, and the IAEA determines that Iraq is in full compliance with its safeguards agreement with that agency (article 3.vi).''

	The plan, and the annexes thereto, which constitute an integral part of the plan, were adopted by the Security Council as document number S/22872/Rev.1. This unprecedented document, drafted with Iraq's specific case in mind, is in fact the first legally binding document in which all activities prone to nuclear weapon proliferation are clearly and comprehensively defined. It is the first time that in an official document, the many ambiguous activities which broadly come under the name of ``peaceful nuclear activities,'' as well as those which are generally considered as non-military scientific research activities, are explicitly recognized as important for the acquisition or development of nuclear weapons. Similarly, by clearly defining those applications of nuclear physics and nuclear energy that are useful for ``medical, agricultural or industrial purposes,'' this document also defines which kind of ``peaceful nuclear activities'' are really benign from the point of view of nuclear weapon proliferation.
	In practice, in obliging Iraq ``not to acquire or develop nuclear weapons or nuclear-weapons-usable material or any subsystems or components or any research, development, support or manufacturing facilities related to the above'' (article 12 of resolution 687), the UN Security Council developed and accepted a document unambiguously defining what in essence is a \emph{nuclear free zone} and created a legal precedent which makes Iraq the first example of such a zone. This precedent is particularly significant because it included the development of procedures and equipments for ongoing monitoring and verification, which are now applied in Iraq.

	While many would object to the idea that resolutions 687 and 707 (or more precisely the plan making Iraq a nuclear free zone and thus a de facto nuclear weapon free zone) could be applied to or adopted by all nations, this idea merits much consideration in the light of the danger that nuclear weaponry represents for the world.

	Annex 1 of document S/22872 defines activities prohibited or permitted under resolutions 687 and 707. It is completed by Annex 3, a fifteen page long list of all items specifically prohibited, or may be prohibited if used in prohibited activities, and by Annex 4 which details permitted activities. Significantly, Annex 1 and Annex 3 are much more explicit and comprehensive than any previous official documents listing equipment and materials subject to nuclear export controls, including the so-called ``Zangger list'' \cite{AA.83}. Considering the importance of these annexes for future discussions on nuclear weapon free zones and an eventual nuclear free world, Annex 1 is reproduced in the Appendix 1 of this report.

	A salient feature of Annex 1 is that it makes a clear distinction between activities prohibited by Resolutions 687 and 707. Activities prohibited by resolution 687 (paragraphs 2.1-2.9 of Annex 1) are those which are clearly prohibited to non-nuclear-weapon states by the Non-proliferation treaty and those constituting a direct short-term nuclear weapon proliferation threat in case of diversion or misuse, and therefore put under IAEA safeguards.

	The prohibition of nuclear activities by Resolution 707 (paragraphs 2.10-2.18 of Annex 1) is much more comprehensive; it comprises all possible nuclear activities except applications of isotopes to agriculture, industry and medicine. While the activities put under IAEA safeguards are essentially those related to nuclear power generation by means of fission reactors, Resolution 707 prohibition extends to \emph{nuclear fusion} based on magnetic or inertial confinement (paragraph 2.15), \emph{production of isotopes} of any kind (paragraph 2.16) and \emph{particle accelerators} of all types (paragraph 2.17).

	In other words, Resolution 707 is a legal implementation of the suggestions made in 1980 that international safeguard measures should be extended to particle accelerator and fusion technologies \cite{AA.1,AA.2,AA.3}, and an explicit recognition of the fact that these technologies constitute a direct threat for nuclear weapon proliferation.

	A second important feature of Annex 1 is that it prohibits not only ``design, manufacturing, import of systems, equipment and components, pilot plant construction, commissioning and operation, or utilization,'' but also ``research and development'' on the specified activities. This is a very important novelty because, until Resolutions 687 and 707, research and development activities have always been excluded in arms control agreements \cite{AA.84}. In the case of the future Comprehensive Test Ban Treaty, only a very limited range of research activities, i.e., those in which a \emph{fission} chain reaction is started, will be prohibited. In particular, there will be no prohibition on thermonuclear fusion, including inertial confinement fusion, even though such research is ultimately motivated by the possibility of triggering large scale thermonuclear explosions without needing a fission primer. The potential of inertial confinement fusion for studying thermonuclear weapons physics and effects has been discussed in open literature since 1975 (see \cite{AA.2,AA.3,AA.84} and references therein).

	The necessity of arms control and disarmament measures at the research and development stage has been repeatedly stated albeit but by a few \cite{AA.3,AA.47,AA.84,AA.85,AA.86}, in contradiction with the prevailing opinion in the scientific community \cite{AA.87}. These measures should constrain both ``civilian'' and ``military'' research activities, including fundamental research as it was deemed necessary for example, in the case of antimatter \cite{AA.86}. In the instance of a ban of inertial confinement fusion research, the construction of very large laser facilities for fundamental research in astrophysics, hydrodynamics, high-pressure physics or plasma physics \cite{AA.88} would be forbidden.

\subsection{Intelligence failure or staging for counter-proliferation?}
\label{Sec.2.5.4}

	Until now, the most comprehensive coverage of Iraq's nuclear weapons program is provided in a series of articles by David Albright and Mark Hibbs in \emph{The Bulletin of the Atomic Scientists} and \emph{Arms Control Today}. These articles also provide a good introduction to some of the most disturbing problems relating to Iraq's case, such as the question of the ``intelligence failure'' and the prospect of a ``counter-proliferation'' policy replacing the current non-proliferation regime. A proper assessment of this ``failure'' is essential precisely because it is often quoted today in the context of either proliferation and nuclear terrorism \cite{AA.89} or counter-proliferation \cite{AA.90}.

	On the intelligence failure question, Albright and Hibbs' opinion in 1992 was categoric:
\begin{quote}
	{\it ``After the Osiraq bombing, Iraq simultaneously pursued several means of producing highly enriched uranium. Postwar revelations of Iraq's most developed enrichment route, based on archaic calutron electromagnetic separation technology, startled the world. Western intelligence agencies had been fully aware that Iraq was attempting to develop the means to enrich uranium, but had focussed on the gas centrifuge effort --- which matched current approaches used in several developed nations --- and missed the calutron effort completely''} \cite[p.5]{AA.91}.
\end{quote}

In 1993 however, Albright's opinion was somewhat more cautious:
\begin{quote}
	{\it ``Iraq is considered an intelligence failure because large-scale nuclear activities were not discovered by the IAEA or Western intelligence agencies'' (before the end of the Gulf War). The reasons for this failure, however, involve more than a deficiency in either safeguards or intelligence collection methods.

	Western governments did not aggressively pursue leads about Iraqi nuclear efforts or seriously impede Iraq's nuclear program during the 1980s. (...)

	Despite a failure to detect the full scope of Iraq's program, intelligence agencies knew enough before the Iraqi invasion of Kuwait to have justified some sort of intervention. (...)''} \cite[p.15]{AA.92}.
\end{quote}

It can safely be assumed that there are inconsistencies between what was known to Western intelligence agencies and what was done with that information. This was highlighted in the response of IAEA officials to international media criticism of the IAEA when Iraq's nuclear weapons program was uncovered. Jon Jennekens, retired Deputy Director General for Safeguards of the IAEA, in his first public statement since retiring, said for instance: 
\begin{quote}
	{\it ``The Americans and the British knew what (Saddam) Hussein was up to because they pulled a sting operation when the Iraqis were trying to import very high-precision timing devices from the U.S., through the U.K. They pulled a sting operation and arrested people so it was clear why the Iraqis wanted these instruments, but the information was never divulged to the IAEA secretariat''} \cite{AA.93}.
\end{quote}

In fact, the whole process which after the end of the Gulf War led to the uncovering by IAEA inspectors of the details of Iraq's nuclear weapons program, can be seen as a cover-up operation. One in which the IAEA was used to hide the extent of the knowledge that Western intelligence had already obtained before the war. For example, is it credible that it was only on the basis of information provided by two defectors, that the most important discoveries about Iraq's nuclear weapons program were made?  Firstly, in June 1991, a defector provided the information that led to the huge facility in which uranium was enriched with calutrons.  Secondly, in September 1991, another defector informed Western intelligence on the location of 25'000 documents.  These included design information definitely confirming that research and development in several key areas specifically related to nuclear weapons had been done in Iraq \cite{AA.94}.

	Moreover, it is difficult to believe that Iraq's enormous nuclear effort had remained undetected for so many years, an effort which cost over 10 billion dollars during the 1980s and employed 10'000 or more scientists, technicians and others. It is improbable that only a few isolated individuals like Gsponer, were aware of Iraq's long time interest in calutron technology. There are also several hints indicating that the full extent of Iraq's effort was already known before the Gulf War. For instance, this knowledge would certainly have been necessary to convince some countries to join or accept operation `Desert Storm', and President Bush's repeated suggestion before the war, that Iraq's bomb was only months away \cite{AA.95}. 

	A possible explanation of these events could be that from the early 1980s there was a gradual shift from a \emph{non-proliferation} to a \emph{counter-proliferation} policy \cite{AA.90,AA.96,AA.97}. If this is so, the lessons of Israel's precedent, the bombing of Osiraq in 1981, must have been taken into account.

	The bombing of Osiraq by the Israelis took place when the reactor was almost ready to produce plutonium. By comparison, Iraq's pilot calutron plant started in 1981. Assuming that intelligence services knew this, counter-proliferation action would still have been difficult from a political point of view. The military significance was not sufficient at the time: it was a small experimental facility.

	The installation of the first large alpha calutrons in the Tarmyia production plant only started in 1989. Not until 1989--1990 therefore could Israel or the USA have begun to put political or military pressure on Iraq: A counter-proliferation policy needs a clear cut proliferating situation in order to justify intervention. Otherwise the political cost, which was quite high in Israels' precedent of 1981, would be prohibitively high. 

	But Iraq invaded Kuwait in August 1990. Was this invasion encouraged by Saddam Hussein's belief that he was very close to having the atomic bomb in his hands? Was this just the kind of mistake the Western powers were waiting for in order to neutralize Iraq's growing military strength? Were the Western powers expecting such a mistake and did not intervene earlier by political or economical means because of a shift from non-proliferation to counter-proliferation? This would explain why Iraq was allowed to come so close to having an atomic bomb.

	In any event, 1991 was not only the year of the Gulf War --- which could be seen as the first major act of counter-proliferation --- but also the year in which France and China finally joined the NPT. An event that paved the way to the unconditional permanent extension of the NPT, decided in New York in 1995. Not all the motivations of France and China to join the NPT are known, but it may well have been that the assurance that deliberate force could be legitimized to keep nuclear weapons in the hands of the superpowers, must have been an important one. In declaring on the 31th of January 1992, that the \emph{proliferation} of weapons of mass destruction was a threat to international peace and security, the Security Council has in effect authorized the use of force against any proliferating state, including those which are not party to any international treaty.

\section{References}
\label{Sec.2.6}

\chapter{UN Security Council resolutions 687, 707, 
and 715, and their implications for a halt of all  
proliferation prone nuclear activities
--- a technical and legal assessment}
\pagestyle{myheadings}
\markboth{Report}{ISRI-96-06}
\label{Chap.3}

by {\bf Andre Gsponer}, {\bf Jean-Pierre Hurni}, and {\bf Stephan Klement}\\ ~\\ Report ISRI-96-06 ~~ --- ~~ 12 February 1997\\


\section{Introduction}
\label{Sec.3.1}

In paragraph 13 of United Nations Security Council (UNSC) resolution 687 \cite{BB.1,BB.2} adopted on 3 April 1991, the International Atomic Energy Agency (IAEA) was requested by the Security Council to carry out immediate on-site inspection of Iraq's nuclear capabilities and carry out a plan for the destruction, removal or rendering harmless of items prohibited to Iraq under paragraph 12 of the resolution 687. On 15 August 1991 the Security Council adopted a further resolution, number 707 \cite{BB.3}, obliging Iraq to ``\emph{halt all nuclear activities of any kind, except for use of isotopes for medical, agricultural or industrial purposes}, until the Council determines that Iraq is in full compliance with the present resolution and with paragraphs 12 and 13 of resolution 687 (1991) and the Agency determines that Iraq is in full compliance with its safeguards agreement with the Agency.''

	The plan, and the annexes thereto, which constitute an integral part of the plan, were approved by the UNSC in resolution 715 \cite{BB.4}  as  document number S/22872/Rev.1, 20 September 1991 \cite{BB.5}. This unprecedented document, drafted with Iraq's specific case in mind, is in fact the first legally binding document in which all activities prone to nuclear weapon proliferation are clearly and comprehensively defined. It is the first time that in an official document, the many ambiguous activities which broadly come under the name of 'peaceful nuclear activities', as well as those which are generally considered as non-military scientific research activities, are explicitly recognized as important for the acquisition or development of nuclear weapons. Similarly, by clearly defining those applications of nuclear physics and nuclear energy that are useful for 'medical, agricultural or industrial purposes', this document also defines which kind of 'peaceful nuclear activities' are really benign from the point of view of nuclear weapon proliferation.

	In practice, in obliging Iraq ``not to acquire or develop nuclear weapons or nuclear-weapons-usable material or any subsystems or components or any research, development, support or manufacturing facilities related to the above'' (article 12 of resolution 687), the UNSC developed and accepted a document unambiguously defining what in essence is a \emph{nuclear free zone} and created a legal precedent which makes Iraq the first example of such a zone. This precedent is particularly significant because it included the development of procedures and equipments for ongoing monitoring and verification, which are now applied in Iraq.

\section{Technical assessment}
\label{Sec.3.2}

Annex 1 of document S/22872 defines activities prohibited or permitted under resolutions 687 and 707. It is completed by Annex 3, a fifteen page long list of all items specifically prohibited, or may be prohibited if used in prohibited activities, and by Annex 4 which details permitted activities. Significantly, Annex 1 and Annex 3 are much more encompassing and comprehensive than any previous official documents listing equipment and materials subject to nuclear export controls, including the so-called `Zangger list' or London Guidelines \cite{BB.6}. {\bf Considering the importance of these annexes for future discussions, Annex 1 is reproduced in the Appendix 1 of this report.}

	A salient feature of Annex 1 is that it makes a clear distinction between activities prohibited by resolutions 687 and 707. Activities prohibited by resolution 687 (paragraphs 2.1-2.9 of Annex 1) are those which are clearly prohibited to non-nuclear-weapon states by the Nuclear Non-Proliferation Treaty (NPT) and those constituting a direct short-term nuclear weapon proliferation threat in case of diversion or misuse, and therefore put under IAEA safeguards.

	The prohibition of nuclear activities by resolution 707 (paragraphs 2.10-2.18 of Annex 1) is much more comprehensive; it comprises all possible nuclear activities except applications of isotopes to agriculture, industry and medicine (paragraphs 2.19-2.21). While the activities put under IAEA safeguards are essentially those related to nuclear power generation by means of fission reactors, resolution 707 prohibition extends to \emph{nuclear fusion}  based on magnetic or inertial confinement (paragraph 2.15), \emph{production of isotopes} of any kind (paragraph 2.16) and the use of \emph{particle accelerators} of all types (paragraph 2.17).

	In other words, resolution 707 is a legal implementation of the suggestions made in 1980 that international safeguard measures should be extended to particle accelerator and fusion technologies \cite{BB.7,BB.8}, and an explicit recognition of the fact that these technologies constitute a direct threat for nuclear weapon proliferation. A first step in that direction is that tritium and lithium-6 are now included in the list of nuclear-related dual-use equipment and material and related technology mentioned in the Warsaw Guidelines \cite{BB.9}. But, 'fusion' and 'accelerators' are not even mentioned as dual-use technologies, even though both technologies can be used to breed fissile material \cite{BB.8}, tritium \cite{BB.10} or antimatter \cite{BB.11} for military purposes. Similarly, in the amended London Guidelines \cite{BB.12}, the only change has been to include the full range of enrichment technologies without any reference to fusion materials or technologies. Finally, the Wassenaar Arrangement --- which in April 1996 succeeded to the Coordinating Committee on Export Controls (COCOM) two years after it ceased to exist --- includes several lists of sensitive dual-use equipment. These lists, however, only refer to particle accelerator and fusion technologies in the context of directed energy weapons, and to lasers as a controlled technology, without any explicit reference to their use in the nuclear fuel cycle of for thermonuclear fusion \cite{BB.13}.

	A second important feature of Annex 1 is that it prohibits not only ``design, manufacturing, import of systems, equipment and components, pilot plant construction, commissioning and operation, or utilization,'' but also \emph{research and development} on all nuclear activities excepted those permitted by paragraphs 2.19-2.21. This is a very important novelty because, until resolutions 687 and 707, research and development activities have always been excluded in arms control agreements \cite{BB.14}. For example, Article II of the NPT only forbids the \emph{manufacture} of nuclear weapons by non-nuclear-weapon States.

	In more recent agreements not specifically related to Iraq, the scope of the control of research and development is less generously defined. For example, research and development is explicitly mentioned in the Warsaw guidelines \cite{BB.9}. In particular, ``suppliers should not authorize transfers of equipment, material, or related technology (...) for use in a non-nuclear-weapon state in a nuclear explosive activity (where) 'nuclear explosive activity' includes research on or development (...) on any nuclear device or components or subsystems of such a device'' \cite{BB.9}. But the limited range of technologies covered by these guidelines leaves open the possibility of conducting proliferation prone research activities using particle accelerators and fusion technologies. Such activities are not possible in the African Nuclear-Weapon-Free Zone where Article 3 of the Treaty of Pelindaba explicitly prohibits conducting ``research on (...) any nuclear explosive device by any meansanywhere'' \cite{BB.15}. However, as in the Warsaw guidelines, this prohibition only applies to \emph{nuclear explosives}. In resolutions 687 and 707 prohibition of research applies to \emph{all nuclear activities} (including fundamental research) except research on application of radiation and isotopes in food and agriculture, medicine and industrial processes.

\section{Legal assessment}
\label{Sec.3.3}

The political implications and the legal status of resolutions adopted by the UNSC has been an issue of extensive discussions since the foundation of the UN more than fifty years ago. Whereas this was more or less of purely academic interest during the period of the Cold War due to the political limitations of the UNSC, the changed global political climate has allowed for new developments in this field that are of importance for practical purposes \cite{BB.16}.

	Today, it is mostly recognized that the UNSC has quasi-legislative and quasi-judicial authority with regard to the implementations of the provisions in Chapter VII of the UN Charter, i.e., when there is a threat to the international peace, breach of the peace or act of aggression \cite{BB.17,BB.18}. In the case of UNSC resolutions 687, 707 and 715 that were adopted in connection with Iraq's aggression against Kuwait the UNSC intended to reach several objectives, e.g., the implementation of a cease-fire agreement and an arms control regime (comprising the elimination of nuclear, chemical and biological weapons) upon Iraq (comprising the total elimination of nuclear, chemical and biological weapons including the corresponding production capabilities), a solution to the boundary dispute between the two states and the settlement of economic issues related to the former occupation of Kuwait by Iraq.

	The legal nature of UNSC resolutions and the legislative authority of the UNSC within the UN setting \cite{BB.17} can be characterized in the following way:

\begin{itemize}

\item legal acts are generated in a unilateral form, namely through the adoption of resolutions by the UNSC,

\item a special legal norm is created or a general legal norm is modified, in our case given by a significant extension of Iraq's existing safeguards obligations.

\end{itemize}

	Therefore, when we consider the legal status of UNSC resolutions 687, 707, 715 and the corresponding plan for their implementation that is given by the document S/22872/Rev.1, and following the respective legal expertise \cite{BB.17,BB.18}, we arrive at a new comprehensive and legally binding definition of all proliferation prone activities in the context of mastering the dismantling of the clandestine Iraqi nuclear weapons programme. Clearly, these developments can have substantial implications on further political and legal activities in this field, in particular in connection with current efforts to strengthen the IAEA safeguards regime.

	Of course, there are a variety of possible objections that can be made to our line of argumentation. From a political point of view it can be argued that the UNSC, because of its limited and biased membership \cite{BB.19}, cannot be accepted as a `lawmaking' authority. One can very reasonably support the opinion that the UNSC is not empowered to impose upon UN members such binding rules that are not restricted in their application to a specific case the UNSC has explicitly dealt with before.

	However, we have to take into consideration that the UNSC by adopting the resolutions mentioned above did not exclusively act with regard to the provisions of Chapter VII of the UN Charter but also in order to enforce Iraq's obligations under the NPT and to secure its future compliance after having broken crucial provisions of this treaty. Since the NPT is an international legal instrument with nearly global adherence, the legal consequences for the NPT regime as a whole cannot be neglected. In this sense the UNSC resolutions and the corresponding action plan should also be seen as implementation measures of the nuclear non-proliferation regime that had to be adopted in order to deal with with a situation where basic obligations of the NPT were violated. It cannot be denied that the UNSC has created in this way a new interpretation with regard to an effective implementation of the corresponding NPT provisions. The revelation of an extensive clandestine nuclear weapons research and development programme in Iraq has forced the UNSC to expand the scope of the existing IAEA safeguards system via adoption of resolution 707 and 715 in order to ensure Iraq's compliance with Art.II of the NPT.

\section{Conclusion}
\label{Sec.3.4}

In this article we have tried to give an assessment of the non-proliferation implications of the UNSC resolutions aiming at the containment of Iraq's nuclear ambitions. By necessity, it was important to rely on a technical and legal analysis.

	Because of the increasing complexity of technological development, it will be more and more important to approach emerging proliferation concerns at the earliest possible stage. In order to meet this objective, the necessary legal framework has to be continuously adapted. In this perspective, the case studied in this paper shows several directions for possible future development:

\begin{itemize}

  \item first, the scope of international treaties and arrangements should include all materials and technologies related to fission, fusion \cite{BB.20}, acceleration and annihilation processes;

  \item second, the treaties and arrangements should include effective measures of \emph{preventive arms control}  \cite{BB.21}, such as legally binding restrictions in all relevant areas of research and development, whether they are claimed to be for military or civilian purpose;

  \item third, the concept of 'peaceful nuclear activities' should be given a well accepted and unambiguous meaning by making explicit which materials, technologies and activities are really benign from the point of view of nuclear weapon proliferation. This is of crucial importance for the creation of genuine \emph{nuclear weapon free zones} and for serious discussions on a total abolition of nuclear weapons.

\end{itemize}

	In all three of these points the UNSC resolutions considered in this paper provide useful legal precedents. Moreover, our analysis indicates that the corresponding UNSC resolutions 687, 707 and 715 could be used in the future as the basis for a legally binding definition of all proliferation prone nuclear activities. Finally, not only did the events in Iraq partly trigger the still continuing 93+2 discussions on the IAEA safeguards system, but they obliged the international community to create what in essence is a \emph{nuclear free zone} and to develop adequate monitoring and verification procedures and equipment.

	It is therefore likely that future developments will show that the case studied in this paper will help to put additional pressure on the continuing process of expanding the scope of existing safeguards agreements.

\section{Notes and references}
\label{Sec.3.5}

 Dr.\ Andre Gsponer is a physicist and the director of the Independent Scientific Research Institute (ISRI) Box 30, 1211 Geneva 12, Switzerland.

\noindent Dr.\ Jean-Pierre Hurni is a physicist and Senior scientist at ISRI.

\noindent  DDr.\ Stephan Klement is physicist and lawyer working at the Mountbatten Center for International Studies, University of Southampton, Highfield, Southampton SO17 1BJ, United Kingdom. His work is supported by the `Fonds zur F\"orderung der wissenschaftlichen Forschung in Oesterreich', Project No. J01258-SOZ.

\chapter{Iraq's calutrons: 1991 - 2001}
\pagestyle{myheadings}
\markboth{Iraq's calutrons:}{1991 - 2001}
\label{Chap.4}


by {\bf Andre Gsponer}.\\ ~\\ Posted on internet on 31 July 2001.

\section{Introduction}
\label{Sec.4.1}

   In July 1991, shortly after the Gulf War, news from Iraq confirmed what I had concluded twelve years earlier:  That Iraq had decided to use the calutron electromagnetic isotope separation (EMIS) process to produce highly enriched uranium, i.e., the very same process that was actually used to produce the uranium-235 that was fissioned in the atomic bomb that exploded over Hiroshima in August 1945.

   At that time I was still recovering from a cancer treatment, and I had to wait until 1995 to find the energy to write a report on my discovery in 1979 of Iraq's intention to use the calutron technology, and my discomfort with the official thesis, namely that nobody knew before the Gulf War that Iraq had built a gigantic calutron enrichment plant and was very close to assemble its first atomic bomb.

   For a number of reasons the report I wrote in collaboration with Dr.\ Jean-Pierre Hurni is still not published.  In particular, I had great hopes that I would get additional information from the IAEA that would help putting the report into a final form.  But that did not happen with the unfortunate death of Dr.\ Maurizio Zifferero, late head of the UNSC 687 Action Team from 1991 to 1997, and his replacement by Mr.\ Garry Dillon who may have decided to cut off contact with me and not to keep Dr.\ Zifferero's commitments.

   Only thirty copies of the draft report, ISRI-95-03, were printed, and only a few additional photocopies were made.  Now, ten years after the Gulf War, the time has come for this document to be made publically available.  It is a matter of truth and justice that the facts I discovered in 1979 while working at CERN (the European Center for Nuclear Research, Geneva, Switzerland) are published.  At the least, this  report should help professional historians and political analysts to review what happened in Iraq before and during the Gulf War, e.g., in order to better understand the origin and the spectacular growth of the Iraqi nuclear weapon program, independently of the official statements made by the U.N. and the Allies, or by the U.S. and Iraqi governments.  Moreover, the correspondence related to this report with the IAEA and the reviewers, as well as the additional information on calutrons, EMIS, accelerator technology, CERN, and the origin of Iraq's calutron program collected between 1995 and 2001, should be a useful complement to the web pages opened by the IAEA Iraq Action Team on the Agency's {\it WorldAtom}\footnote{ http://www.iaea.org/worldatom/Programmes/ActionTeam/index.html .} web site in June/July 2001.

   Although the 1995 report could have been rewritten and a number of improvements made, I have decided to leave it exactly in its original form, apart from a few minor modifications such as a few spelling or typographical mistakes.  In the same spirit, the written commentaries from various sources are presented in their original form as copies of the letters I have received.  Of particular interest is the correspondence with Dr.\ Maurizio Zifferero of the IAEA, who was as much interested in knowning the truth about the origin of the Iraqi calutron program, than in making sure that whatever would be published would not contribute to the further proliferation of nuclear weapons.

   Consequently, the form of the following presentation of the documents on EMIS technology and Iraq's calutrons is that of a chronology, interleaved with some personal comments that I hope will be useful to the reader.  It begins with Suren Erkman's paper in {\it Le Journal de Gen\`eve} of April 1995 and the reviewing of the main report, ISRI-95-03, that started in October 1995.  There are then two sections on two attempts, one to present at a conference, the other to publish in an academic journal, some of the material contained in the report ISRI-95-03. This is followed by a section presenting some more recent documents that would have been mentioned in an updated version of the report, as well as links to related web sites.  Finally, I added two sections providing independent confirmation of my 1979 findings at CERN, as well as a conclusion emphasizing some lessons I learned from this more than twenty years long experience.

\section{Suren Erkman's article in  {\it Le Journal de Gen\`eve}
         of 22-23 April 1995}
\label{Sec.4.2}

   When I returned to Geneva in January 1995 I met with a number of friends and former colleagues who encouraged me to publish what I knew about the origin of Iraq's calutron program.  In particular, Suren Erkman, a science journalist at  {\it Le Journal de Gen\`eve} (which from an intellectual and political standpoint was considered as the ``best'' newspaper in Geneva, and the only one with an international audience) wrote a long article based on a careful inquiry comprising a number of interviews including the two main CERN witnesses, i.e., myself, and Dr.\ Klaus Freudenreich who was repeatedly approached by Iraqi engineers in early 1979.  This article was announced on a full size poster [Fig.~\ref{fig:Poster}] and published in the weekend edition of 22-23 April 1995  of  {\it Le Journal de Gen\`eve} [Figs.~\ref{fig:Page 1} and \ref{fig:Page 5}], see Paper~\ref{Chap.1}.

   The following Monday there were a number of emotional reactions in Geneva and especially at CERN, and six people made the effort to write down their reaction to Suren Erkman's paper:

\begin{enumerate}

\item 24 April 1995 : Dr.\ Med. Denis Dupont, Geneva

\item 24 April 1995 : Dr.\ Reinhard Budde, CERN staff member

\item 25 April 1995 : Prof.\ C.H. Llewellyn Smith, Director General of CERN 

\item 27 April 1995 : Prof.\ Jack Steinberger, Nobel laureate and CERN staff member

\item 28 April 1995 : Dr.\ Gerd Harigel, CERN staff member and GIPRI board member

\item 28 April 1995 : Dr.\ Jean-Pierre Stroot, CERN staff member and GIPRI board president

\end{enumerate}

   The only positive reaction was by Dr.\ Denis Dupont, M.D., who asked to receive a copy of my forthcoming report (entitled at the time ``{\it Calutrons: 1945-1995}'') that was announced in Suren Erkman's paper.

   The other five reactions were negative, and very emotional except for that by Prof.\ C.H. Llewellyn Smith, Director General of CERN.  This official reaction, which is basically an attempt to dismiss CERN's responsibility, is reproduced for the record as [Letter~\ref{Doc.03-Llewellyn-Smith-25Apr95}]. Since {\it Le Journal de Gen\`eve} article was written in French, all these reactions were in French, except for the one by Prof.\ Jack Steinberger, a member of Pugwash who co-edited with Joseph Rotblat the book: A Nuclear-free World: Desirable? Feasible? (Westview Press, Boulder Colorado, 1996).

   The four reactions by current or former CERN staff members are all in the same vein and it is fortunate that the one by Nobel laureate Jack Steinberger [Letter~\ref{Doc.04-Steinberger-27Apr95}] is a good summary of the other three.  In particular, it is claimed that ``{\it Gsponer has become psychopathic on the issue of alleged military involvements of CERN [...] His stories of military use of CERN are pure invention,}'' whereas a number of positive details are given on Jafar Dhia Jafar, who after working at Imperial College spent a four-year period at CERN before becoming the head of Iraq's nuclear weapons program.

   For example, it is said in Prof.\ Steinberger's letter ``{\it that at this time Jafar had no Iraqi military motives or connections.}''  This is wrong since it was well known to all members of the CERN collaboration in which Jafar Dhia Jafar worked in the early 1970s that there were two high-ranking military officers in that collaboration: one Israeli and on Iraqi!  Moreover, it is now abundantly confirmed that Dr.\ Jafar was actively involved in Iraq's clandestine nuclear weapon program since its very beginning, which includes all the time he had spent at CERN.

   Otherwise, in agreement with all recollections by CERN staff members that I have spoken to and who have personally known Jafar Dhia Jafar, it seems that he was very good at building lasting relationships, and was much appreciated as a colleague and friend.  For instance, senior CERN staff member Douglas R.O. Morrison remembered him very well as a an excellent bridge player...

   Strikingly, all five reactions by CERN staff members did not show any interest in knowning more about the facts: none of them asked to receive the announced report ``{\it Calutrons: 1945-1995}.''  On the contrary, little concerning the substance or the facts is disputed:  All criticisms are directed at me as a person, as if I was a traitor, and Jafar Dhia Jafar is  presented as a loyal CERN user.

   On the other hand, the IAEA was immediately interested and quick to react: the head of UNSC 687 Action Team in Iraq, Dr.\ Maurizio Zifferero, called Suren Erkman at the beginning of May already [Letter~\ref{Doc.05-To-Zifferero-5May95}].  Moreover, the joint IAEA/UNIDO library sent a purchase order for ``{\it Calutrons: 1945-1995}'' on 17 May 1995.

\section{``Iraq's Calutrons'' report [ISRI-95-03] and its reviewing}
\label{Sec.4.3}

   Work on ``{\it Calutrons: 1945-1995}'' in collaboration with Dr.\ Jean-Pierre Hurni spanned over the Summer and Autumn 1995.  I wanted the document to be useful as much for understanding the past as for preventing further proliferation of particle accelerator-based nuclear technology such as electromagnetic isotope separators, plutonium or tritium breeders, antimatter production, particle beam weapons, missile defence, etc.  A big effort was therefore made to be as pedagogical and comprehensive as possible, both from the technical and the political standpoints.  The final document was larger than expected, and its final title became ``{\it Iraq's calutrons --- Electromagnetic isotope separation, beam technology, and nuclear weapon proliferation}'' in order to emphasize its broad focus on the nuclear weapon proliferation implications of particle beam technology.

   Thirty copies of the report, i.e., Paper~\ref{Chap.2}, were printed and the reviewing started by sending copies to the IAEA/UNIDO library and to the Action Team [Letter~\ref{Doc.08-To-Zifferero-8Nov95}].  Dr.\ Zifferero answered by asking permission to copy it to two experts [Letter~\ref{Doc.09-Zifferero-15Nov95}], one in LLNL (Lawrence Livermore National Laboratory, U.S.A.) the other in ORNL (Oak Ridge National Laboratory, U.S.A.).  I replied by sending two extra copies of the report [Letter~\ref{Doc.10-To-Zifferero-20Nov95}].

   The reviewing took place between 8 November 1995 and 18 June 1996 and consisted of inviting commentaries from 22 people: 3 science journalists, 1 international lawyer, and 18 physicists.   

\begin{enumerate}

\item Mr.\ Suren Erkman, Journal de Gen\`eve, Geneva

\item Mr.\ William Broad, New York Times, New York

\item Mr.\ Mark Hibbs, Nucleonics Week, Bonn

\item DDr.\ Stephan Klement, MCIS, Southampton

\item Dr.\ Jafar D. Jafar, IAEC, Baghdad

\item Dr.\ Maurizio Zifferero, IAEA, Vienna

\item Prof.\ Jack Steinberger, CERN, Geneva

\item Dr.\ Klaus Freudenreich, CERN, Geneva

\item Dr.\ Douglas R.O. Morrison, CERN, Geneva

\item Dr.\ Richard L. Garwin, IBM, Yorktown

\item Dr.\ Christopher C. Paine, NRDC, Washington

\item Dr.\ Harry Bernas, CSNSM, Orsay

\item Dr.\ Martin Kalinowski, INESAP, Darmstadt

\item Dr.\ Wolfgang Liebert, IANUS, Darmstadt

\item Dr.\ Annette Schaper, PRIF, Frankfurt

\item Dr.\ David Albright, ISIS, Washington

\item Prof.\ Robert F. Mozley, CISAC, Stanford

\item Dr.\ Bhupendra Jasani, Kings College, London

\item Dr.\ Frank Barnaby, SIPRI, Stockholm

\item Dr.\ Richard Kokoski, SIPRI, Stockholm

\item Dr.\ Paul J. Persiani, ANL, Argonne

\item Dr.\ Isaac Chavet,  SOREQ NRC, Yavne, Israel

\end{enumerate}

   The only person who was invited to make commentaries after the 18 June 1996 was Dr.\ Isaac Chavet, a leading member of Israel's electromagnetic separation program, who wrote to me spontaneously.

   As will be seen, only four people supplied substantial comments: Dr.\ Maurizio Zifferero, Prof.\ Robert F. Mozley, DDr.\ Stephan Klement, and Dr.\ Isaac Chavet.

   Trying to correspond with Dr.\ Jafar was a special case because of the sanctions imposed on Iraq.  I had first to write to the Iraqi permanent mission in Geneva [Letter~\ref{Doc.11-To-Al-Tikriti-20Nov95}] which answered that I had to send any correspondence for Dr.\ Jafar to the IAEA who would decide if the mail could be forwarded, and whether he could be allowed to comment or not upon ISRI's report [Letter~\ref{Doc.12-Al-Tikriti-16Jan96}].

   I therefore wrote to Dr.\ Zifferero [Letter~\ref{Doc.13-To-Zifferero-22jan96}] with a long letter to Dr.\ Jafar [Letter~\ref{Doc.14-To-Jafar-22jan96}].  The answer of Dr.\ Zifferero came very soon [Letter~\ref{Doc.15-Zifferero-26Jan96}].  It was negative, but at the same time appreciative of the quality of the questions I had raised since some of them were to be added to the list of questions the Action Team had about Iraq's calutron program. Moreover, the answer was also very encouraging since Dr.\ Zifferero suggested a compromise: that I should be given access to at least part of the ``Full, Final and Complete Declaration'' (FFCD) of the activities carried out under the clandestine nuclear program when this declaration would be completed.

   Another encouraging event happened a few months later:  I received a phone call from Iraq's permanent mission in Geneva and a secretary told that {\it ``Jafar Dhia Jafar sends you his regards and encouragements for your efforts to achieve nuclear disarmament.''}

   Meanwhile, I received the first substantial comment, that of Prof.\ Robert F. Mozley [Letter~\ref{Doc.16-Mozley-28Feb96}] who had written a report that I did not know at the time I wrote my report with Jean-Pierre Hurni:
\begin{itemize}

\item  Robert F. Mozley, \emph{Uranium enrichment and other technical problems relating to nuclear weapons proliferation,} Center for International Security and Arms Control (CISAC), Stanford University, 320 Galvez Street, Stanford CA 94305-6165, U.S.A., (July 1994) 65 pp.

 \end{itemize}

   In his letter, Prof.\ Mozley referred to a paper by Dr.\ Paul J. Persiani of the Argonne National Laboratory that I had also missed in my literature search with Jean-Pierre Hurni:
\begin{itemize}

\item  Paul J. Persiani, \emph{Non-Proliferation Aspects of  Commercial Nuclear Fuel Cycles,} report to the  Institute  of  Nuclear  Materials  Management, 33rd Annual Meeting, Orlando, Florida, (July 1992) 4 pp. Available from P.J. Persiani, TD 207, Argonne National Laboratory, 9700 South Cass Av., Argonne, IL 60439, U.S.A.

 \end{itemize}

   As the University of Geneva library was not able to get hold of this paper, I wrote to Dr.\ Persiani [Letter~\ref{Doc.17-To-Persiani-17Jun96}] who was kind enough to send me a copy of his paper.

   The report of Prof.\ Mozley, and even more specifically the paper of Dr.\ Persiani, emphasize that unpublished studies made in the U.S.A. in the late 1970s (i.e., precisely at the time when Iraq started to be actively interested in the calutron technology) had concluded that:
\begin{quote}
{\it ``a potential proliferation path, using calutrons for isotopic enrichment of low-enriched uranium, LEU, to highly enriched uranium, HEU, could be developed with mature technology by countries pursuing a clandestine weapons program.''}
\end{quote}

   Therefore, since the documents of Prof.\ Mozley and Dr.\ Persiani have been published after the Gulf War, its seem that I was the only one to have consistently tried to warn publically about this proliferation path during the 1980s.  In this respect, as a complement to what is explained in the report ISRI-95-03, I should add that on 13 July 1982 I submitted with Bhupendra Jasani to Frank Blackaby, then Director of SIPRI,  a research proposal entitled ``Military applications of high energy beams'' that included a chapter on ``Enrichment of uranium and plutonium using accelerators and lasers'' in which we planed to discuss the proliferation potential of the calutron technology.

   To end the reviewing process I sent on 18 June 1996 a letter with Dr.\ Jean-Pierre Hurni to all the above mentioned people, but we did not receive any further comments.

\section{Attempt to present a paper at the EMIS-13 conference
         and spontaneous contribution of Dr.\ I. Chavet }
\label{Sec.4.4}

   The reviewing being essentially terminated, I thought it would be a good idea  to present a summary of the report ISRI-95-03 at the forthcoming international conference of world experts of electromagnetic isotope separation, EMIS-13, that was to take place on September 23-27, 1996, at Bad D\"urkheim in Germany.  I therefore wrote to Prof.\ G. M\"unzenberg, Chairman of EMIS-13, and submitted the abstract of a contribution [Letter~\ref{Doc.19-To-emis13-19Jun96}].

   A few days later I had the good surprise to receive a letter from Dr.\ Isaac Chavet who had been mainly active in Israel's mass separation systems [Letter~\ref{Doc.20-Chavet-25Jun96}]. (See Sec.~\ref{Sec.2.3.10} of ISRI-95-03: ``EMIS in Israel.'')  In my answer I asked him to send his comments on our report and on Israel's EMIS facilities [Letter~\ref{Doc.21-To-Chavet-1Jul96}]. I also took this opportunity to write to Dr.\ Zifferero to remind him about our interest in receiving the information he had promised and about the possibility to meet some members of his Team at EMIS-13 [Letter~\ref{Doc.22-To-Zifferero-1Jul96}].

   In his answer [Letter~\ref{Doc.23-Zifferero-2Aug96}] Dr.\ Zifferero confirmed his commitment to send us a copy of the EMIS part of the FFCD as soon as possible.  However, concerning EMIS-13, he explained that nobody of his team would have the time to attend the conference.

   In any event, I still had no reaction from EMIS-13 and time was running out to write the paper for this conference.  I therefore decided to call Prof.\ M\"unzenberg on the telephone in order to ask him whether our contribution had been accepted or not.  He told me {\it ``that our abstract had been discussed, but that it was felt to deal mostly with old technologies while EMIS conferences are discussing concepts that are on the frontier in the field.''}  The answer was a polite {\it ``no.''}

   As a consolation, I received a few days later Dr.\ Chavet's commentary in the form of a letter of three pages with many interesting details, as well as a positive introductory remark on my efforts at trying to stop the proliferation of nuclear weapons [Letter~\ref{Doc.24-Chavet-6Aug96}].

\section{Attempts to publish a paper [ISRI-96-06] on the legal
         implications of the UNSC resolutions related to Iraq}
\label{Sec.4.5}

   In June 1996 I had the good fortune to meet with DDr.\ Stephan Klement, a physicist and international lawyer working at the Mountbatten Center for International Studies at University of Southampton.  Discussing the question of the legal implications of the U.N. Security Council resolutions related to Iraq's nuclear weapon program we reached the conclusion that it would be a good idea to write a specific paper expanding the ideas sketched in Sec.~\ref{Sec.2.5.3} of ISRI-95-03 and submitting it to a journal of international law.  This required adding a legal assessment to the essentially technical assessment done by Dr.\ Jean-Pierre Hurni and myself.

   The draft of the paper was circulate for review to a few experts in international law (i.e., Alyn Ware and members of the board of the Lawyer's Committee for Nuclear Policy, LCNP) and then sent to {\it The International and Comparative Law Quarterly} which did not accept the paper for publication [Letter~\ref{Doc.25-ICLQ-28Aug96}].

   Before sending the paper to another journal, I wanted to inquire at the IAEA about the fate of the FFCD, because it could have contained information affecting our paper with DDr.\ Stephan Klement.  The pretext came in the form of the discovery of a very interesting recent paper by R. Meunier, a French EMIS specialist who with others built the  ``Sidonie''  electromagnetic  separator of Orsay in the 1960-70s (see Sec.\ref{Sec.2.3.7} of ISRI-95-03: ``EMIS in France''): 
\begin{itemize}

\item  R. Meunier, \emph{Faisabilit\'e et evaluation exploratoire d'un projet pour la s\'epa\-ra\-tion isotopique du caesium issu de la fission,} report  CSNSM-94-30 (Octobre 1994). Available from R. Meunier, C.S.N.S.M., IN2P3-CNRS, Bat. 108, F-91405 Orsay Campus, France.

\end{itemize}

   As I mentioned in my letter of 30 December 1996 to Dr.\ Zifferero [Letter~\ref{Doc.26-To-Zifferero-30Dec96}], this document describes  an industrial scale electromagnetic  separation  system  which could handle 4.3*238/135 = 7.6  tons  of  natural  uranium  per  year.  It  is a very detailed study by somebody who has  spent his life building high-current electromagnetic separators. The study  is especially interesting because it gives precise estimates for  the  construction and operating costs of an industrial scale EMIS plant.

   Unfortunately, this letter may have never reached Dr.\ Zifferero because I later learned that he died rather suddenly from a cancer.

   I therefore decided with Dr.\ Jean-Pierre Hurni and DDr.\ Stephan Klement to finalize our paper and to send it on 12 February 1997 to {\it Security Dialogue}, which rejected it on 20 May 1997 [Letter~\ref{Doc.27-SecurityDialogue-20May97}].  In his letter of rejection, the editor stressed the facts that the resolutions in question were quite old at the time (i.e., more than five years) and that they actually had little legal weight.  While both of these facts are true, it turns out that these Security Council resolutions were nevertheless providing the only legal basis for what UNSCOM was doing in Iraq.  If international lawyer agree that they have little or no value, then what about the legal status of the ongoing U.N. and other coercive operations in Iraq?

\section{The loss of contact with the IAEA}
\label{Sec.4.6}

   While the death of Dr.\ Zifferero in 1997 was an unfortunate incident of life, the whole political situation concerning Iraq, UNSCOM, and the sanctions regime changed very dramatically in the first half of 1997 as a consequence of the implementation by the U.S. of new policies against Iraq.

   Concerning UNSCOM, U.N. Secretary-General Kofi Annan announced on 1st May 1997 that the chairman will be replaced by the Australian Richard Butler, who was soon to assume a much more anti-Iraq position than his predecessor, the Swede Rolf Ekeus.  Similarly, the successor of Dr.\ Zifferero,  Mr.\ Garry Dillon, was apparently chosen as much more for his political correctness than for his professional credentials. (Mr.\ Garry Dillon was the chief inspector of most inspection campaigns in Iraq since 1993 until they were interrupted in 1997.)  Consequently, my hopes of positive collaboration with the IAEA and UNSCOM faded out.

   What I had been promised by Dr.\ Zifferero was access to part of the ``Full, Final and Complete Declaration.''  The first version of this FFCD had been supplied to UNSCOM on 12 March 1992, and a 1019 pages long revised version on 1 March 1995. In 1996, the only part of the FFCD that was publically available was the table of content of a draft version attached to document S/1996/261.  This table of content indicated that EMIS technology was as expected a substantial part of the FFCD and that the information it contained was certainly much more detailed than what I needed to answer the questions I had raised in my letter of 22 January 1996 to Dr.\ Jafar.  In fact, according to the more recent IAEA documents S/1988/312 and S/1999/127, the final version of the FFCD (dated 25 March 1998) is essentially a consolidated version of the text dated 7 September 1996 supplemented by written revisions and additions.  Therefore, by early 1997, Dr.\ Zifferero, his deputy, or later his successor, could have fulfilled the promise made in the letters I had received in 1995 and 1996.

   Since I was not receiving any response from UNSCOM, I made a last attempt when I learned that Dr.\ Richard Hooper (the director of IAEA's Department of Safeguards who had also been involved in UNSCOM inspections) was going to give a lecture on 1st December 1997 at a conference of nuclear weapon proliferation experts to which I had been invited.  I wrote him a letter [Letter~\ref{Doc.28-To-Hooper-21Nov97}] that he did not see because he was not in Vienna before the conference.  However, at the conference, where I gave him a copy of my letter, he told me that I should not write to the successor of Dr.\ Zifferero since he would personally transmit my letter to Mr.\ Garry Dillon.  Unfortunately, I never received any answer from him.

\section{Additional technical papers and sources of information}
\label{Sec.4.7}

   Since 1995 I have become aware of a number of technical reports and books that would have been included in the bibliography of an updated version of the 1995 report on Iraq's calutrons. Three of those (by R. Mozley, P. Persiani and R. Meunier) have already been mentioned but I repeat them below for completeness.  As can be seen, the first one is dated October 1945: it should have been included in the 1995 version of the report already because it contains the first open technical discussion of the wartime EMIS effort in the U.S.A.

\begin{itemize}

\item H.D.\ Smyth, \emph{Atomic energy for military purposes}, Rev. Mod. Phys. {\bf 17} (October 1945) 351--470. Chapter IX (pp. 422--429) is a general discussion of separation of isotopes, and chapter XI (pp. 437--444) discusses electromagnetic separation of uranium isotopes.  This article is a republication of a report first issued by the  ``Manhattan District,'' U.S. Corps of Engineers, and constitutes the scientific basis of the book: Atomic Energy for Military Purposes (Princeton University Press, 1945).   

\item S.\ Villani, ed., Uranium Enrichment, Topics is applied physics, Vol. 35 (Springer Verlag, Berlin, 1979) 322 pp.   

\item Paul J.\ Persiani, \emph{Non-Proliferation Aspects of  Commercial Nuclear Fuel Cycles}, report to the  Institute  of  Nuclear  Materials  Management, 33rd Annual Meeting, Orlando, Florida (July 1992) 4~pp. Available from P.J. Persiani, TD 207, Argonne National Laboratory, 9700 South Cass Av., Argonne, IL 60439, U.S.A.   

\item Robert F.\ Mozley, \emph{Uranium enrichment and other technical   problems relating to nuclear weapons proliferation}, Center for International Security and Arms Control, Stanford University, 320 Galvez Street, Stanford CA 94305-6165, U.S.A.\ (July 1994) 65~pp.   

\item Robert W.\ Selden, \emph{Accelerators and national security --- The evolution of science policy for high-energy physics, 1947-1967}, History and Technology, {\bf 11} (1994) 361--391.   

\item R.\ Meunier, \emph{Faisabilit\'e et evaluation exploratoire d'un projet pour la s\'epara\-tion isotopique du caesium issu de la fission}, report  CSNSM-94-30  (Octobre 1994) 26 pp. Available from R. Meunier, C.S.N.S.M., IN2P3-CNRS, Bat.\ 108, F-91405 Orsay Campus, France.   

\item Richard Kokoski, Technology and the Proliferation of Nuclear Weapons, SIPRI monograph (Oxford University Press, 1995) 351 pp.   

\item Joseph Magill et al., \emph{Accelerators and (non)-proliferation}, Paper presented at the ANS Winter Meeting in Albuquerque, New Mexico (Nov.\ 16-20, 1997) 7 pp. Available form J.\ Magill, Institute for Transuranium Elements, P.O. Box 2340, D-76125 Karlsruhe, Germany, or in the publication section of ITU's website.\footnote{ http://ituwebpage.fzk.de .}   

\item J.\ Magill and P.\ Peerani, \emph{(Non-) Proliferation aspects of accelerator driven systems}, J.\ Phys.\ IV France {\bf 9}, Pr7-167 (1999) 19 pp.  Available form J. Magill, Institute for Transuranium Elements, P.O. Box 2340, D-76125 Karlsruhe, Germany, or in the publication section of of ITU's website.\footnote{ http://ituwebpage.fzk.de .}   

\end{itemize}

   Of course, there is also a lot of new published technical information related to calutron technology and electromagnetic isotope separation.  In particular, such information can be found in the proceedings of the last EMIS conference (EMIS-13), or the next one (EMIS-14, May 6-10, 2002, in Canada)\footnote{ http://emis14.triumf.ca/info/ .}, as well as of the latest magnet technology conferences, the next one of which (MT-17) taking place on 24-28 September 2001 at CERN.\footnote{ http://www.cern.ch/MT-17/ .}

   Moreover, there are a number of web sites which collect information on Iraq's former nuclear weapons program and related disarmament questions:

\begin{itemize}

\item {\bf Iraq Watch}, Wisconsin Project on Nuclear Arms Control, Washington:\\  http://www.iraqwatch.org/ .

\item {\bf Special Collection: Iraq}, The Bulletin of Atomic Scientists, Chicago:\\ http://www.bullatomsci.org/research/collections/iraq.html .

\item {\bf Iraq Special Collection}, Center for Nonproliferation Studies, Monterey:\\ http://cns.miis.edu/research/iraq/index.htm .

\item {\bf Country Assessment: Iraq}, Institute for Science and International Security, Washington:\\ http://www.isis-online.org/publications/iraq/ .

\item {\bf IAEA Iraq Action Team web pages}, Vienna:\\ http://www.iaea.org/worldatom/Programmes/ActionTeam/index.html . 

\end{itemize}

   The availability of the latest documents, reports, and activities of IAEA's Iraq Action Team on the Agency's {\it WorldAtom} web site was announced by the IAEA in March 2001.  Possibly the most useful document of general interest on this site is the fourth consolidated report S/1997/779 which provides an overview of the activities undertaken by the IAEA since it began the implementation of its obligations to carry out on-site inspection of Iraq's nuclear capabilities. However, this and the other documents give only little information on the origin of Iraq's nuclear weapons program, on how they acquired information on nuclear technology and nuclear weapons, and on other aspects that are essential for devising and implementing truly effective disarmament and preventive arms control policies.  In particular, the {\it Action Team Fact Sheet}\footnote{ http://www.iaea.org/worldatom/Programmes/ActionTeam/nwp2.html .}  of 13 July 2001is purely descriptive and totally neutral on these matters, as if the IAEA which had been created to promote atomic energy had no responsibility in the proliferation of nuclear technology.

   Finally, there is certainly some insight to be gained from a study of the differences and similarities between the 1991 Gulf War, and the 1999 intervention in Kosovo and previous events in former Yugoslavia.  For instance, during the war over Kosovo, two Serb scientists were asked to leave CERN, a decision which followed a couple of years old controversy about what CERN should do with the scientists from ex-Yugoslavia [See, {\it Tribune de Gen\`eve} (7 f\'evrier 1993, page 5; 8 f\'evrier 1993, page 13).]  The facts (related to the present subject) are that a cyclotron was under construction at the Vinca Institute for Nuclear Research, and that Serbia has a long standing expertise in calutron technology [see, B. Dunjic, \emph{Multi-ion source electromagnetic isotopic separator,} Nucl. Inst. and Methods, {\bf 38} (1965) 109--122)].

\section{Independent confirmation of Iraq's early interest in EMIS technology}
\label{Sec.4.8}

   In Autumn 1997 I met a retired Algerian professor living in Paris named Mohamed Larbi Bouguerra.  He told me that he had met a CERN physicist named Wilson who knew Jafar Dhia Jafar personally, and who knew a number of things about Iraq's nuclear weapons program.  Since Wilson is quite a common name (there are eight ``Wilsons'' just in the CERN telephone directory), it was rather difficult to find out which Wilson it was, especially since Prof.\ Bouguerra himself was not too sure about his recollections. 

   Fortunately, Jean-Pierre Hurni finally discovered an article by Richard Wilson, Mallinckrodt Professor of Physics at Harvard University, who had published an article entitled {\it Iraq's uranium separation: The huge surprise} in the Middle East monthly {\it New Outlook}, Vol.34, No.5 (Tel Aviv, September/October 1991) 36--40.  On page 38, Prof.\ Wilson writes:
\begin{quote}
   {\it ``The scientific world is international.  Scientists are inquisitive.  In the 1970s Iraqi scientists collaborated with Russians, who automatically kept an eye on them. In 1981 the Iraqi wanted Western collaboration.  In 1982 I made a personal visit myself that they were not making bombs with their publically known facilities.  Their best scientist, Dr Jafaar, wanted to study inelastic neutron scattering on the ellipsoidal nuclei of the rare earths using separated isotopes which were known to be available from Oak Ridge.  I carried to U.S. DoE a request  for a loan of the isotopes, a loan which could have been accompanied by a scientific collaborator reporting to any intelligence agency he wished.  Although these isotopes were of no conceivable use in a bomb program, no U.S. government official wanted to appear to help the Iraqis, and the request was denied.''}
\end{quote}

   With hindsight, we can interpret this initiative of Dr.\ Jafar as an attempt to establish a scientific link with Oak Ridge, the laboratory where the World War II calutron plant was built, and where a couple of wartime calutrons were still in daily operation, despite the fact that they were nearly forty years old.

\section{Independent confirmation of Iraq's use of CERN 
as a source of information}
\label{Sec.4.9}

   David Albright and Kevin O'Neill of the {Institute for Science and International Security} (ISIS) in Washington posted on the ISIS web site a document entitled \emph{Iraq's Efforts to Acquire Information about Nuclear Weapons and Nuclear-Related Technologies from the United States} dated  November 12, 1999.\footnote{ http://www.isis-online.org/publications/iraq/infogather.html .}  This document, based on a report and interviews of Khidhir Hamza, a senior Iraqi nuclear scientist who held several high-level positions in Iraq's pre-Gulf War nuclear weapons program, shows how Iraq obtained information from various sources including CERN.  In particular, in the section entitled \emph{Observation: Magnet designs were obtained from abroad} one reads:
\begin{quote}
   {\it In one interview, Hamza discussed how Iraq sought computer programs in the 1970s to design magnets for the EMIS program. Many of the programs obtained by Iraq were of U.S.-origin. According to Hamza:

         ``We tried to get some standard U.S. magnet design programs. We found two or three at CERN. These were originally American programs, but CERN used them to design their own magnets. Jaffar brought with him the entire CERN library when he came back to Iraq in 1974. We also sent some people back to CERN and they also brought back some tapes.

         Another source of programs was a library at Belfast, in Northern Ireland. There is a library there, which has an international repository of [computer] programs. You can join in and gain access for a couple of thousand dollars a year, so we joined that library -- I was the corresponding member (...) we [also] purchased a professional package from some companies that designed magnets for CERN. One of them was a Swedish company. I forget the name. The other was German. I don't remember its name, either. According to Jaffar, these two contracts would result in an EMIS magnet, but each, by itself, would not'' (May 27, 1999 interview of Khidhir Hamza).

   Iraq had difficulty in finding programs compatible to its computer systems. According to Hamza, ``the only problem [with the CERN magnets] is that they were installed on CDC computers in Geneva and we had to install them on IBM. We had to do a lot of changes in the command systems in the program, which we did'' (May 27, 1999 interview of Khidhir Hamza).}
\end{quote}

   Another publication of ISIS by David Albright, Corey Gay and Khidhir Hamza entitled `\emph{Development of the Al-Tuwaitha site: What if the public or the IAEA had overhead imagery?}\footnote{ http://www.isis-online.org/publications/iraq/tuwaitha.html .} is to my knowledge the only document apart from ISRI-95-03 which is highlighting the contradiction between the huge Iraqi nuclear weapon program, the fantastic information gathering capabilities available to the intelligence agencies of the great powers, and the claim that Iraq's effort was discovered only \underline{after} the Gulf War.

\section{Conclusions}
\label{Sec.4.10}

   Over the twenty-two years that began in 1979 with my analysis of the reasons why Iraqi engineers could be interested by the constructions details of a large magnet such as the one of the NA10 experiment at CERN, everything that I had concluded has been confirmed.

   Today, I can even claim that if I would have been successful in carrying out the research program that I had drafted for the {Geneva International Peace Research Institute} (GIPRI) which I created at the end of 1979, and been able to  publish its conclusions, history might have been different:  Without the prospect of soon being a nuclear power, Iraq might have been less tempted to invade Kuwait, a major war would have been avoided, and the Middle East situation would be very different now.

   However, since I was not alone in these efforts to avoid further nuclear proliferation and to eliminate nuclear weapons, and since many of these efforts were made by scientists much more competent, famous, or influential than me, the main unanswered question is why so little progress has been made so far. 

   I think that the main elements of an answer to this question are illustrated by the documents presented in these web pages, which is why I feel it was so important to post them.  These elements fall in two main categories: political and scientific.

   First, if I may start with the political elements, I think that the shift from a non-proliferation to a counter-proliferation policy which I suspected in my 1995 report with Dr.\ Jean-Pierre Hurni is now confirmed, and that this shift is the natural consequence of the nuclear weapon States's policies of the past fifty years.  For instance, the current drive towards the abolition of the ABM treaty and the development of a full-fledged missile defense system is the direct consequence of these policies, which are based on the premise that the original nuclear weapon States's will never renounce nuclear weapons --- which give them a major military edge over all other nations.  It is therefore unavoidable that all efforts towards genuine nuclear disarmament, and in particular all professional-level research efforts of the kind I tried to make in the past twenty years, were and still are doomed to fail:  Nuclear disarmament is simply not a priority of any scientific funding agency in any country...

   Second, concerning the reasons why it has not been possible to interest and associate the scientific community to the efforts of the kind I made, a major reason is certainly the tremendous feeling of guilt which drives the majority of scientists to avoid working, reading, and even thinking about the military (and other negative) implications of their work.

   In this respect, it is symptomatic that the only organizations which have some influence, e.g., {Pugwash}, do not do any real professional-level work (either scientific or political) on the issues they talk about.  Their role is apparently to express concerns and to reassure the scientific community by suggesting that scientists in either military or civilian institutions have no fundamental responsibility for the consequences of their work and blindness.  This attitude explains to a large extent the very emotional reactions of CERN staff members to the 22-23 April 1995 article in  {\it Le Journal de Gen\`eve}, and why these same people show no interest in knowing more about the facts.

   In contrast, people working in organizations having a direct contact with the military implications of nuclear science and technology generally show a very different and much more rational attitude.  After twenty years of relations with both ``academic scientists'' and ``nuclear weaponeers,'' there is no real surprise that most encouragements for my efforts come from people like Maurizio Zifferero, Jafar Dhia Jafar, or Isaac Chavet, rather than from my peers.

   Finally, a few words about my personal experience as a ``whistle-blower.''  Doing what I did was unavoidable in order to preserve my intellectual and moral integrity.  However, the personal and professional consequencies were disastrous: becoming a ``whistle-blower'' means leaving the herd, and there is no reward for breaking the rule --- the awkward silence about science and warfare.

\section{Acknowledgements and invitation for comments}
\label{Sec.4.11}

   I am indebted to Dr.\ Jean-Pierre Hurni and Prof.\ Gilles Falquet for their help in assembling these web pages, and to Carey Sublette for providing space on {\bf The High Energy Weapons Archive} web site.

   I would greatly appreciate any comments, suggestions or criticism, as well as any additional information, material, or personal recollection regarding the facts and questions raised in these web pages.  Please address these contributions by mail to ISRI, P.O.\ Box 30, CH-1211 Geneva-12, Switzerland, or by e-mail to gsponer@vtx.ch.

\section{Letters}
\label{Sec.4.12}

This section contains the letters which are referred to in the previous sections.  They have been typeseted and slightly edited in order to have the same format.  They are also available as scanned documents on the {\bf The High Energy Weapons Archive} web site.\footnote{ http://nuclearweaponarchive.org/Iraq/Calutron.html .}

\subsection{~~From C. Llewellyn Smith, CERN, 25 April 1995}
\label{Doc.03-Llewellyn-Smith-25Apr95}

{\bf CERN --- EUROPEAN ORGANIZATION FOR NUCLEAR RESEARCH\\}
Professeur C.H. Llewellyn Smith\\
Directeur g\'en\'eral\\
CH-1211 Gen\`eve 23, Suisse

\noindent Notre R\'ef.: DG/mnd/1099

\noindent\rule{7cm}{0mm} Monsieur\\
\noindent\rule{7cm}{0mm} Antoine Maurice\\
\noindent\rule{7cm}{0mm} R\'edacteur en Chef\\
\noindent\rule{7cm}{0mm} Journal de Gen\`eve\\
\noindent\rule{7cm}{0mm} rue de Hesse 12\\
\noindent\rule{7cm}{0mm} 1211 Gen\`eve 11\\
 ~\\
\noindent\rule{7cm}{0mm} Gen\`eve, le 25 avril 1995

\noindent Monsieur le R\'edacteur en Chef,

Votre journal a fait para\^itre en premi\`ere page de son no 95, des 22-23 avril 1995, un article intitul\'e ``Bombe atomique: l'Irak est pass\'e par le CERN,'' suivi en page int\'erieure d'un retentissant: ``Gen\`eve berceau de la bombe irakienn.''  J'ai \'et\'e surpris par ces titres excessifs, qui me paraissent sans raport tant avec la r\'ealit\'e qu'avec le contenu même de l'article.

Puis-je rappeler que, selon la convention internationale qui le cr\'ee, le CERN ``s'abstient de tout activit\'e \`a fins militaires'' et qu'il se conforme strictement \`a une \'ethique de rescherche civile et pacifique.  Les r\'esultats de ses recherches sont publi\'es; il n'y a rien de secret ou de confidentiel dans les r\'esultats des travaux de notre institution.

En l'occurence, l'article \'elabor\'e, semble-t-il sur la base d'une opinion individuelle, n'\'etablit pas la d\'emonstration d'un lien direct entre les activit\'es du CERN et la bombe irakienne.  Il semble même \'etablir le contraire.

En effet, il souligne d'abord que ``finalement, les Irakiens ont choisi un design diff\'erent de celui du CERN pour les aimants de leurs calutrons.''  Il rappelle ensuite que le future responsable de la bombe irakienne avait travaill\'e au CERN durant les ann\'ees 70, en tant que rattach\'e \`a l'Imperial College de Londres, et qu'ult\'erieurement il avait envoy\'e au CERN un ing\'enieur irakien pour rechercher des informations sur les aimants construits par ce dernier.  Cependant, il constate aussi, de bonne source, que les ``informations que cet ing\'enieur a reçues se trouvaient dans la litt\'erature scientifique publique et ... qu'il n'a pas eu acc\`es aux plans d'ing\'enieur originaux, ni \`a certains proc\'ed\'es particuliers d\'evelopp\'es par le CERN pour construire cet aimant.''  Il est difficile de mieux souligner que la technologie choisie n'\'etait pas celle utilis\'ee par le CERN.

Je suis navr\'e que votre journal, pour lequel j'ai de la consid\'eration, se soit laiss\'e entra\^iner \`a une conclusion qu'aucun fait ne corrobore.  Je vous serai reconnaissant d'informer vos lecteurs de la pr\'esente r\'eaction du CERN \`a votre article.

Veuillez agr\'eer, Monsieur le R\'edacteur en Chef, l'expression de mes sentiments distingu\'es.

\noindent\rule{7cm}{0mm} C.H. Llewellyn Smith

\subsection{~~From J. Steinberger, CERN, 27 April 1995}
\label{Doc.04-Steinberger-27Apr95}

Jack Steinberger\\
CERN\\
CH-1211 Geneva 23\\
Switzerland

\noindent\rule{7cm}{0mm} M. Antoine Maurice\\
\noindent\rule{7cm}{0mm} R\'edacteur en Chef\\
\noindent\rule{7cm}{0mm} Journal de Gen\`eve\\
\noindent\rule{7cm}{0mm} rue de Hesse 12\\
\noindent\rule{7cm}{0mm} 1211 Gen\`eve 11

\noindent \rule{7cm}{0mm} Geneva, 27.4.1995

\noindent Dear M. Maurice,

The 24/4 issue of the Journal de Gen\`eve distinguishes itself by a front page headline and an article signed Suren Erkman, both referring to CERN, and both remarkable for their misleading and misinforming character.  I think I know what goes on in CERN pretty well, since I have participated in physics experiments at CERN since the middle 50's, have been a staff member from '68 till my retirement in '86, was director of research for a while in the seventies, and still work there full time.

I KNOW that CERN has no involvement with any military.  Neither the physics we do nor the technology we develop, has excited any particular interest by the military, European or otherwise.  In all of my time at CERN I have not see any sign of interest by any military organization in anything we do here.

For his information regarding the alleged Iraqian use of CERN, Erkman bases himself on statements by A.\ Gsponer, who worked here as a young physicist in the late 70's.  I know the history of Gsponer, who has become psychopathic on the issue of alleged military involvement of CERN.  He went on to GIPRI, and made life difficult there until he left.  His stories of military use of CERN are pure invention.

I know the N10 magnets which are referred to.  They are a straightforward application of principles known since Ampere and Maxwell. Being aircore toroids, the basic design is of no conceivable interest for calutrons.  Neither could their technology have been of interest, since there is nothing there which is in anyway new.  The connection with the Iraq calutron is the invention of the mind of Gsponer.

The physicist Jafar, referred to as the patron of bomb development in Iraq, was in the seventies the associate at a postdoctoral level of a professor at Imperial College whom I know very well and trust.  They were working together in particle physics experiments.  According to my colleague, Jafar was excellent, and was proposed to a permanent faculty at Imperial, but this proposal did not go through.  As physicist at Imperial, Jafar participated in a normal way in experiments at CERN with my colleague, who believes that at this time Jafar had no Iraqi military motives or connections.

There is no doubt that many technologically underdevelopped countries, in their military development, have used and continue to use, persons trained in the West.  Does that mean that the West should refuse to teach foreign students?

It would be most interesting for me to understand why Mr. Erkman saw fit to write such dribble, and why your paper saw fit to publish such disinformation, certainly no service to your readers.  I would be most grateful to you for any information which might help me to understand this action of your paper.  A telephonic attempt to reach Mr.\ Erkman was without response.

\noindent\rule{7cm}{0mm} Sincerely yours,

\noindent\rule{7cm}{0mm} Jack Steinberger\\
\noindent\rule{7cm}{0mm} Nobel Laureate in Physics

\subsection{~~To M. Zifferero, IAEA, 5 May 1995}
\label{Doc.05-To-Zifferero-5May95}
 
{\bf ISRI --- Independent Scientific Research Institute\\}
P.O. Box 30\\
Ch-1211 Geneva-12\\
Switzerland

\noindent\rule{7cm}{0mm} Dr Maurizio ZIFFERERO\\
\noindent\rule{7cm}{0mm} Leader UNSC 687 action team\\
\noindent\rule{7cm}{0mm} International Atomic Energy Agency\\
\noindent\rule{7cm}{0mm} P.O. Box 100\\
\noindent\rule{7cm}{0mm} A-1400 Vienna\\
\noindent\rule{7cm}{0mm} Austria

\noindent\rule{7cm}{0mm} 5 May 1995

\noindent Dear Dr Zifferero, 
 
 Suren Erkman from the Journal de Gen\`eve has informed me that you called him in connection with his article on my discovery of Iraq's strong interest in calutron technology while I was working at CERN in 1979. 
 
 In the coming weeks I will be working out of my office most of the time. If you want to call me, the best would be to use my home telephone number (Geneva 312'22'91) between 18h00 and 21h00. 
 
 Enclosed are a few articles on the kind of work done at ISRI.

\noindent\rule{7cm}{0mm} Yours sincerely, 

\noindent\rule{7cm}{0mm} Dr A. Gsponer

\subsection{~~To M. Zifferero, IAEA, 8 November 1995}
\label{Doc.08-To-Zifferero-8Nov95}
 
{\bf ISRI --- Independent Scientific Research Institute\\}
P.O. Box 30\\
Ch-1211 Geneva-12\\
Switzerland

\noindent\rule{7cm}{0mm} Dr Maurizio ZIFFERERO\\
\noindent\rule{7cm}{0mm} Leader UNSC 687 action team\\
\noindent\rule{7cm}{0mm} International Atomic Energy Agency\\
\noindent\rule{7cm}{0mm} P.O. Box 100\\
\noindent\rule{7cm}{0mm} A-1400 Vienna\\
\noindent\rule{7cm}{0mm} Austria 
 
\noindent\rule{7cm}{0mm} 8 November 1995

\noindent Dear Dr Zifferero, 
 
 Please find enclosed the report on Iraq's calutrons that was announced in the Journal de Gen\`eve edition of 22-23 April 1995. 
 
 I would greatly appreciate yours comments and could, if it is of interest, come to Vienna to give a seminar on this topic. 
 
\noindent\rule{7cm}{0mm} Yours sincerely, 
 
\noindent\rule{7cm}{0mm} Dr A. Gsponer

\subsection{~~From M. Zifferero, IAEA, 15 November 1995}
\label{Doc.09-Zifferero-15Nov95}   

{\bf IAEA - INTERNATIONAL ATOMIC ENERGY AGENCY}\\
Wagrammerstrasse 5\\
P.O. Box 100\\
A-1400 Vienna\\
Austria

\noindent\rule{7cm}{0mm} Dr.\ A.\ Gsponer\\
\noindent\rule{7cm}{0mm} ISRI\\
\noindent\rule{7cm}{0mm} Box 30\\
\noindent\rule{7cm}{0mm} 1211 Geneva 12\\
\noindent\rule{7cm}{0mm} Switzerland\\

\noindent\rule{7cm}{0mm} 15 November 1995

\noindent Dear Dr.\ Gsponer,

We just received your report on Iraq's calutrons.  Thank you for sending it.  We plan to study it and would like to ask you permission to copy it to two experts, one in LLNL and one in ORNL, who assisted us in evaluating the Iraqi effort on EMIS.

Once the analysis is finished, I would like to get back to you with our comments.

Thanks again for your courtesy.

\noindent\rule{7cm}{0mm} Yours sincerely,

\noindent\rule{7cm}{0mm} Maurizio Zifferero\\
\noindent\rule{7cm}{0mm} Leader\\
\noindent\rule{7cm}{0mm} UNSC 687 Action Team

\subsection{~~To M. Zifferero, IAEA, 20 November 1995}
\label{Doc.10-To-Zifferero-20Nov95}
 
{\bf ISRI --- Independent Scientific Research Institute\\}
P.O. Box 30\\
Ch-1211 Geneva-12\\
Switzerland

\noindent\rule{7cm}{0mm} Dr Maurizio ZIFFERERO\\
\noindent\rule{7cm}{0mm} Leader UNSC 687 action team\\
\noindent\rule{7cm}{0mm} International Atomic Energy Agency\\
\noindent\rule{7cm}{0mm} P.O. Box 100\\
\noindent\rule{7cm}{0mm} A-1400 Vienna\\
\noindent\rule{7cm}{0mm} Austria 

\noindent\rule{7cm}{0mm} 20 November 1995

\noindent Dear Dr Zifferero,

     Please find enclosed, for your experts at LLNL and ORNL, two extra copies of the report ISRI-95-03 on Iraq's calutrons.

\noindent\rule{7cm}{0mm} Yours sincerely, 
 
\noindent\rule{7cm}{0mm} Dr A. Gsponer

\subsection{~~To B. Al-Tikriti, Mission of Iraq, 20 November 1995}
\label{Doc.11-To-Al-Tikriti-20Nov95}
 
{\bf ISRI --- Independent Scientific Research Institute\\}
P.O. Box 30\\
Ch-1211 Geneva-12\\
Switzerland

\noindent\rule{7cm}{0mm} Mr Barzam Al-Tikriti\\
\noindent\rule{7cm}{0mm} Ambassador\\
\noindent\rule{7cm}{0mm} Permanent Mission of the\\
\noindent\rule{7cm}{0mm} Republic of Iraq\\
\noindent\rule{7cm}{0mm} 28a, Ch. du Petit-Saconnex \\
\noindent\rule{7cm}{0mm} 1209 GENEVA 

\noindent\rule{7cm}{0mm} 20 November 1995

\noindent Your excellency, 
 
 Please forgive my writing you without having been introduced. 
 
 I am a physicist who worked at CERN at a time when Dr Jafar Dhia Jafar, Vice Chairman of the Iraq Atomic Energy Commission, was also there. I have written a study on Iraq's electromagnetic enrichment program and would very much like sending a copy of the report to Dr Jafar for his comments. 
 
 You will find enclosed a copy of a Journal de Gen\`eve article in which the report in question was announced. 
 
 Since I don't have the proper address of Dr Jafar, and would like to make sure that my report gets into his hands as soon as possible, it may be that the best might be to sent it by diplomatic mail, unless you think otherwise. 
 
 In either case, I would greatly appreciate that you let me know Dr Jafar's address, or whether you would be ready to forward a letter and a 50 pages document to Dr Jafar through your embassy.

\noindent\rule{7cm}{0mm} Yours sincerely,

\noindent\rule{7cm}{0mm} Dr Andr\'e Gsponer

\noindent Encl: Journal de Gen\`eve 22-23 avril 1995

\subsection{~~From Mission of Iraq, 16 January 1996}
\label{Doc.12-Al-Tikriti-16Jan96}

{\bf Mission permanent de la R\'epublique d'Irak\\
Aupr\`es de l'office des Nations unies \`a Gen\`eve}

\noindent {\bf Recommand\'e}

\noindent Ref.: MP/013/1996

\noindent\rule{7cm}{0mm} Dr.\ Andr\'e GSPONER\\
\noindent\rule{7cm}{0mm} ISRI\\
\noindent\rule{7cm}{0mm} Bo\^ite postale 30\\
\noindent\rule{7cm}{0mm} 1211 Gen\`eve 12

\noindent Monsieur,

Comme suite \`a votre lettre du 02/11/95, adress\'ee \`a S.E.\ Monsieur Barzan AL-TIKRITI, Ambassadeur, Repr\'esentant permanent de la R\'epublique d'Irak aupr\`es de l'ONU \`a Gen\`eve, nous vous avisons que votre courrier a \'et\'e transmis aux autorit\'es irakiennes concern\'ees.  Ces derni\`eres tiennent \`a vous informer que, pour recevoir une r\'eponse favorable \`a votre demande, vous devez obtenir une autorisation de l'Agence internationale de l'\'energie atomique (AIEA) \`a Vienne, permettant \`a Monsieur Jafar Dhia Jafar de commenter l'\'etude mentionn\'ee dans votre lettre.

Par ailleurs, vous pouvez aussi envoyer votre \'etude \`a Monsieur Jafar Dhia Jafar par le bias de l'AIEA \`a Vienne et le destinataire pourra \'egalement vous adresser ses remarques par l'interm\'ediare de l'Agence susmentionn\'ee.

Nous vous prions d'agr\'eer, Monsieur, nos salutations distingu\'ees.

\noindent\rule{7cm}{0mm} Gen\`eve, le 16 janvier 1996

\subsection{~~To M. Zifferero, IAEA, 22 January 1996}
\label{Doc.13-To-Zifferero-22jan96}
 
{\bf ISRI --- Independent Scientific Research Institute\\}
P.O. Box 30\\
Ch-1211 Geneva-12\\
Switzerland

\noindent\rule{7cm}{0mm} {\bf Registered}\\
\noindent\rule{7cm}{0mm} Dr Maurizio ZIFFERERO\\
\noindent\rule{7cm}{0mm} Leader UNSC 687 action team\\
\noindent\rule{7cm}{0mm} International Atomic Energy Agency\\
\noindent\rule{7cm}{0mm} P.O. Box 100\\
\noindent\rule{7cm}{0mm} A-1400 Vienna\\
\noindent\rule{7cm}{0mm} Austria 
 
\noindent\rule{7cm}{0mm} 22 January 1996 

\noindent Dear Dr Zifferero,

 Please find enclosed a copy of the report ISRI-95-03 on Iraq's calutrons and a letter addressed to Dr Jafar Dhia Jafar. 
 
 I am sending you these documents at the suggestion of the Permanent mission of Iraq in Geneva which stated that in order to send a copy to Dr Jafar Dhia Jafar and allow him to give comments, I should transit through your office. 
 
 I would therefore greatly appreciate if you could forward the enclosed letter and report to Dr Jafar Dhia Jafar, and grant him the permission to answer my questions and send his comments back to me.

 Thank you very much for your courtesy.

\noindent\rule{7cm}{0mm} Yours sincerely, 
 
\noindent\rule{7cm}{0mm} Dr A. Gsponer

\noindent Encl.: 
 
\indent - Letter to Dr Jafar Dhia Jafar 
 
\indent - Report ISRI-95-03 

\subsection{~~To J.D. Jafar, IAEC, 22 January 1996}
\label{Doc.14-To-Jafar-22jan96}
 
{\bf ISRI --- Independent Scientific Research Institute\\}
P.O. Box 30\\
Ch-1211 Geneva-12\\
Switzerland
 
\noindent\rule{7cm}{0mm}  Dr Jafar Dhia Jafar\\
\noindent\rule{7cm}{0mm}  Vice Chairman of the\\ 
\noindent\rule{7cm}{0mm}  Iraq Atomic Energy Commission \\
\noindent\rule{7cm}{0mm}  c/o International Atomic Energy Agency \\
\noindent\rule{7cm}{0mm}  Vienna 
 
\noindent\rule{7cm}{0mm} 22 January 1996 

\noindent Dear Dr Jafar Dhia Jafar, 
 
 In the early 1970s, while you were working at CERN on the  measurement of polarization effects in various reactions using a  polarized target, I was involved in similar experiments, possibly using the same South hall beam line (e.g., L. Dick et al., Phys.  Lett. {\bf 57B} (1975) 93-96). 
 
 We never met, and I would not be writing the present letter  if you had not sent somebody to CERN in 1979 in order to enquire  about the magnet of the NA10 experiment on which I was working at  the time. This event, which is described in some detail in chapter 2 of the report ISRI-95-03, together with the discovery in 1978 of the existence of research on particle beam weapons (to be later expanded in the SDI program), was one of the reasons why I decided in 1980 to leave high energy physics to work full time on disarmament. 
 
 The enclosed report, ISRI-95-03, that I am sending you for review and comments, is a typical example of the kind of work I am presently doing. It is an independent assessment of EMIS technology and an attempt to draw the main disarmament conclusions. 
 
 This report has been sent for review and comments to more than fifteen researchers, including specialists at CERN and at the IAEA. In fact, I expect that there will be substantial differencies in the final version of the report, which I hope to be ready in March or April. 
 
 A major weakness of this report is our discussion of the development of EMIS technology in Iraq. In particular, several aspects of our interpretation of the published information on the design principles of the EMIS plant that was under construction at Tarmiya are rather speculative. In fact, as you can see from section 2.4 of our report, our interest is not to document in all possible detail each part of the enrichment plant, but to describe the general scientific principles of the final design of the electromagnetic separation system, and to give the main general characteristics such as sizes, beam currents, number and position the ion sources and collectors, etc. Another aspect we would like to document with some precision is the reasoning which lead to the choice of the mass analyzing field system, and in particular the role of the NA10 magnet in this context.

 I would therefore greatly appreciate it if you could send us any information that could help us to answer the following questions: 
 
A) Concerning the design under construction at Tarmiya (section 2.4, pages 24-27 of ISRI-95-03):

1) Is the description we give in section 2.4 globally correct, or have we made any gross mistakes in our interpretation - and if so what are they? 
 
2) What was the focusing angle in the alpha-calutrons? In the case where the focusing angle was slightly less than 180$^{\rm{o}}$, could it be that they were two independent sets of beams in each half of the separation chambers? 
 
3) Did the alpha-stage use multiple overlapping beams of similar radii (like the two- or four-beam units used at Oak Ridge), multiple overlapping beams with different radii, or non-overlapping beams like the two-beam units studied by the French (our reference number [26]). 
 
4) How many ion-sources and collectors were there in the alpha-calutrons? How were they positioned? What were the overall dimensions of each of these sources? What were the lengths of the ion extraction slits? What were the design and measured ion source currents? What were the measured ion currents at the collectors? 
 
5) Is the schematic of the calutron track shown in Fig. 6 correct? In particular, could you confirm that the minimum spacing between the magnets was about 100 cm. 
 
6) Is our interpretation of Fig.7a correct (i.e., that this chamber was designed to contain three concentric and non-overlapping double-beams)? If our interpretation is wrong, could you tell us how the ion-sources and ion-collectors were positioned? How were the pumps connected to the chamber: Are the three ``channels'' pump-vibrations absorbers? 
 
7) Is it correct that the beta calutrons were to use the 255$^{\rm{o}}$ method? In this case, how many beams were there to be in each chamber? Is Fig.7b a beta calutron chamber or just a prototype chamber?

B) Concerning the NA10 magnet and other analyzing field systems (sections 2.1 to 2.3, pages 19-23 of ISRI-95-03): 
 
1) What was the main reason for not using simple magnets of the kind used at Oak Ridge? 
 
2) Did you consider in detail the possibility of using a $1/r$ field such as the one produced by the S-2 separator at Arzamas-16? Did you consider the possibility of using a Bernas-type sector separator? 
 
3) Did you consider in detail the possibility of using a magnet similar to the NA10 magnet? What was the main reason for not using such a magnet?

 Obtaining answers to some, or all of the above questions would greatly help us to finalize the report ISRI-95-03. I would also very much appreciate it if you could send us any comments you may have on our report. 
 
 Thank you very much for your collaboration. 
 
\noindent\rule{7cm}{0mm} Sincerely,

\noindent\rule{7cm}{0mm} Dr Andr\'e Gsponer

\noindent Encl.:  Andr\'e Gsponer and Jean-Pierre Hurni, ``Iraq's calutrons ---  Electromagnetic separation, beam technology and nuclear weapon proliferation,'' report ISRI-95-03 (19 October 1995).

\subsection{~~From M. Zifferero, IAEA, 26 January 1996}
\label{Doc.15-Zifferero-26Jan96}

{\bf IAEA - INTERNATIONAL ATOMIC ENERGY AGENCY}\\
Wagrammerstrasse 5\\
P.O. Box 100\\
A-1400 Vienna\\
Austria

\noindent\rule{7cm}{0mm} Dr.\ A.\ Gsponer\\
\noindent\rule{7cm}{0mm} ISRI\\
\noindent\rule{7cm}{0mm} Box 30\\
\noindent\rule{7cm}{0mm} 1211 Geneva 12\\
\noindent\rule{7cm}{0mm} Switzerland

\noindent\rule{7cm}{0mm} 26 January 1996

\noindent Dear Dr.\ Gsponer,

Please refer to your request dated 22 January 1996 concerning the set of questions you would like to be answered by Dr.\ Jafar.  Having gone through these questions in some detail, I am afraid we cannot assist you in this matter, since most of these questions look rather tutorial in nature in the sense that could assist Iraqi scientists and engineers by providing them food for new thoughts, should they decide to start again clandestine activity on EMIS.

I would propose an alternative way to meet, at least in part, your request.  Our Iraqi counterpart is preparing, at the request of the IAEA, what is called a Full, Final, Complete Declaration (FFCD) of the activities carried out under their past clandestine programme.  We expect a first draft of the FFCD, which should include a substantive chapter covering their efforts on EMIS, to be ready for discussion within a couple of months or so.  We plan to submit each chapter of this report to the scrutiny of experts in each relevant area.  This examination is likely to result in a number of questions to be put to the counterpart.  In this conjunction, some of your questions, excluding those tutorial in nature could be added to the list.  I hope this proposal will meet, albeit partially your request.

\noindent\rule{7cm}{0mm} Sincerely,

\noindent\rule{7cm}{0mm} Maurizio Zifferero\\
\noindent\rule{7cm}{0mm} Leader, UNSC 687\\
\noindent\rule{7cm}{0mm} Action Team

\subsection{~~From R.F. Mozley, CISAC, 28 February 1996}
\label{Doc.16-Mozley-28Feb96}

601 Laurel Ave.\\
Menlo Park, CA 94025\\
U.S.A.

\noindent\rule{7cm}{0mm} Dr.\ Andr\'e Gsponer\\
\noindent\rule{7cm}{0mm} ISRI\\
\noindent\rule{7cm}{0mm} Box 30\\
\noindent\rule{7cm}{0mm} 1211 Geneva 12\\
\noindent\rule{7cm}{0mm} Switzerland

\noindent\rule{7cm}{0mm} February 28, 1996

\noindent Dear Dr. Gsponer,

Thank you for sending me your report on Iraq's calutrons.  I found it very useful and interesting.  (...)

Although you were early in this, you are not alone in your lack of success in warning about the possible use of calutrons as a method of enrichment by proliferants.  As you may have seen in my enrichment paper, p 48, Paul Persiani also had no success.

I don't share your view that governments and physicists have conspired to prevent such publication.  The physicists I know delight in showing governmental policy to be wrong.  I think you have observed the reluctance of publishers to accept new ideas, especially if they don't come from an internationally famous scientist.

My comments which you requested are:

You have written two papers, one on the Iraqi calutron program, the other on your difficulties in publication.  I believe that the first paper will be readily accepted for publication.  The combined paper that you sent will either
be refused publication or the difficult part will be edited out.  I think it would be useful to write about the difficulties in a separate article.

I would enjoy receiving other reports from your institute.

\noindent\rule{7cm}{0mm} Yours sincerely,

\noindent\rule{7cm}{0mm} Robert F. Mozley

\subsection{~~To P.J. Persiani, ANL, 17 June 1996}
\label{Doc.17-To-Persiani-17Jun96}
 
{\bf ISRI --- Independent Scientific Research Institute\\}
P.O. Box 30\\
Ch-1211 Geneva-12\\
Switzerland
 
\noindent\rule{7cm}{0mm}  Dr Paul J. Persiani\\
\noindent\rule{7cm}{0mm}  Argonne National Laboratory\\
\noindent\rule{7cm}{0mm}  9700 South Cass Av.\\
\noindent\rule{7cm}{0mm}  Argonne, IL 60439\\
\noindent\rule{7cm}{0mm}  U.S.A.

\noindent\rule{7cm}{0mm} 17 June 1996

\noindent Dear Dr Persiani,

 I would greatly appreciate receiving a copy of your document entitled \emph{Non-Proliferation Aspects of Commercial Nuclear Fuel Cycles}, report to the Institute of Nuclear Materials Management, 33rd Annual Meeting, Orlando, Florida, July 1992.

 I am very much concerned by the question on non-proliferation and would like to draw your attention to the fact that I have discovered in 1979 already Iraq's interest in calutron technology. (J.D. Jafar, the head of Iraq's atomic bomb program worked at CERN for many years.) With a colleague at ISRI we have written a report on the proliferation problems associated with various beam-related technologies.

 This report (the abstract of which is enclosed) has been sent to a number of individuals at several institutions for reviewing (IAEA, ORNL, LLNL, CERN, etc). I would very much like having your comments on this report, provided you have the time and interest to do it. We plan to finalize our report in September this year. If you are interesting receiving a copy of our report, please let me know.
 
\noindent\rule{7cm}{0mm} Yours sincerly,
 
\noindent\rule{7cm}{0mm} Dr Andr\'e Gsponer

\subsection{~~To G. M\"unzenberg, GSI, 19 June 1996}
\label{Doc.19-To-emis13-19Jun96}
 
{\bf ISRI --- Independent Scientific Research Institute\\}
P.O. Box 30\\
Ch-1211 Geneva-12\\
Switzerland

\noindent\rule{7cm}{0mm}  Professor G. M\"unzenberg \\
\noindent\rule{7cm}{0mm}  Chairman of EMIS-13 \\
\noindent\rule{7cm}{0mm}  GSI \\
\noindent\rule{7cm}{0mm}  Planckstrasse 1 \\
\noindent\rule{7cm}{0mm}  D-64291 Darmstadt 
 
\noindent\rule{7cm}{0mm} 19 June 1996 
 

\noindent Dear Professor M\"unzenberg,

 Please find in attachment the abstract of a contribution we are submitting to the organizing committee for a short presentation at the EMIS-13 conference. 
 
 We realize that this contribution, written by two nuclear physicists, is special in the sense that it is discussing the nuclear weapon proliferation implication of EMIS rather than a technical point related to the development of the technology. 
 
 Our contribution will be a summary of the final version of the report ISRI-95-03 released on 19 October 1995. This preliminary report, of which we include the title page and table of contents on pages 3 and 4, has since been extensively reviewed at a number of places including IAEA, ORNL, LLNL, CERN, etc. 
 
 We think that such a contribution would be important, as much for the historical and technical analysis we present, as for the kind of background it provides to a scientific community which is evolving in a more and more complex world. 
 
\noindent\rule{7cm}{0mm} Yours sincerely, 
 
\noindent\rule{7cm}{0mm} Dr Andr\'e Gsponer

\begin{quote} 
\begin{center}

 {\bf Electromagnetic isotope separation, beam technology and nuclear weapon proliferation} 
  
 by 
 
 Andr\'e Gsponer and Jean-Pierre Hurni 
 
 Independent Scientific Research Institute \\
 P.O. Box 30, 1211 Geneva 12 \\
 Switzerland \\
 
 ~\\

 \underline{Abstract }

 The past and present status of high-current electromagnetic isotope separation (EMIS) technology for uranium and plutonium enrichment (i.e., calutrons) is reviewed in the five nuclear weapons States and in four critical States; Japan, India, Israel and Iraq.

 The circumstances and significance of the  discovery in 1979 of Iraq's definite interest in  calutron technology is discussed in some details. 
 
 The conclusion stresses the potential of ``old'' technologies such as calutrons (and the plasma separation process), e.g., for uranium or plutonium enrichment, and of particle accelerators for the efficient production of plutonium or tritium. 

 ~\\
 
 Submitted to EMIS-13 \\
 September 23--27, 1996 \\
 Bad D\"urkheim 

\end{center}
\end{quote}

\subsection{~~From I. Chavet, SOREQ, 25 June 1996}
\label{Doc.20-Chavet-25Jun96}

{\bf SOREQ}\\
Soreq NRC\\
Yavne 81800\\
Israel

\noindent\rule{7cm}{0mm} Mr Andre Gsponer\\
\noindent\rule{7cm}{0mm} ISRI, Box 30\\
\noindent\rule{7cm}{0mm} 1211 Geneva 12\\
\noindent\rule{7cm}{0mm} Switzerland

\noindent\rule{7cm}{0mm} June, 25 1996

\noindent Sir,

A friend of mine from the I.A.E.C.\ showed me your very interesting paper, written in collaboration with Jea-Pierre Hurni:

\begin{quote}
{\bf Iraq's Calutrons --- Electromagnetic isotope separation, beam technology and nuclear weapon proliferation (ISRI-95-03) }
\end{quote}

This paper appealed to me since I have been mainly active in ion optics and especially in mass separation systems.

However there was no accompanying bibliography.

Would you kindly send me the bibliography of this important paper in order to fully appreciate its scientific message.

\noindent\rule{7cm}{0mm} Thankfully yours,

\noindent\rule{7cm}{0mm} Dr I.\ Chavet

\subsection{~~To I. Chavet, SOREQ, 1 July 1996}
\label{Doc.21-To-Chavet-1Jul96}
 
{\bf ISRI --- Independent Scientific Research Institute\\}
P.O. Box 30\\
Ch-1211 Geneva-12\\
Switzerland

\noindent\rule{7cm}{0mm} Dr I. Chavet \\
\noindent\rule{7cm}{0mm} SOREQ NRC \\
\noindent\rule{7cm}{0mm} Applied radiation \\
\noindent\rule{7cm}{0mm} technology division \\
\noindent\rule{7cm}{0mm} Yavne 81800 \\
\noindent\rule{7cm}{0mm} Israel 
 
\noindent\rule{7cm}{0mm} 1 July 1996 

\noindent Dear Dr Chavet, 
 
 Thank you very much for your letter of June 25. According to your request, I am sending you the bibliography and the figures of the report ISRI-95-03. 
 
 In fact, this report is preliminary and we have have sent it to a number of people and institutions for reviewing. (See attached copy of letter of June 18.) 
 
 Since you have now a complete copy of our report, I would greatly appreciate any comments or suggestions you may have.

 Also, I would greatly appreciate if you could send us any recent publication concerning your activities in the field of mass separation systems, and - if there is - a recent annual report in which the activities of SOREQ in this field are described. 
 
 On a personal level, I can tel you that I am a good friend of Harry Bernas, the nephew of Rene Bernas. 
 
 Hoping to meet you in September at EMIS-13,

\noindent\rule{7cm}{0mm} Yours sincerely, 
 
\noindent\rule{7cm}{0mm} Dr Andre Gsponer

\subsection{~~To M. Zifferero, IAEA, 1 July 1996}
\label{Doc.22-To-Zifferero-1Jul96}
 
{\bf ISRI --- Independent Scientific Research Institute\\}
P.O. Box 30\\
Ch-1211 Geneva-12\\
Switzerland

\noindent\rule{7cm}{0mm} Dr Maurizio ZIFFERERO\\
\noindent\rule{7cm}{0mm} Leader UNSC 687 action team\\
\noindent\rule{7cm}{0mm} International Atomic Energy Agency\\
\noindent\rule{7cm}{0mm} P.O. Box 100\\
\noindent\rule{7cm}{0mm} A-1400 Vienna\\
\noindent\rule{7cm}{0mm} Austria 

\noindent\rule{7cm}{0mm} 1 July 1996

\noindent Dear Dr Zifferero,

 Please find enclosed the copy of a letter of Dr. I Chavet, the leader of EMIS activities at SOREQ NRC, Israel. I have sent him the part of our report on Iraq's Calutrons he was missing, and asked him to send us any comments he may have. 
 
 The irony of the introductory sentence of Dr Chavet's letter is that it is not clear whether he got part of our report form the ``Israel Atomic Energy Commission,'' the ``Iraq Atomic Energy Commission'' or possibly the ``IAEA'' if the last letter is mistyped! 
 
 I would like to take this opportunity to remind you of our interest in receiving any comments and suggestions from your Team on our report, as well as all possible answers to my questions to Dr Jafar Dhia Jafar along the lines of your letter of January 26. 
 
 Finally, I would like to suggest that the EMIS-13 Conference, September 23--27, 1996, might provide a forum for informal discussions on these questions and other points of common interest. In this perspective, I am enclosing an announcement of EMIS-13, hoping that some members of your team will attend this conference.

\noindent\rule{7cm}{0mm} Yours sincerely, 
 
\noindent\rule{7cm}{0mm} Dr A. Gsponer

\subsection{~~From M. Zifferero, IAEA, 2 August 1996}
\label{Doc.23-Zifferero-2Aug96}

{\bf IAEA - INTERNATIONAL ATOMIC ENERGY AGENCY}\\
Wagrammerstrasse 5\\
P.O. Box 100\\
A-1400 Vienna\\
Austria

\noindent\rule{7cm}{0mm} Dr.\ A.\ Gsponer\\
\noindent\rule{7cm}{0mm} ISRI\\
\noindent\rule{7cm}{0mm} Box 30\\
\noindent\rule{7cm}{0mm} 1211 Geneva 12\\
\noindent\rule{7cm}{0mm} Switzerland

\noindent\rule{7cm}{0mm} 2 August 1996

\noindent Dear Dr.\ Gsponer,

Please accept my apologies for the late reply to your kind letter of 1 July 1996.  I have read your report for a second time and I believe it contains a good review of the subject.  The Iraqi ``Full, final and Complete Declaration'' (FFCD) of their past programme was delivered to the Agency on 1 March 1996 in draft form.  It contained an extensive description (over 1000 pages) of their programme and EMIS part of it.

As you can imagine it took some time to go through this document and identify ambiguities, inconsistencies and contradictions, discuss them with the Iraqi counterpart and ask for adequate revisions.  The EMIS part of FFCD was extensively discussed in the second half of June and we are now waiting for the corrected and probably final version.  As promised in my letter to you dated 26 January 1996, I will provide you with a copy of the EMIS part of the FFCD as soon as I receive it and hope you will find in it replies to some of your questions.  I hope that this will occur before the EMIS-13 Conference of 23 September.  Attendance of representatives from the Action Team is unfortunately problematic since most of us will be in Iraq at that time.

\noindent\rule{7cm}{0mm} Sincerely,

\noindent\rule{7cm}{0mm} Maurizio Zifferero\\
\noindent\rule{7cm}{0mm} Leader, UNSC 687\\
\noindent\rule{7cm}{0mm} Action Team

\subsection{~~From I. Chavet, SOREQ, 6 August 1996}
\label{Doc.24-Chavet-6Aug96}

{\bf SOREQ}\\
Soreq NRC\\
Yavne 81800\\
Israel

\noindent\rule{7cm}{0mm} Dr Andre Gsponer\\
\noindent\rule{7cm}{0mm} ISRI, Box 30\\
\noindent\rule{7cm}{0mm} 1211 Geneva 12\\
\noindent\rule{7cm}{0mm} Switzerland

\noindent\rule{7cm}{0mm} 6 August 1996

\noindent Dear Dr Gsponer,

Thank you for your answer of the 1st July.  I apologize for this delay.  I was somewhat busy and I had to search for some of the references.

I personally congratulate you for devoting your time at trying to stop, or at least to reduce, the proliferation of nuclear weapons, although I do not expect substantial results.  This perspective only enhances the merit of your efforts and those of your collaborators.

To answer your request of comments on your paper, I really cannot say much on the core of the topic, but I can only express my modest opinion concerning some technical subjects mentioned there.

1) You describe in some detail the ``orange'' Ris\o spectrometer and the NA10 one you helped to build at CERN.  Both are indeed ingeniously conceived and beautifully designed.  However, one must distinguish between instruments intended for measurements and those intended  for material separation with a view to reasonable output.  The requirements and problems are different and so are their design principles.

2) You stress the importance of the focusing in the transverse (or vertical) direction in addition to the lateral (or horizontal) focusing.  This vertical focusing which is very important for spectrometers is not really necessary for production mass separators, since these machines have forcibly long emission slits and it does not matter where the ions land on the collector, as long as they do not depart laterally from the image line.  This last condition should preferably be chosen as part of the optical requirements.

3) Without detailing here the merits and drawbacks of inhomogeneous fields, I would remind that the control of the ion trajectory in the transverse (or vertical) direction can also be achieved with adequately sloped field boundaries.  An added bonus of this last method is a much lighter and cheaper magnet.

4) With all my admiration for the NA10 spectrometer design, I think that it would be very difficult to apply a similar design to isotopes separators, for reasons too long to discuss here.  Even if it were possible, it could be very awkward and expensive to operate such a battery.  The closed magnetic loop of the Oak Ridge calutrons (or its modification used in Iraq) seems still the simplest and best design.

5) In the field of isotope separation we prefer the term ``enrichment enhancement factor'' instead of mass separation power. 
It is defined as the ratio of the abundance of two isotopes of a given element after a separation, divided by the same ratio before the separation.  You were quite right in stating that this factor depends on the dispersion and the width of the peaks at the collector, but it depends also of the magnitude of the ``tails'' of the peaks which are due to various contamination factors well studied by workers in the field.
This enrichment factor is strictly defined for only two isotopes of a given element.  However, it is also loosely used for the enrichment of a single isotope relative to all isotopes of an element.

6) You are right in saying that a large beam current and high enrichment factor are antagonistic.  This is well illustrated by the American calutrons versus the European laboratory separators.  However, as in light optics, where a sophisticated compound lens can achieve both wide aperture and high resolution, we were convinced, when first planning MEIRA, that a well designed ion-optics could achieve both high beam current and high enhancement factor, thus combining the qualities of both types of separators.  The challenge was indeed met as proved for instance in the separation of Tellurium-124 (natural abundance 4.75\%).  From the data of your reference [39], namely a purity of 99.77\% for Te-124 obtained in production operating conditions (beam current 42.5~mA), one can infer an enhancement factor of 8700 which compares honorably with the best European machines.
This performance was achieved by an elaborate design, eliminating all practical aberrations of the second and third order (heterogeneous terms included).  So it is not quite correct to say that MEIRA was built on the model of the Orsay separator, which was not corrected even for the second order aperture aberration term.  Nevertheless, this last machine represented at its time a tremendous leap forward in the technology of isotopes separators and it did a splendid job in the early sixties (we used to operate it continuously in three shifts 24 hours a day).  I was very luck to be trained at Orsay under the late Ren\'e Bernas and will never forget his thorough understanding and openmindness as as his tutorial qualities.

Incidentally, may I mention that MEIRA has been copied at the Amersham Laboratories (Amersham, England) and successfully operated for industrial use.

Concerning your question about my recent work, I am retired since 1989 and have stopped doing any significant work worth publishing since then.
Anyhow, my last papers were:

\noindent \emph{Effective aberrations in electromagnetic isotopes separators} in collaboration with J.\ Camplan, N.I.M. Phys. Res. {\bf 226} (1980) 250.

\noindent \emph{Imaging properties of a convex spherical surface emitting a space charge limited ion current}, N.I.M. Phys. Res. {\bf A251} (1986) 201.

\noindent \emph{Size of virtual source behind a convex spherical surface emitting a space charge limited ion current}, N.I.M. Phys. Res. {\bf A258} (1987) 58.

\noindent \emph{Principles of unattended operation of isotope separators}, N.I.M. Phys. Res. {\bf B26} (1987) 44.

Concerning the last work done on MEIRA, in fact, all activities linked with isotope separation were stopped more than ten years ago for lack of financial support and the machine has been used as low voltage accelerator for research in the following topics, the results of which were published in various journals:

a) Ion implantation to improve surface quality of steels.

b) Bombardment of various materials with low voltage oxygen ions to assess their suitability for use in space projects.  For this work, a deceleration system was designed and adapted to the separator.

c) Development of diamond-like films by carbon implantation at adequate energies.  For this work a precise two-dimensional mechanical scanning system was designed.

In all these activities I only helped to develop the necessary accessories and fit them to the separator.

I am happy that you meet Harry Bernas.  Please give him my best regards.

I am sorry that, for various reasons, I would not be able to attend the EMIS-13 Conference.  I can only wish it to be a very successful one.

\noindent\rule{7cm}{0mm} Truly yours,

\noindent\rule{7cm}{0mm} Dr I. Chavet

\subsection{~~From I. Langermann, ICLQ, 28 August 1996}
\label{Doc.25-ICLQ-28Aug96}

{\bf THE INTERNATIONAL AND COMPARATIVE LAW QUARTERLY}\\
{\bf The British Institute of International and Comparative Law Quarterly}\\
17 Russel Square\\
London QC1B 5DR\\

\noindent\rule{7cm}{0mm} Dr Andre Gsponer\\
\noindent\rule{7cm}{0mm} ISRI\\
\noindent\rule{7cm}{0mm} Box 30\\
\noindent\rule{7cm}{0mm} CH-1211 Geneva 12\\
\noindent\rule{7cm}{0mm} Switzerland

\noindent\rule{7cm}{0mm} 28 August 1996

\noindent Dear Dr Gsponer,

Thank you very much for submitting the article by yourself, Dr Hurni and Dr Clement entitled ``UN Security Council Resolutions 687, 707, and 715 and their Implications for a Halt of all Proliferation Prone Nuclear Activities --- a Technical and Legal Assessment'' for consideration for publication in \emph{ICLQ}.

However, I regret to inform you that, after very careful consideration, the editors have decided not to accept your article for publication.

I appreciate that this will come as a disappointment and would like to thank you for letting the \emph{ICLQ} consider your work.

Please let me know if you wish the article to be returned to you.

\noindent\rule{7cm}{0mm} Yours sincerely,

\noindent\rule{7cm}{0mm} Ida Langermann\\
\noindent\rule{7cm}{0mm} Assistant General Editor

\subsection{~~To M. Zifferero, IAEA, 30 December 1996}
\label{Doc.26-To-Zifferero-30Dec96}
 
{\bf ISRI --- Independent Scientific Research Institute\\}
P.O. Box 30\\
Ch-1211 Geneva-12\\
Switzerland

\noindent\rule{7cm}{0mm} Dr Maurizio ZIFFERERO\\
\noindent\rule{7cm}{0mm} Leader UNSC 687 action team\\
\noindent\rule{7cm}{0mm} International Atomic Energy Agency\\
\noindent\rule{7cm}{0mm} P.O. Box 100\\
\noindent\rule{7cm}{0mm} A-1400 Vienna\\
\noindent\rule{7cm}{0mm} Austria 

\noindent\rule{7cm}{0mm} 30 December 1996

\noindent Dear Dr Zifferero,

 I recently got across a remarquable document by Robert Meunier who with others built the ``Sidonie'' electromagnetic separator of Orsay in the 1960-70s. This document CSNSM-94-30 describes an industrial scale electromagnetic separation system for cesium which could handle $4.3 \times 238 / 135 = 7.6$ tons of natural uranium per year. It is a very detailed study by somebody who has spent his life building high-current 
electromagnetic separators. The study is especially interesting because it gives precise estimates for the construction and operating costs of an industrial scale EMIS plant.

 It is quite possible that you already have this report but I wanted to make sure that the IAEA had seen it.

 I was recently at a Wilton Park conference entitled ``Preventing the proliferation of weapons of mass destruction: Is it an achievable goal?'' (December 16--20, 1996). I was rather surprised by the number of 
references made to UNSCOM and Iraq. At one time I had to make the point that UNSCOM was an INDEPENDENT commission and that one had to wait for its final report before drawing too many conclusions.

 I look forward the EMIS part of the FFCD so we can put our report on Iraq's calutron into final form and publish it in 1977.

 Best wishes to you and to UNSCOM for 1997.
 
\noindent\rule{7cm}{0mm} Yours sincerely,
 
\noindent\rule{7cm}{0mm} Dr Andr\'e Gsponer

\subsection{~~From A. McDermott, Security Dialogue, 20 May 1997}
\label{Doc.27-SecurityDialogue-20May97}

{\bf SECURITY DIALOGUE}\\
{\bf International Peace Research Institute, Oslo}\\
Fuglehauggt. 11\\
N-0260 Oslo\\
Norway

\noindent\rule{7cm}{0mm} Andre Gsponer\\
\noindent\rule{7cm}{0mm} Independent Scientific\\
\noindent\rule{7cm}{0mm} Research Institute\\
\noindent\rule{7cm}{0mm} Box 30\\
\noindent\rule{7cm}{0mm} 1211 Geneva 12\\
\noindent\rule{7cm}{0mm} Sveits

\noindent\rule{7cm}{0mm} 20 May 1997

\noindent Dear Andre Gsponer,

We have now completed the evaluation of your article ``UN Security Council Resolutions 687, 707, and 715 ...''  I am sorry that this took such a long time to complete.  As you know, one of our referees became ill and it took a bit longer than we had counted on to get his review.  However, two referees have now commented on your article and I have also read it with interest.  Unfortunately, I have to agree with the reviewers, both of whom mention that the 1991 resolutions of the UN Security Council (UNSC) referred to are some years old now. There would have been acute problems enforcing the resolution, particularly since neither China nor France were as yet parties of the Nuclear Non-Proliferation Treaty (NPT).  The quasi-legislative and quasi-judicial autority of the UNSC was regarded as fairly tenuous and without sufficient authority to force compliance out of Iraq.

Again, our apologies that it has taken us this long to come to a negative conclusion!

\noindent\rule{7cm}{0mm} Sincerely,

\noindent\rule{7cm}{0mm} Anthony McDermott\\
\noindent\rule{7cm}{0mm} Editor, \emph{Security Dialogue}

\subsection{~~To R. Hooper, IAEA, 21 November 1997}
\label{Doc.28-To-Hooper-21Nov97}
 
{\bf ISRI --- Independent Scientific Research Institute\\}
P.O. Box 30\\
Ch-1211 Geneva-12\\
Switzerland

\noindent\rule{7cm}{0mm} Dr Richard HOOPER\\
\noindent\rule{7cm}{0mm} Director\\
\noindent\rule{7cm}{0mm} Department of Safeguards\\
\noindent\rule{7cm}{0mm} International Atomic Energy Agency \\
\noindent\rule{7cm}{0mm} P.O. Box 100 \\
\noindent\rule{7cm}{0mm} A-1400 Vienna

\noindent\rule{7cm}{0mm} 21 November 1997


\noindent Dear Dr Hooper,

Please find enclosed the title page of the report ISRI-95-03 on ``Iraq's Calutrons'' that we have circulated for review starting October 1995.

It is quite possible that you have seen a copy of this report. It was sent, in particular, to late Maurizio ZIFFERERO, then leader of the UNSC 687 action team in Iraq. In case of need, you should find a copy of it in the IAEA Library.

The purpose of this letter is connected with the fact that you will be giving a lecture on Monday December 1st at the 515th Wilton Park Conference: ``Building an effective international consensus against nuclear proliferation.'' Since I will be attending this conference, I would greatly appreciate to have a brief conversation with you on the correspondence I had with Dr Maurizio ZIFFERERO about our report on Iraq's calutron effort.

In effect, since Octobre 1995, we had quite a substantial correspondence about our report. This results in many important improvements and we would like to put this document into a final form, hopefully in the beginning or Spring of 1998. In particular, I would like to know when we may expect to receive the EMIS part of the Iraqi ``Full, Final and Complete Declaration'' that was promised by Dr Maurizio ZIFFERERO in his letters to us of 26 January and 2 August 1996 (enclosed).

I believe that a conversation at Wilton Park about this question might be useful before I write to the new leader of the UNSC 687 action team.

I would greatly appreciate that you help me on this point.

\noindent\rule{7cm}{0mm} Yours sincerely,

\noindent\rule{7cm}{0mm} Dr Andr\'e Gsponer\\
\noindent\rule{7cm}{0mm} Director

\noindent Encl.:

\indent  - Title page and abstract of ISRI-95-03, 19 October 1995\\
\indent  - Letter of 2 August 1996 by Dr Maurizio ZIFFERERO to ISRI

\chapter{Iraq's calutrons: 2002 - 2005}
\pagestyle{myheadings}
\markboth{Iraq's calutrons:}{2002 - 2005}
\label{Chap.5}

by {\bf Andre Gsponer}.

Many events related to Iraq's former nuclear weapons program, including a second war on Iraq,  occurred after the paper \emph{Iraq's calutrons: 1991 - 2001}, reproduced in Part \ref{Chap.4} of this report, was posted on internet on July 31, 2001, that is only a few weeks before September 11, 2001.  This is because the US Government immediately started to claim that there could be a link between Al Quaida's terror attack on the United States and Saddam Hussein's regime in Iraq  --- which led to the false claim that Saddam Hussein was trying to restart his nuclear program.\footnote{The false claims that Iraq had resumed its nuclear, chemical, and biological weapons programs were used to justify Gulf War II.  The invasion of Iraq began on 20 March 2003,  Bagdhad fell on 9 April 2003, and the war was officially declared to be ``over'' on 30 April 2003.}

In order to remain brief, only four events will be reported here: The unsuccessful attempt in April 2003 by Dr.\ Jafar Dhia Jafar (who ran Iraq's nuclear program from its beginning to its end) to come to Geneva to tell his version of the facts right after the beginning of the second Gulf War; the contribution in August 2003 of a comment to Lord Hutton's inquiry on Dr.\ David Kelly's suicide; the publication in January 2005 of J.D.\ Jafar's book ``The Assignment;'' and finally in May 2005 the publication of a conspicuous paper on US calutron technology on the occasion of the 60th anniversary of the production of the 50 kg of enriched uranium used in the atomic bomb that destroyed Hiroshima.

\section{J.D.\ Jafar's aborted visit to Geneva during the 2003 invasion of Iraq}
\label{Sec.5.1}

Although he had been approached by US intelligence to defect before the 20 March 2003 US-led invasion of Iraq, Jafar Dhia Jafar remained in Iraq until he fled to the United Arab Emirates just two days before Baghdad fell to coalition forces, on 9 April 2003.  This is how, three weeks later, I received the following e-mail:
\begin{quote}
{\tt \footnotesize
Date: Sat, 26 Apr 2003 20:09:27 +0400\\
From: ajafar@emirates.net.ae\\
Subject: Attn: Professor Andre Gsponer\\
To: gsponer@vtx.ch\\
~\\
Dear Professor Gsponer,\\
~\\
My father, Dr. Jafar Dhia Jafar, would like to speak to you confidentially 
and has requested my help to obtain your private telephone number.\\
~\\
Since he does not currently have access to email, please can you 
respond to this address.\\
~\\
ajafar@emirates.net.ae\\
~\\
Thank you for your assistance.\\
~\\
Yours sincerely,\\
~\\
Amin Jafar
}\end{quote} 
\noindent To which I immediately replied:
\begin{quote}{\tt \footnotesize
Date: Sat, 26 Apr 2003 18:24:01 +0100\\
From: gsponer <gsponer@vtx.ch>\\
Subject: Telephone number\\
To: ajafar@emirates.net.ae\\
~\\
Dear Amin Jafar,\\
~\\
Please inform your father that he is most welcome to call me at the number below which is currently both my professional and private telephone number.  He may call me on this direct line any time (except tomorrow) between 9 am and 9 pm Swiss time.\\
~\\
Sincerely,\\
~\\
Andre Gsponer\\
}\end{quote} 

Jafar called me on the 4th of May at 6 pm.  After greeting he said: \emph{``I come right to the point: I need an invitation to get a visa for coming to Geneva.  I have a valid passport and I can travel.  But I need an invitation and I thought that you could help me with that.''}  Jafar further explained that he wanted to come to Geneva to explain that Iraq destroyed its nuclear weapons program in 1991 and never restarted it, so that there was nothing to be found in Iraq, contrary to the claims made to justify the invasion of Iraq.

I responded that I was ready to invite him at ISRI so he could do that, but that  I had no direct experience with inviting somebody like him under the current circumstances. I would therefore first have to enquire on how to proceed in order to avoid a categorical \emph{``no''} from the Swiss administration.

Consequently, right after we convened of the time of our next telephone conversation, I called some diplomats I knew with whom I could speak in full confidence.  After some brainstorming, I decided to take the advice of trying to get the support of the Swiss Ambassador to the UN.

It was not possible to speak right away to Christian Faessler, the Swiss Ambassador to the UN responsible for disarmament matters.  But I could immediately speak to his military counselor, Dr.\ Ren\'e Haug.  Fortunately, I knew Haug since 1980, i.e., the time when after having left CERN I was learning something about international relations, strategy and international security, and diplomacy, by taking classes as student at large at University of Geneva's Graduate Institute of International Studies (IUHEI\footnote{Institut Universitaire de Hautes Etudes Internationales)}), where Haug was an assistant.  I could therefore casually explain him that I was seeking the Ambassador's support for insuring that Jafar could get a visa to enter Switzerland and spend some time at ISRI to express his point of view, something I thought should be good for Switzerland's diplomacy and reputation as a neutral country.

Ren\'e Haug talked to the Ambassador, who said to be ready to support Jafar's visit to Switzerland.  The next step was to enquire at the federal office in Berne responsible for granting visas, to which Haug sent an e-mail on 6th of May.  I could therefore tell to Jafar that within a week or so he should know whether his visa application would be accepted or not.  Unfortunately, when the answer came, on 13 May, it was negative: 
\begin{quote}
{\it ``No personality of the former Baghdad regime will receive a visa 
following a decision of the Federal Council of February 2003.''}
\end{quote}

  This answer was given in writing to Ren\'e Haug,\footnote{\it ``Les personalit\'es de l'ancien r\'egime de Baghdad ne recevront aucun visa suite \`a la d\'ecison du Conseil F\'ed\'eral de f\'evrier 2003.''} who told me informally that the Ambassador and himself were surprised since they had not been informed that such a decision was taking before the war in Iraq had started. 

When I reported this negative answer to Jafar he was very disappointed as well: \emph{``I have never been a member of the Baath party, I have never been in the government, I am not in the list of the 55 most-wanted Iraqi},\footnote{It was later told by US officials, in July 2003, that besides the list of 55 there was also a longer, unpublished list of Iraqi leaders earmarked by the US government.} \emph{I am completely free to move...''}  In response to this I suggested that we may wait a couple of months to make another try, and asked for his mailing address so that I could send him a letter:
\begin{quote}

\emph{
\noindent\rule{7cm}{0mm} May 19, 2003
~\\
Subject: Visit to Geneva\\
~\\
Dear Dr.\ Jafar,\\
~\\
First of all I wish to thank you once again for having taken the initiative to get in touch with me.  This is a recognition of the fact that I have always worked for a truly universal nuclear disarmament, and that I have always kept a respectful attitude towards all parties.\\
~\\
In this letter I would like to reiterate what I said on the telephone, namely that you should be able to come to a place like Geneva in order to express what you have to say.  This is very important because the current political situation is extremely bad and thus only truly competent and well informed people like you may have an influence on the course of events.  For instance, it was mainly because senior scientists who took part in the Manhattan Project were willing and able to express themselves (in the political arena and in journals like {\bf The Bulletin of the Atomic Scientists}) that some of the most terrifying possibilities did not realize in the past fifty years.  These people have now passed away, and we are witnessing a come back of the worse diseases which plague affluent societies: imperialism, militarism, racism, and colonialism. (In annex I am sending you two examples which confirm that I am not alone in making such a pessimistic assessment.) \\
~\\
As I have said to the Swiss ambassador to the U.N., the danger of not granting a visa to personalities like you is that scientists and technician who have worked on military technology in your (and other) countries may ultimately decide to disappear and work in clandestinity.  He agreed with me and it is therefore very annoying that you could not come.  However, the main reason for not granting the visa is apparently purely administrative: It is too soon to challenge a decision made by the Federal Council less than three months ago.\\
~\\
Therefore, if you can wait a few more months, it might become much easier to get a visa.  Whenever you will feel like making another try I shall be ready to help you again.  In the mean time, if you start working on some papers or a report on your experience I am also ready to help you if I can.  Feel free to ask me if you need some documents, or to send me your drafts for review, etc.\\
~\\
In conclusion, I very much hope that you will be able to come to Geneva in a few months time, which will give me the opportunity to meet and work work with you on questions of common interest.\\
~\\
Best regards,\\
~\\
\noindent\rule{7cm}{0mm} Dr.\ Andre Gsponer\\
~\\
P.S.: Please let me know if you have received my letter.\\
~\\
Encl.:
}
       \begin{itemize}
       \item J. Scheffran, \emph{Editorial: The new American dream is a nightmare}, INESAP Information Bulletin, No.\ 21 (April 2003) 2.
       \item \emph{E-mail exchange between Dr.\ Daniel Amit, a leading Jewish physicist, and Dr.\ Martin Blume, editor-in-chief of the Physical Review} (March 21, April 8, April 9, and April 14, 2003).
       \end{itemize}

\end{quote}
To which Jafar responded:
\begin{quote}
{\tt \footnotesize
Date: Mon, 26 May 2003 11:45:56 +0400\\
From: Jafar DA Jafar <jafdjaf@hotmail.com>\\
Subject: Visit to Geneva\\
To: gsponer@vtx.ch\\
~\\
Dear Dr. Gsponer\\
~\\
Thank you for your letter dated May 19, 2003 which I received yesterday.\\
~\\
Also, I read with interest the two attachments to your letter. It is unfortunate that opinions like  these are not well publicized, in fact contrary opinions find their way, far more easily, in the press of widest circulation. However, it is still very important to learn the lessons of the more recent past in order to make it harder for a repeat performance on some other unfortunate people. I am glad to see that you are as much concerned as some of us who were directly affected by the more recent events.\\
~\\
Thank you again for your efforts and I shall await your signal for the proper time.\\
~\\
With best wishes,\\
~\\
Dr. J. D. Jafar
}\end{quote} 

Unfortunately, I could never send a positive signal to Dr.\ Jafar, who in the meantime turned to other countries, where he met similar problems.   For instance, in November 2003, he was prevented from flying to Britain after his entry visa was cancelled just hours before he was due to board a plane to Heathrow.  It was only the following year that he was finally able to go to England, where he could give interviews, including his first broadcast interview aired on BBC on the 11 August 2004.

\section{Comment: ``Farewell to a fellow whistle-blower''}
\label{Sec.5.2222}

Comment sent to the BBC on 2003/08/21 at 21:15 GMT in relation to the Hutton inquiry on Dr.\ David Kelly's suicide, with the request of forwarding the full commentary to Lord Hutton, which was done.  Excerpts of the comment were published on the BBC website on 2003/08/28 at 21:17 GMT.
\begin{quote}

{\it

As the Hutton inquiry proceeds it becomes more and more clear that what has killed Dr.\ David Kelly is the collapse of his system of values.  After having worked for more than twenty years for the cause of true peace and disarmament, he could not live anymore with the evidence that these goals were nothing but words in the mouth of the most senior political leaders.

As today's testimony of Ambassador David Broucher confirmed, Dr.\ Kelly could not imagine that a democracy such as the United Kingdom would take part in the open aggression of a sovereign State.  In the previous days we heard how much facts --- simple undisputable technical facts --- were distorted and reworded in order to fit the `political judgement' of the top politicians.  It would have seemed that such  Middle-Age casuistry should be anachronistic in modern democracy...  But that is not all:  There are more than twenty years that facts about Iraq's (and other countries's) weapons of mass destruction (WMD) are systematically distorted by Western governments.

Like Dr.\ Kelly, who happens to be only a couple of years older than me, I have accumulated much evidence over the 24 years that I have been working on Iraq's WMD, see {http://arXiv.org/abs/physics/0512268}.  Already before the 1991 Gulf War everything was done to make sure that WMD could become a pretext for attacking Iraq in case of disgrace.  At no point between 1979 and 1991 did Western governments really object to Iraq's development and use of WMD.  Neither did Britain object to Iraq's invasion of Koweit after having been duly informed of its imminence.

What Dr.\ Kelly is recalling us is that words about human rights, rule of law, democracy (in short, `top Western values') should be put into practice in international relations.  Otherwise, we will only add more weight to people like Oussama Bin Laden whose main criticism of the West is precisely its lack of sincerity, and its systematic use of double standards.

Dr.\ Andre Gsponer

}

\end{quote}

\section{J.D. Jafar's book: ``The Assignment''}
\label{Sec.5.333}

In January 2005 the Norwegian publisher \emph{SPARTACUS Verlag} announced the publication of a book based on a manuscript by J.D. Jafar on Iraq's nuclear weapons program.  According to \emph{SPARTACUS Verlag}'s web-site:\footnote{{http://www.spartacus.no/index.php?Books=\&ID=Artikkel\&ID2=Vis\&counter=71} }


\begin{quote}
\begin{center}
{\bf The Assignment}\\
The inside Story of the Iraqi WMD-Programme\\
by\\
Jafar Dhia Jafar, Numan Saadaldin Al-Niaimi, Lars Sigurd Sunnan\aa \\
240 pages\\
ISBN 82-430-0333-9
\end{center}
~\\
{\it 
This book presents the complete story of the Iraqi programme 
for aquiring WMD, told by those who directed it on behalf of Saddam Hussein. 

Dr.\ Jafar Dhia Jafar, born in 1942, educated in the UK
and married to a British woman, was the
initiator and scientific leader of the
Iraqi nuclear programme, and later the
chief liaison between the Iraqi regime and
the various international bodies working to
disarm Iraq. He fled Baghdad shortly before
the collapse of Sadddams regime in April
2003. Together with his close colleague
Numan Saadaldin Al-Niaimi, also a
distinguished leader of the Iraqi
WMD-programme, and the Norwegian
journalist Lars S. Sunnan\aa, he has written
a story from within, both personal and
political, viewing the unfolding events
from the perspective of the engaged
scientist as well as that of a citizen and
father. In addition to thrilling accounts
on the progress of the development of an
Iraqi nuclear bomb, the authors present
extraordinary glimpses from the inner
circles of Iraqi political life, including
several encounters with Saddam Hussein
himself. 

Engaging in their criticism of the regime,
however, the authors offer their most
severe criticism of US foreign politics and
the UN-inspectors, convincingly
demonstrating that Saddam closed down his
WMD-programme already in 1991. One of their
points in case, is that the ongoing bloody
war could have been avoided had the
inspectors done their job without looking
to political concerns.

Lars Sigurd Sunnan\aa ~is an acclaimed
journalist, and was Norwegian
Broadcasting's Middle East correspondent
from 1999 to 2003.

World rights held by Spartacus Publishing
House in Oslo, Norway, except Arabic
language territories.
}
\end{quote} 

\begin{figure}
\begin{center}
\resizebox{8cm}{!}{ \includegraphics{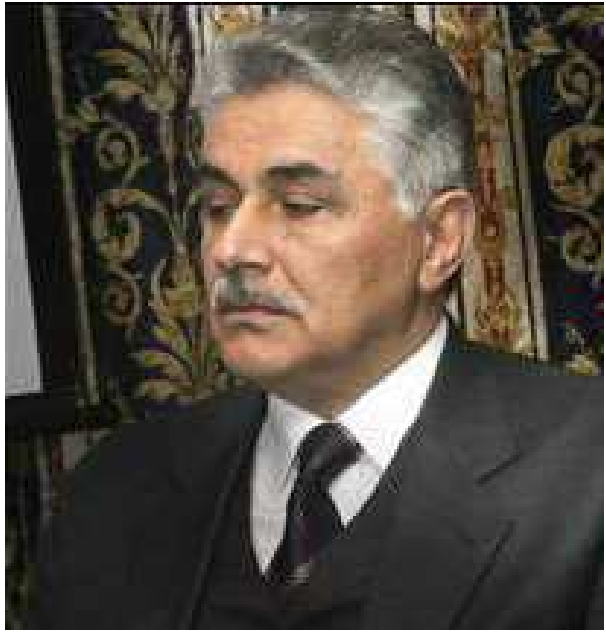}}
\end{center}
\caption[J.D. Jafar in January 2005]{\emph{J.D. Jafar at the 27 January 2005 press conference in Oslo presenting his book:} The Assignment --- {\it The inside Story of the Iraqi WMD-Programme.}}
\label{fig:Jafar-2005}
\end{figure}

As this book is in Norwegian its audience is of course somewhat limited, although an English version is expected to be published sometimes in the future.

Fortunately, Prof.\ Richard Wilson at Harvard University has a copy of Jafar's original text in English, and was kind enough to suggest an excerpt which is of special interest in the context of the present report:\footnote{Richard Wilson, private communication, 19 January 2006.}

\begin{quote}
{\it
A comparative study of three proven enrichment 
technologies, conducted during the second half of 
1981 came to the following conclusions:

\begin{itemize}
\item       The electromagnetic Isotope Separation 
(EMIS) method had a number of advantages:
   \begin{enumerate}
   \item       It is reasonably well documented in unclassified literature;
   \item       The design and manufacture of prototype 
equipment could be achieved with available resources and skills; and
   \item       HEU could be achieved after only two enrichment stages.
   \end{enumerate}
The main disadvantages of this method were: it 
requires bulky and weighty components, intricate 
chemical processing and is an uneconomical 
proposition when a large output is required.
\item       The gaseous diffusion method had inherent disadvantages:
   \begin{enumerate}
   \item       The barrier has to be invented from basic 
principles, thereby requiring an intensive R \& D 
effort with no guarantees of success.
   \item       The method requires a large number of 
compressors that could neither be purchased easily nor manufactured locally.
   \item       Gaseous diffusion demands a large cascade 
of approximately 3500 stages to produce HEU.
   \end{enumerate}
\item       The gaseous centrifuge method is 
technologically more advanced than gaseous 
diffusion and, apart from an intensive R \& D 
effort would need more than one invention to 
harness magnetic centrifuge technology.
\end{itemize}
}
~\\
Uranium enrichment programme\\

{\it 
Before 1981 came to a close, a decision was taken 
to establish an enrichment technology as a 
precursor to the production of HEU in accordance with the following strategy:
\begin{itemize}
\item       To adopt EMIS as a primary option and to 
proceed with its development in three phases:
   \begin{enumerate}
   \item       To build research units in the first phase.
   \item       To build demonstration units in the second phase.
   \item       To build production units to achieve 15 
kg U/year of 93\% enrichment with natural uranium 
as feed. Ultimately using LEU as feed material 
would enhance output by fivefold.
   \end{enumerate}
\item       To adopt gaseous diffusion as a parallel 
option with the following objectives:
   \begin{enumerate}
   \item       To develop a suitable barrier.
   \item       To develop the specifications of suitable 
compressors and blowers and to operate them singly and in short cascades.
   \item       To build a cascade to produce LEU of 
around 3-4\% enrichment and about 5 ton U/year to provide the feed for EMIS.
   \end{enumerate}
\item       Activities were organised in such a 
fashion so as to ensure the following criteria:
   \begin{enumerate}
   \item       A thorough assessment of published 
information had to precede any project.
   \item       Research and development projects were 
adopted so that adequate understanding of the 
know-how in the main and supporting technologies could be achieved.
   \item       Work had to start in the laboratory 
phase, the results of which were then utilised in 
the pilot phase. The final production phase was 
to be implemented on the basis of the outcome of 
the preceding two phases. These phases could be 
partially overlapped whenever possible. In other 
words, at the time of initiating the first phase 
of experimental activity a pilot project could 
also be commenced. Inherent in this mode of work 
was the deployment of assumptions and parameters 
that had not yet been adequately tested or 
verified experimentally. In many cases work on 
the production phase could begin at approximately 
the same time as experimental work relating to 
the previous phase was initiated. But in such 
cases, the margin for error is wider and that had 
to be set against the `time saving' 
consideration. As the work progressed, 
modifications and changes would have to be 
introduced into the designs as and when such 
alterations were deemed advantageous.
   \end{enumerate}
\item       All research and development activities 
were to be performed at the Tuwaitha site, while 
production plants and ancillary fabrication 
plants were to be constructed at new, carefully 
selected, sites. The main criteria for selection 
of those sensitive sites were:
   \begin{enumerate}
   \item       They had to be topographically fortified 
against aerial attack or sabotage.
   \item       They had to be near large communities so 
that local manpower could be recruited and trained.
   \item       Sites had to be geographically scattered 
so that a simultaneous attack on all would 
require immense military resources on the part of a potential enemy.
   \end{enumerate}
\end{itemize}
}
\end{quote}

\section{Sixty years after Hiroshima}
\label{Sec.5.444}

In May 2005, {\bf Physics Today}, the flagship magazine of the \emph{American Physical Society}, published a seven page-long apologetic paper by William E. Parkins, a participating scientist in the World War II calutron project that succeeded  in producing the fissionable material for the Hiroshima bomb during World War II:
\begin{quote}
 
W.E. Parkins, 
\emph{The Uranium Bomb, the Calutron and the Space-Charge Problem,} 
{\bf Physics Today}, Vol.\ 58 (May 2005) 45--51.

\end{quote}

Let me just quote the opening sentence: 
\begin{quote}

{\it ``Uranium! Uranium! Uranium!'' A voice shouted out into
 the night from the second floor of a dormitory in 
Oak Ridge, Tennessee.  It was 6 August 1945.  That day, 
President Harry S Truman had announced to the world 
that the US had dropped a new weapon, a uranium bomb, 
on the city of Hiroshima, Japan''} [p.~45].

\end{quote}
and the closing sentence:
\begin{quote}

   {\it ``The development and use of the calutron to produce 
enriched uranium for the first atomic bomb that was 
exploded in warfare, and then to produce the full 
spectrum of separated isotopes for uses in peacetime, 
is the greatest example of beating swords into 
plowshares in the history of mankind.  For its 
contribution to both wartime and peacetime, the physics 
profession can be proud''} [p.~51].

\end{quote}

   To write in such glorious terms about the science that enabled one of the most tragic events in History, an act that is qualified by many as a crime against humanity, only one year after Gulf War II began, is most disturbing, not just to myself but to many people as well, including yesterday's Japanese victims, and today's Iraqi victims of the development of calutron technology.

If war is the mother of all necessities, as is sometimes said, there is no reason for publishing such apologetic sentences in a prominent scientific journal, especially since people like William E. Parkins were very young at the time, and probably totally blinded by years of propaganda in one of the most overtly patriotic countries in the world.

But if that type of blindness (i.e., that lack of historical and philosophical distance and judgment with regards to ongoing events) could be excused to some extent at the end of a major world war, there is no excuse for presenting these events in heroic terms at a time when an all powerful Western coalition is trying to use its overwhelming military power to convert to democracy a country that is wrongly accused to have secretly resumed its calutron program.

Moreover, it is highly disturbing that this article contains a number of scientific claims that are only partially correct, which suggests that it was more important to the Editors of {\bf Physics Today} to print a very positive and politically correct apology of the wartime calutron effort on the occasion of the 60th anniversary of its technically successful end, than to make sure that the paper was factually and historically correct as far as its scientific and technical contents are concerned.

Indeed.  While on the one hand Parkins's paper does not contain any striking new technical information, it does give a unique qualitative and quantitative appreciation of the relative importance of the key elements and ideas which made the US calutrons feasible.  In particular, on page 50, Parkins claims that the impact of space-charge neutralization was a factor of 400 on productivity, much more than I would have thought.  On the other hand, however, Parkins claims on page 46 to deserve \emph{``credit for having discovered and explained the automatic self-neutralization of intense ion beams''} propagating through the residual air left in the calutron chambers.  Moreover, most of Parkins paper is more or less structured around the pretense that the development of the calutron was a major contribution to basic science (e.g., plasma physics), as well as a major technological development with numerous unique applications in the non-military domain.

As a matter of facts, as I got interested in the question of particle beam weapons\footnote{See Ref.~\cite{AA.71} on page \pageref{ISRI-82-04}, available at { http://arXiv.org/pdf/physics/0409157 }.} in 1978 already\footnote{See the postface, \emph{An afterword --- twenty years after} by Robert Jungk, \emph{in:} Robert Jungk, ``Die Grosse Maschine --- Auf dem Weg in eine andere Welt'' (Goldmann Verlag, Second edition with a 20 page-long postface, ISBN 3-442-11373-3, 1986) 251--284; French translation available at\\ { http://cui.unige.ch/isi/ssc/phys/Jungk-postface.html }.} (one year before the question of calutrons), I have always been aware of the importance of space-charge neutralization for overcoming detrimental beam self-space-charge effects.  I therefore knew that this effect (for both electron and ions beams) was already known before World War II, as is confirmed (in particular) by theoretical papers such as by W.H.\ Bennett,\footnote{W.H. Bennett, \emph{Magnetically self-focusing streams}, Phys. Rev. {\bf 45} (1934) 890--897.} as well as by photographs such as those taken on 26 March 1936 of a high current cyclotron beam injected into the atmosphere at E.O.\ Lawrence's laboratory at Berkeley.\footnote{See Ref.~\cite{AA.8} cited on page \pageref{Heilbron}. } As for the space-charge neutralization of ion beams in calutrons, the last chapter of Koch's book (pages 274--300 of Ref.~\cite{AA.23} cited on page \pageref{Koch}) is discussing both the work of Parkins {\it et al.}~and the wartime contributions by German scientists (e.g., F.G.\ Houtermans,\footnote{Professor Houtermans, who encouraged Robert Jungk to write his book ``Brighter than Thousands Suns'' on the political and moral problems faced by those who built the first atomic bombs, is mentioned in the preface of ``La Quadrature du CERN,'' Ref.~\cite{AA.47} cited on page \pageref{Grinevald}.} K.H.\ Riewe,\ and W.\ Walcher) which were published in 1941 and 1943 already.

But, when I read Parkins's paper in June 2005, I was less concerned by these technical inconsistencies than by the possibility that this paper could offer to Dr.\ Jafar an opportunity to express his views in the form of a letter to the Editors of {\bf Physics Today}.  I therefore sent him, on June 24, a paper copy of Parkins's article, together with as a cover letter saying: 
\begin{quote}
{\it ``The main purpose of my letter is to suggest that the publication of Parkins's paper in {\bf Physics Today} gives you an opportunity to respond to it, and therefore to put on record some of the facts that you think should be made available to physicists and historians about what happened in Iraq.  I know that you are writing a book on this, but it may also be useful to state in a shorter form some of the most important facts.  In particular, as a pacifist, I do not see why the US calutron effort should be presented as a success sorry, while what was done in Iraq is actually not very different from what was done in Switzerland (before and after becoming a party to the NPT), not to mention other countries.''}
\end{quote} 

   Jafar chose not to write such a letter to the Editors. But some American readers did, and three letters were published together with a reply by Parkins on pages 13 to 15 of the November 2005 issue of {\bf Physics Today}.  Unfortunately, the purpose of all three letters was just to correct (or comment on) some factual details, such as Parkins claim to have discovered the self-pinch effect, which according to H.F.\ Ivey's letter\footnote{Physics Today (November 2005) 13.} was apparently known in the 1920s already.   In other words, just like Parkins's paper, the three letters were purely ``scientistic,'' i.e., typical of the dominant quasi-religious attitude of contemporary scientists, who believe that science and technology should be blindly pursued and developed, irrespective of their social, political, and philosophical consequencies.

\backmatter

\chapter{Postface: Nuclear weapons and Western democracy}
\pagestyle{myheadings}
\markboth{Postface}{Postface}
\label{Postface}

The main document in this collection, Paper \ref{Chap.2}, was written in 1995 because that year Andre Gsponer restarted working in Geneva after a long convalescence for a cancer treatment.\footnote{Although written in the third person, this postface is by {\bf Andre Gsponer}.}  That same year Gsponer postulated for a job opening at the Swiss Federal Office of Energy (OFEN\footnote{Office f\'ed\'eral de l'\'energie.}): The position of Vice-director for Non-proliferation affairs.

Gsponer thought he was almost the ideal person for that job, since he could not think of any Swiss citizen with better training (in nuclear physics and in international affairs), experience, and motivation than him.  Moreover, he easily obtained the support of personalities such as Prof.\ Maurice Cosandey, former President of the council of the Swiss Federal Schools of Technology, and Prof.\ Claude Zangger, best known in the world as the initiator of the ``Zangger list'' of dual-purpose nuclear equipments put under international control by the so-called London Club of nuclear suppliers.

Gsponer was therefore interviewed by Dr.\ Eduard Kiener, Director of OFEN, as well as by Dr.\ Alec-Jean Bear, the retiring Vice-director that he hoped to replace.  To Gsponer's surprise, Baer asked him a number of embarrassing and vicious questions.  In particular, Baer wanted to know how, as a pacifist and well-known outspoken nuclear-skeptic, Gsponer could be in favor of the abolition of nuclear weapons while at the same time working in an Office whose duty is to promote nuclear energy as one of the main sources of Switzerland's energy supply.

Gsponer replied that he did not see any contradiction. As Vice-director for Non-proliferation affairs his duty was to promote the non-proliferation of nuclear weapons, and to represent the interests of Switzerland at the Board of Governors of the International Atomic Energy Agency (IAEA) in Vienna, two things that as a convinced democrat he could perfectly do, since his opposition to nuclear energy was subordinated to his duties as a Swiss citizen.

But Dr.\ Baer kept similing and asking nasty questions, implying that Gsponer was naive for the job, and that Gsponer did not appreciate that \emph{one cannot defend the case for nuclear energy if one is {\bf not} in favor of nuclear weapons as well!} For this, and most probably other reasons, Gsponer was not selected for the job. 

The following year, on 25 April 1996, at 14.30 precisely, the Swiss Federal Council declassified a previously secret report: The official history of the Swiss nuclear weapons program, from its initiation on February 5, 1946, until its termination on November 1, 1988 --- i.e., eight years before its mere existence was formally recognized and its termination announced to the Swiss parliament and people.\footnote{Gsponer was among the people who strongly suspected the existence of this program, and wrote several papers about it.  See, A.\ Gsponer, \emph{La Suisse et la bombe}, Le Rebrousse Poil, No.\ 20, 21, 23, 25, 26 (Lausanne, 1979, 1980) \emph{ca.}~21~pp., and \emph{Die Schweiz und die Atombombe}, Virus, No. 31, 32, 33 (Zurich, 1980) \emph{ca.}~14~pp.   The first in-depth academic investigation of the full motivations of the development of nuclear energy in Switzerland is by Peter Hug, \emph{Geschichte der Atomtechnologie\-entwicklung in der Schweiz}, Lizentiarbeit in Neuerer Allgemeiner Geschichte (Historischen Institut der Universit\"at Bern, Bern, April 1987) 390~pp.}  Gsponer immediately wrote to the History section of the Military department, and got a copy of the report.\footnote{J. St\"ussi-Lauterburg, \emph{Historischer Abriss zur Frage einer Schweizer Nuklearbewaffnung} (Swiss Military Department, Federal Military Library and Historical Service, Spring 1995, secret, declassified 25 April 1996) 99 pp.  This report is non-technical and mainly a chronology of the administrative history of the high-level committees supervising the program since 1966. A rough English translation of its introduction and conclusion are available at:\\ { http://nuclearweaponarchive.org/Library/Swissdoc.html }.\label{STUSSI}}

In this report, to his great surprise, Gsponer discovered that Alec-Jean Baer had been, since January 1, 1986, a member of the highly secret AAA\footnote{Arbeitsausschuss f\"ur Atomfragen} study group whose duty was to make sure that Switzerland had the capability to produce nuclear weapons on short notice, if so requested by the Federal Council.\footnote{Actually, A.-J. Baer was only one of several members of AAA that Gsponer unknowingly met during the past 25 years.  The first one was Deputy foreign minister Herbert von Arx, whom he met in 1983 at the Federal palace on the recommendation of Claude Zangger, and who declined to support a non-proliferation project presented by Gsponer, possibly because it was truly motivated by disarmament considerations.} Therefore, the Swiss representative at the IAEA had the doubtful duty to pretend that his country was in favour of non-proliferation, while in fact Switzerland had not truly renounced to nuclear weapons for at least 19 years after it had signed the Non-Proliferation Treaty (in 1969), and 11 years after it had ratified it (in 1977).\footnote{Switzerland was in no way unique in this respect: The first head of IAEA, Dr.\ Sigvard Eklund, was at the same time the head of Sweden's secret nuclear weapons program.}  Even worse, according to the declassified report, Switzerland, among other things, kept a secret stockpile of uranium which was hidden from the IAEA.\footnote{Moreover, Switzerland had, albeit under IAEA safeguards, almost 100~kg of separated weapons-grade plutonium on long-term loan from England (see, R. de Diesbach, \emph{Une bombe A suisse en quelques jours ?}, Tribune - Le Matin, Lausanne, 6 ao\^ut 1980), that it very reluctantly returned to England after considerable delay.} Thus, although safeguards agreements with the IAEA came into force on September 6, 1978, unlawful nuclear weapons activities continued until November 1, 1988.

Apart from these political facts, over the 42 years of existence of the Swiss nuclear weapons program, many nuclear weapon related technologies were indigenously developed\footnote{Switzerland also published what are possibly the most detailed computer simulations ever published of the explosion of an uranium implosion bomb. See A. Pritzger and W. H\"alg, \emph{Radiation dynamics of nuclear explosion}, J.~of Appl.\ Math.\ and\ Phys.\ (ZAMP) {\bf 32}  (1981) 1--11.} and exported to other countries.\footnote{See, in particular, the chapter \emph{Weitere CH-Beihilfen sur Bombe}, in: P. Hug, ed., Sulzers Bombengesch\"aft mit Argentinien --- Schweizer Beihilfe zum Atomkrieg (Arbeitsgemeinschaft gegen Atomexporte, Z\"urich, Juli 1980) 60--66.} For example, heavy water separation plants were exported to India where heavy water is used as moderator in plutonium and tritium breeding reactors; centrifuge enrichment components were exported to Pakistan where they are used in weapons-grade uranium enrichment facilities; and, of course, various sensitive high-technology machine-tools and components were exported to Iraq during the 1980s...

The question therefore arises as to why Switzerland, contrary to Iraq and former Yugoslavia (which like Switzerland had nuclear weapons ambitions), was not invaded by a Western-led armada, or at least sanctioned for having violated its obligations under the NPT?  

{\bf The answer may lie in the naive observation that the Swiss political leadership has in fact abandoned its nearly 200 years old policy of neutrality\footnote{Which lasted since the Treaty of Vienna, 1815.} at about the same time it definitely renounced to nuclear weapons, so that Switzerland does not have any independent foreign policy any more since around the year 1988...}

In the case of Yugoslavia the challenge to the major European countries (and especially to Germany, who was the first state to recognize Slovenia at the beginning of the break-up of Yugoslavia) was that a large and culturally strong country could emerge in southern-eastern Europe --- a country that would not automatically align on the European Union, but rather remain independent from it and collaborate with eastern European countries that were part of the former Soviet Union.  As Yugoslavia was ready to defend its identity, and capable of making nuclear weapons,\footnote{Former Yugoslavia, who fiercely defended its non-alignment on both the Soviet and Western blocks, has always been on the list of countries suspected to have a latent nuclear weapons program, because it had significant nuclear facilities and activities over the full range of possibilities.} the major European countries and the United States felt that there was no other option than to destroy the federal republic of Yugoslavia, rather than to keep it as a single entity...

For this reason, when IAEA officers flanked by US and Russian special forces raided, on 22 August 2002, the nuclear center of Vinca in the suburbs of Belgrade,\footnote{See, R. Stone, \emph{Belgrade Lab Sets New Course After Top-Secret Uranium Grab}, Science (30 August 2002) 1456.}  as soon as they could after the war against Serbia, to remove the 48~kg of highly-enriched uranium that were stored there,\footnote{This was just sufficient for two uranium-bombs.  However, in collaboration with CERN, Serb scientists at Vinca were considering since some years the use of a small accelerator as a sub-critical neutron multiplier, which would have enabled to convert these 48~kg of uranium-235 into enough plutonium-239 for at least twenty plutonium-bombs.  This technique, in which a secondary target made of U-235 is used to multiply the spallation neutrons produced in a primary target bombarded by a proton beam from an accelerator is in use at Los Alamos since more than thirty years. (See A. Gsponer, B. Jasani, and S. Sahin, \emph{Emerging nuclear energy systems and nuclear weapon proliferation},  Atomkernenergie $\cdot$ Kerntechnik {\bf 43} (1983) 169--174. See also end of Sec.~\ref{Sec.4.7}.)} many countries in the world understood that with the collapse of the Soviet Union the world had entered not just an era of denial of self-identity, but also an era of unrestricted military intervention and active counter-proliferation.

In particular, North Korea was among the first to draw the lesson from what happened in Vinca, and soon afterwards resumed operation of its plutonium breeding reactor.

Similarly, now that Iran is having US-led forces in two neighboring countries (Afghanistan and Iraq), the continuation of its independent nuclear energy program becomes more and more suspicious:  It is as if Alec-Jean Baer's message to Andre Gsponer that \emph{``nuclear energy cannot be dissociated from nuclear weapons''} was suddenly discovered by the international community, and more particularly by the three largest European countries: the United Kingdom, France, and Germany!

But this is not the case:  Nuclear energy, nuclear science, and nuclear weapons have always been part of the \emph{same} foreign policy,\footnote{Henry Kissinger's major book, \emph{Nuclear Weapons and Foreign Policy}, should therefore be rewritten and entitled: \emph{Nuclear Power and Foreign Policy.}}  and this at least since US President Dwight Eisenhower's ``Atoms for Peace'' initiative, presented in a speech before the United Nations General Assembly on December 8, of 1953.\footnote{Concerning the lesser known foreign policy implications of nuclear science, see in particular A.\ Gsponer et J.\ Grinevald, \emph{CERN --- La physique des particules pi\'eg\'ee par l'OTAN}, La Recherche, No.~381 (Paris, d\'ecembre 2004) 6, available at\\ { http://www.larecherche.fr/special/courrier/courrier381.html }.}

 Indeed, as is shown by the political analysis of this initiative, and of the Western Europe reaction to it, the atoms-for-peace policy of the United States was a foreign policy concept --- neither an energy policy nor a trade or economic policy.\footnote{See, for example, R. Kollert, \emph{Die Politik der latenten Proliferation. Milit\"arische Nutzung `friedlicher' Kerntechnik in Westeuropa}, Dissertation (Deutscher Universit\"atsverlag, Wiesbaden, 1994) 551 pp., ISBN-3-8244-4156-X; and R. Kollert, \emph{``Atoms-for-peace:'' A foreign policy concept of the Cold War gets into a clue to latent proliferation}, INESAP-Information Bulletin, No.\ 9 (May 1996) 22--24.}  Its first rationale was to demonstrate `world leadership' in order to bind neutral or politically indifferent countries to the West.  But its long-term goal, of which we see the major effects today, was to use nuclear power as a double-edged sword, by which access to nuclear energy would give to non-nuclear-weapon states the impression to get closer to nuclear-weapon-statehood, while at the same time the further development of nuclear technology by such states would become a justification for preventive intervention if at some point it was getting too close to the next step: the manufacture of nuclear weapons.

  The same implicit foreign policy concept was enshrined in the Non-Proliferation Treaty which came into force in 1970.  Those who joined the NPT as non-nuclear-weapon states were still able to pursue some ambiguous nuclear-weapons-related activities (as is remarquably well explained in the declassified 1996 report on Switzerland's nuclear weapons program, \emph{op.~cit.}, footnote \ref{STUSSI}), but as soon as these activities would no more be tolerable to the now \emph{official} nuclear-weapon states, they would become a justification for immediate sanctions, and possibly military intervention.  In other words, if the hidden agenda of the official nuclear-weapon states was to give access to nuclear energy to as many small states as possible in order to be sure that themselves could keep their nuclear weapons for ever, one could say today that they have succeeded!

  In this perspective, it is clear that what happened in Iraq over the past thirty years should be seen much more as a nuclear-armed ``cat and mouse'' game than as a story about a dangerous dictator in search of nuclear weapons to conquer the world.  For those who have read the various parts of this report there should be no doubt that Iraq's nuclear ambitions could have been easily stopped long before Iraq invaded Koweit in 1990:  There is no doubt that if CERN  physicist Andre Gsponer, Harvard professor Richard Wilson, and many others, knew about Iraq's intentions long ago, the secret services of the official nuclear-powers must have known much more for a much longer time.

  As a matter of fact it is mostly Western countries, such as the United Kingdom, France, and the United States, who trained Iraq's scientists and provided Iraq with  all the hardware it needed to try building a nuclear deterrent.  It is also for this reason that these countries knew that Iraq's nuclear weapons program was in fact quite different from that of Israel or Pakistan, for example.  Its purpose, which is amply documented in the UNSCOM\footnote{United Nations's Special Commission in Iraq.} inspection reports written between 1991 and 1996, was not just to make nuclear weapons by the shortest possible path, but also to turn Iraq into a modern industrialized state following the model set forth by the United States, the United Kingdom, and France.  Just like Yugoslavia in south-eastern Europe, such a state in the heart of the Middle East, which after Saddam Hussein could have emerged as a powerful, independent, modern, secular, and \emph{possibly democratic} country, with a high standard of living and plenty of natural resources, was a far bigger danger to Western's interests than Saddam Hussein.

  In conclusion, a non-superficial analysis of the origin and consequencies of Iraq's nuclear weapons program, and its comparison to similar developments in other countries, clearly shows that nuclear energy and independent statehood are more then ever intimately related, as much for the official nuclear-weapon states (and the non-nuclear-weapon states who have definitely renounced to full sovereignty by strictly adhering to the NPT) as for the countries who are struggling to exist as truly independent nation-states.  This means that nuclear energy, whose abolition is the prerequisite to the elimination of nuclear weapons, is the largest single obstacle to the abolition of nuclear weapons --- contrary to the pretense of ``Atoms for Peace,'' NPT, and IAEA, which is the hypocrisy that nuclear energy could in some way bring freedom and peace on Earth.

  The tragedy, in the over fifty-years-long enforcement of this hypocrisy, is that it is the so-called democratic countries which are precisely at its origin. It is also because the leading Western countries are precisely those which most strongly oppose giving up nuclear energy and nuclear weapons, whose mere possession is a crime, that one has many reasons to be pessimistic.  It is therefore very disappointing to discover, after 25 years of work on peace research and nuclear disarmament, that there is no sincere interest in either of these topics in any Western state, and that their dealing with Iraq's calutrons is just one illustration of their hypocrisy with regards to the true meanings of liberty and democracy.

\chapter{Appendix 1: Annex 1 of UN Security Council document S/22872/Rev.1}
\pagestyle{myheadings}
\markboth{Appendix 1}{Appendix 1}
\label{Appendix 1}

\begin{center}
                     {\bf\underline{DEFINITIONS}}
\end{center}

{\bf For the purpose of UN Security Council Resolutions 687 and 707, the 
following definitions will be adopted:}

{\bf NUCLEAR MATERIAL}

{\bf 1.1	``Source material''}\\
Uranium containing the mixture of isotopes occurring in nature; uranium depleted in the isotope 235; thorium; any of the foregoing in the form of metal, alloy, 
chemical compound or concentrate.

{\bf 1.2	``Special fissionable material''}\\
Plutonium-239; uranium-235; uranium-233; uranium enriched in the isotopes 
235 or 233; any material containing one or more of the foregoing.

{\bf 1.3 	``Nuclear-weapon-usable material''}\\
Nuclear material that can be used for the manufacture of nuclear explosive 
components without transmutation or further enrichment, such as plutonium 
containing less than 80 \% plutonium-238, uranium enriched to 20 \% uranium-
235 and uranium-233 or more; any chemical compound or mixture of the 
foregoing. Plutonium, uranium-233 and uranium enriched to less than 20 \% 
uranium-235 contained in irradiated fuel do not fall into this category.

{\bf NUCLEAR ACTIVITIES}

2.1-2.9  (inclusive) refer to activities prohibited under both Resolutions 687 and 707.

Any activity such as research and development, design, manufacturing, import of 
systems, equipment and components, pilot plant and plant construction, commissioning and operation, or utilization in one or more of the following:

{\bf 2.1  	Production of nuclear weapons}

{\bf 2.2  	Production and any use of nuclear-weapon-usable material}

{\bf 2.3  	Production of metals and alloys containing plutonium or uranium}

{\bf 2.4  	Weaponization}\\
	This covers the research, development, manufacturing and testing required to make nuclear explosives from special fissonable material.

{\bf 2.5	Nuclear fuel fabrication} using plutonium, uranium-233, uranium enriched to  20 \% or more in uranium-235.

{\bf 2.6	Import, construction or use of research and power reactors of any kind} utilizing uranium enriched to $\geq$ 20 \% in uranium-235, uranium-233, 
plutonium or MOX as a fuel or any reactor designed specifically for plutonium 
production. This includes critical and subcritical assemblies.

{\bf 2.7	Reprocessing of irradiated fuel}\\
	Including the use of hot cells and the associated equipment

{\bf 2.8	Enrichment of uranium in the isotope 235} and any preparatory steps in this process, including the preparation of UCl$_4$ and UF$_6$.

{\bf 2.9 	Production and separation of the isotopes of plutonium, hydrogen, lithium and boron}

2.10-2.18  (inclusive) refer to activities, permitted under resolution 687 but prohibited under 707.

Any activity such as research and development, design, manufacturing, import of 
systems, equipment and components, pilot plant construction, commissioning and 
operation, or utilization in one or more of the following:

{\bf 2.10	Import, construction or use of research and power reactors of any type} utilizing natural uranium or uranium enriched to less than 20 \% in 
uranium-235 as fuel. This includes critical and sub-critical assemblies, but 
excludes reactors specifically designed for plutonium production.

{\bf 2.11	Prospecting, mining or processing of ores containing uranium 
and/or thorium}

{\bf 2.12 	Preparation of chemical compounds containing uranium enriched 
to less than 20 \% in uranium-235 and thorium}, excluding the preparation of UCl$_4$ and UF$_6$.

{\bf 2.13 	Nuclear fuel fabrication} using natural uranium or uranium enriched to less  than 20 \% in uranium-235.

{\bf 2.14 	Processing and disposal of radioactive wastes}

{\bf 2.15	Nuclear fusion experimental devices based on magnetic or inertial   confinement, including diagnostics}

{\bf 2.16	Production of isotopes} both radioactive and stable. The production of the isotopes of plutonium, hydrogen, lithium, boron and uranium is prohibited.

{\bf 2.17	Import, construction and use of neutron sources, electron 
accelerators, particle accelerators, heavy ion accelerators}

{\bf 2.18	Research on radiation physics and chemistry and on the physical 
and chemical properties of isotopes} except in area relevant to items 2.19, 
2.20 and 2.21

2.19-2.21  (inclusive) refer to activities permitted under resolution 707

{\bf 2.19	Application of radiation and isotopes in food and agriculture}

{\bf 2.20	Applications of radiation and isotopes in medicine}

{\bf 2.21	Application of radiation and isotopes in industrial processes}

\chapter{Appendix 2: Original of Paper 1\\
~\\
Bombe Atomique: L'Irak est pass\'e par le CERN}
\pagestyle{myheadings}
\markboth{Journal de Gen\`eve}{22--23 avril 1995}
\label{Appendix 2}


by {\bf Suren Erkman}, published in the weekend edition of \emph{Le Journal de Gen\`eve}, Saturday 22 - Sunday 23 April 1995, pages 1 and 5.

\begin{figure}
\begin{center}
\resizebox{4cm}{!}{ \includegraphics{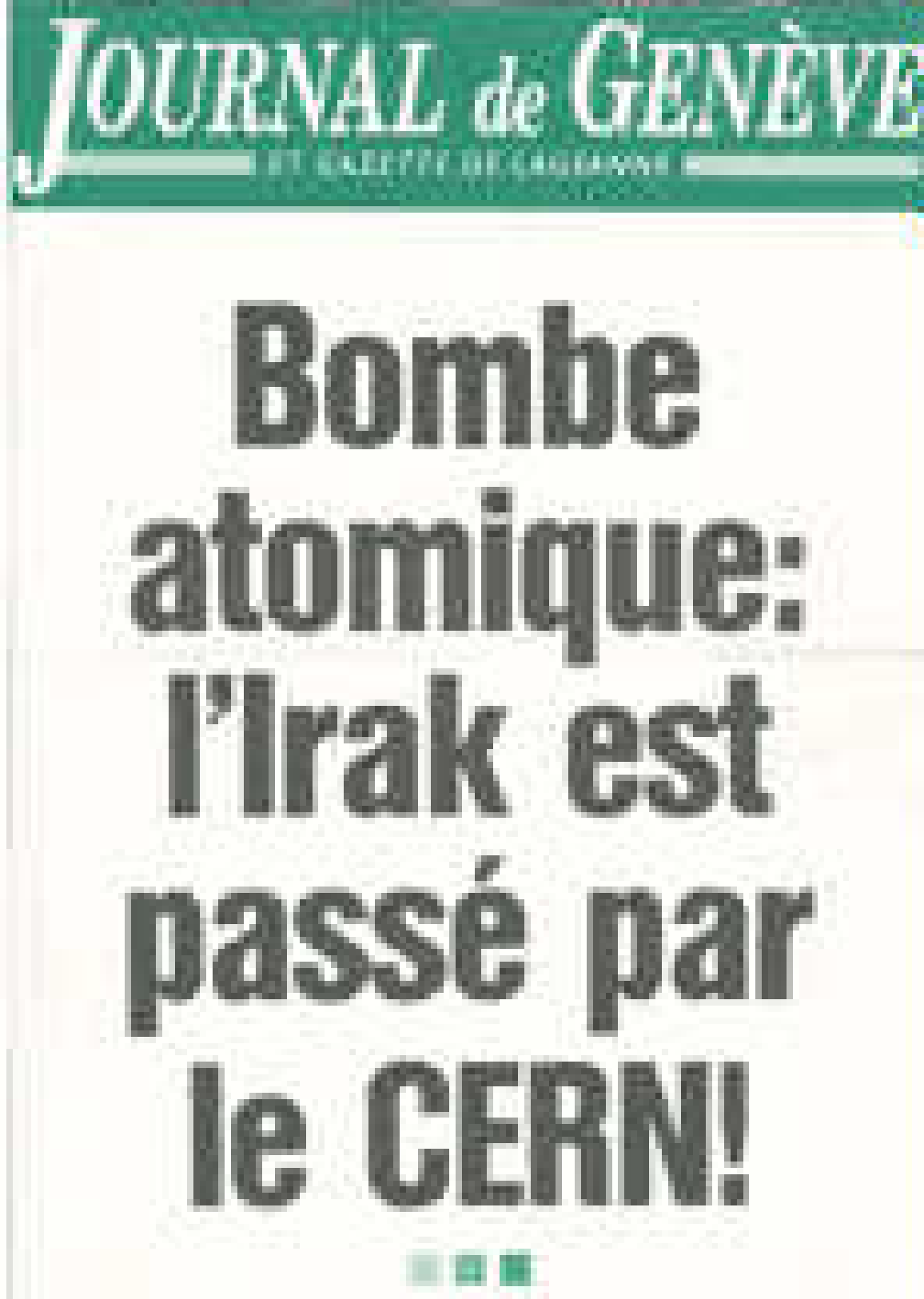}}
\end{center}
\caption[Affiche annon\c{c}ant l'article sur les calutrons irakiens.]{\emph{Affiche annon\c{c}ant l'article sur les calutrons irakiens.}}
\label{figA:Poster}
\end{figure}


\section*{NUCLÉAIRE --- R\'ev\'elations d'un chercheur}
\label{Sec.A2.1}

{\it Dans les ann\'ees 70, des Irakiens se sont rendus \`a plusieurs reprises au CERN pour collecter des informations dans le cadre de leur programme de bombe atomique.}

   Durant les ann\'ees 70, le patron du programme de la bombe atomique irakienne, Jafar Dhia Jafar, a effectu\'e plusieurs s\'ejours au CERN \`a Gen\`eve alors qu'il \'etait rattach\'e \`a l'Imperial College de Londres.  A une reprise au moins, en 1979, il a d\'ep\^ech\'e au CERN l'un de ses ing\'enieurs pour obtenir des renseignements  techniques sur un type d'aimant particulier.  Parmi plusieurs voies possibles pour obtenir de l'uranium enrichi, les Irakiens avaient en effet d\'ecid\'e d'utiliser et de perfectionner les calutrons, une technique qui consiste \`a s\'eparer, \`a l'aide de puissants aimants, les atomes d'uraniun 235 (d'int\'er\^et militaire) des atomes d'uranium 238 (peu fissiles) qui se trouvent m\'elang\'es dans l'uranium naturel. C'est pour cette raison qu'ils s'int\'eressaient pr\'ecis\'ement aux aimants du type de celui d\'evelopp\'e pour une exp\'erience du CERN.  Finalement, les Irakiens ont choisi un design diff\'erent de celui du CERN pour les aimants de leurs calutrons, mais l'\'episode montre qu'ils exploraient cette technique d'enrichissement de l'uranium il y a une quinzaine d'ann\'ees d\'ej\`a.

   Un chercheur genevois, qui avait \`a l'\'epoque tent\'e en vain d'alerter les milieux scientifiques et du d\'esarmement, rend aujourd'hui public cet \'episode, en marge de la ren\'egociation du Trait\'e sur la non-prolif\'eration nucl\'eaire qui se d\'eroule actuellement \`a New York. (R\'ed. {\it  Lire les articles de Suren Erkman en page 5.})

\section*{Gen\`eve, berceau de la bombe irakienne}
\label{Sec.A2.2}

{\it  A Gen\`eve, l'Irak a collect\'e des informations techniques en vue de sa bombe atomique.  Le grand patron du programme militaire nucl\'eaire irakien a lui-m\^eme effectu\'e plusieurs s\'ejours au CERN, le Laboratoire europ\'een pour la physique des particules.}

   Un jour de 1979, un ing\'enieur irakien se rend au CERN, \`a Gen\`eve, le Laboratoire europ\'een pour la physique des particules.  Bien qu'il ne participe pas \`a un programme de recherche du CERN, sa visite n'a rien d'anormal, le laboratoire europ\'een \'etant un lieu d'\'echange scientifique ouvert \`a tous, sans distinction de nationalit\'e.  Il approche l'un des physiciens de l'exp\'erience NA10, et pose des questions sur les d\'etails de construction de l'aimant tout \`a fait particulier qui constitue l'\'el\'ement central de l'appareillage de cette exp\'erience.  Ce physicien, responsable de l'aimant, mentionne lors d'une conversation cette visite \`a l'un de ses coll\`egues, Andre Gsponer, un jeune chercheur qui se pr\'eoccupe du r\^ole de la recherche scientifique dans la course aux armements.

\subsection*{Les relations de Jafar Dhia Jafar}
\label{Sec.A2.2.1}

   Apr\`es une br\`eve analyse, Andre Gsponer parvient \`a la conclusion que, selon toute vraisemblance, les Irakiens sont manifestement int\'eress\'es \`a produire de l'uranium enrichi par le biais d'une technologie connue sous le nom de ``s\'eparation isotopique \'electromagn\'etique''.  Dans son principe, cette technique consiste \`a s\'eparer, \`a l'aide d'un puissant aimant, les atomes d'uranium 235 (d'int\'er\^et militaire) des atomes d'uranium 238 (peu fissiles) qui se trouvent m\'elang\'es dans l'uranium naturel.  Cette technique est connue depuis longtemps, puisque c'est ainsi que les Etats-Unis, entre juin 1944 et juillet 1945, ont fabriqu\'e la charge nucl\'eaire de la bombe l\^ach\'ee sur Hiroshima.  Ce dispositf, mis au point \`a l'Universit\'e de Californie, est connu sous le nom de ``calutron'', contraction de ``California University Cyclotron.''

   Les d\'etails de cette technique, rudimentaire \`a certains \'egards, ont \'et\'e rendus publiques entre 1946 et 1956.  Parmi plusieurs voies possibles pour obtenir de l'uranium enrichi, les Irakiens avaient donc d\'ecid\'es d'utiliser les calutrons mais en les modernisant et en les perfectionnant.  C'est pour cette raison qu'ils s'int\'eressaient pr\'ecis\'ement aux aimants du type de celui d\'evelopp\'e pour l'exp\'eri\-ence NA10 au CERN.

   Les Irakiens ne venaient pas par hasard \`a Gen\`eve.  Le CERN a toujours \'et\'e consid\'er\'e comme le centre d'excellence mondial en mati\`ere de grands aimants.  Surtout, le patron du programme de la bombe irakienne, Jafar Dhia Jafar, avait effectu\'e plusieurs s\'ejours au CERN durant les ann\'ees 70, alors qu'il \'etait rattach\'e \`a l'Imperial College de Londres.  Entre 1967 et 1976, il a publi\'e douze articles scientifiques portant sur des recherches men\'ees aux synchrotrons de Birmingham et du CERN.

   Naturellement, Jafar a utilis\'e ses relations personelles, nou\'ees lors de ses longs s\'ejours en Europe, pour envoyer ses collaborateurs puiser des informations cruciales aux meilleures sources.  C'est ainsi que l'ing\'enieur irakien s'\'etait explicitement recommand\'e de Jafar pour interroger le responsable de l'aimant de l'exp\'erience NA10.  Toutefois, estime ce dernier (qui souhaite conserver l'anonymat), cette visite ne rev\^et pas d'importance particuli\`ere, car ``les informations que cet ing\'enieur \`a re\,cues se trouvaient dans la lit\'erature scientifique publique.  A ma connaissance, il n'a pas eu acc\`es aux plans d'ing\'enieurs originaux, ni \`a certain proc\'ed\'es particuliers d\'evelop\'es par le CERN pour fabriquer cet aimant.''

\begin{figure}
\begin{center}
\resizebox{10cm}{!}{ \includegraphics{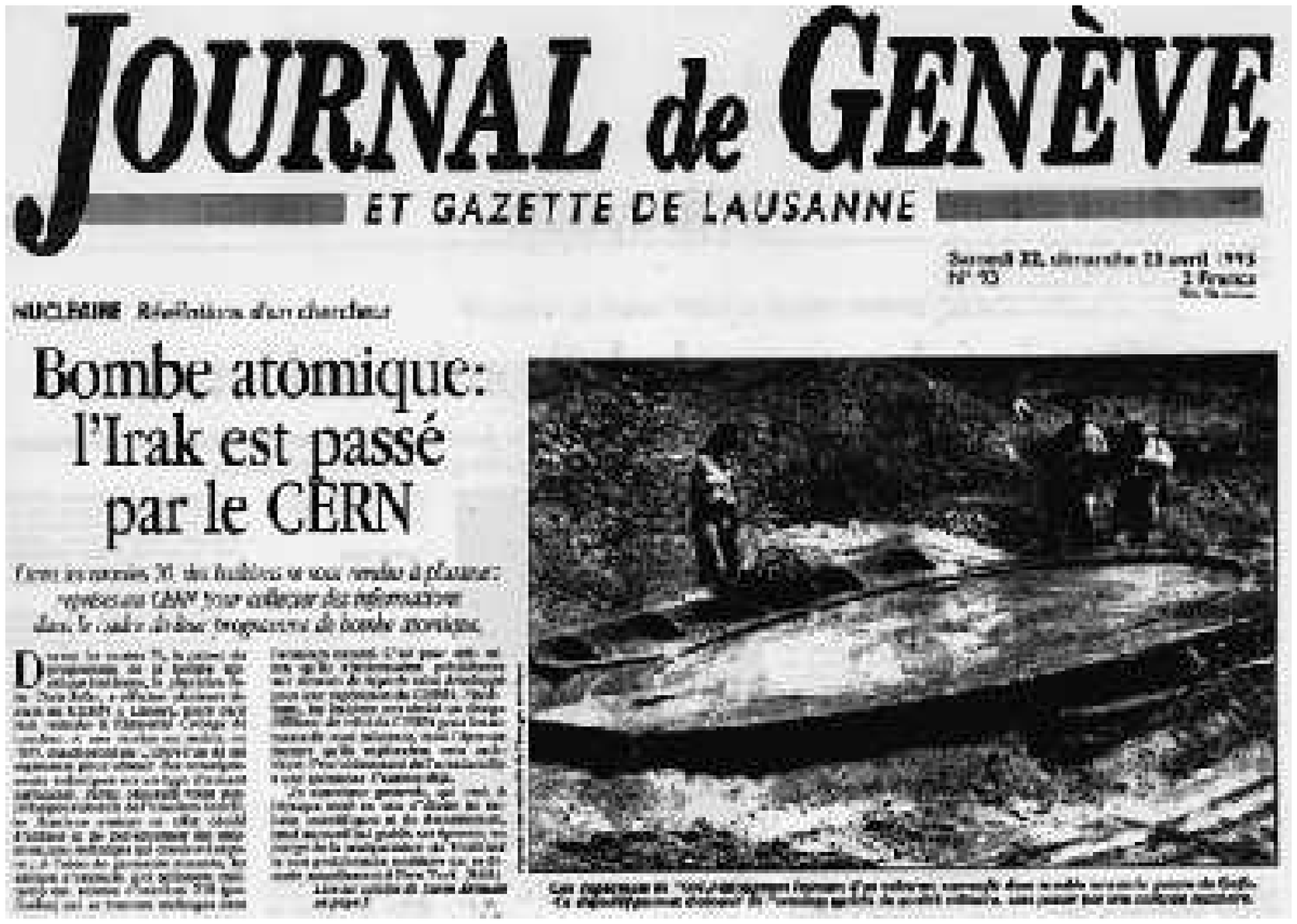}}
\end{center}
\caption[``Bombe Atomique: L'Irak est pass\'e par le CERN.'']{\emph{Journal de Gen\`eve, page 1: ``Bombe Atomique: L'Irak est pass\'e par le CERN.''}}
\label{figA:Page 1}
\end{figure}

\subsection*{Technologie prolif\'erante}
\label{Sec.A2.2.2}

   Lorsque a \'eclat\'e la guerre du Golfe, Jafar Dhia Jafar, qui r\'eside aujourd'hui sous haute protection \`a Bagdad, assumait officiellement les fonctions de ministre adjoint de l'Industrie et de l'Industrialisation militaire, de vice-pr\'esident de la Commission irakienne pour l'\'energie atomique, et de directeur de la physique des r\'eacteurs au Centre de recherches nucl\'eaires de Tuwaitha. [On trouvera une compilation d'informations au sujet de Jafar Dhia Jafar, y compris la mention de ses s\'ejours au CERN, dans l'ouvrage de William E. Burrows \& Robert Windrem: ``Critical Mass, The Dangerous Race for Superweapons in a Fragmenting World'', Simon \& Shuster, New York 1994.  Malheureusement, cet ouvrage comporte de nombreuses erreurs techniques.]

   Pour sa part, Andre Gsponer prend cet incident tr\`es au s\'erieux.  Il quitte le CERN et devient en 1980 le premier directeur du GIPRI, l'Institut international de recherches pour la paix \`a Gen\`eve.  Son objectif est d'\'evaluer, de mani\`ere rigoureuse, les retomb\'ees militaires possibles des recherches effectu\'ees dans le domaine de la physique nucl\'eaire et des particules.  En 1980, \`a l'occasion de la deuxi\`eme Conf\'erence de r\'evision du Trait\'e sur la non-prolif\'eration (TNP), \`a Gen\`eve, il r\'edige un document dans lequel il montre que la technologie de la s\'eparation \'electromagn\'etique, b\'en\'eficiant de nombreux progr\`es techniques, est devenue une technologie potentiellement tr\`es prolif\'erante, et il mentionne m\^eme explicitement l'Irak.

   Apr\`es avoir essay\'e en vain d'\'eveiller l'int\'er\^et des milieux du d\'esarmement sur ce probl\`eme, et apr\`es avoir \'et\'e averti par des sp\'ecialistes des dangers que pouvaient repr\'esenter pour lui et ses anciens coll\`egues du CERN la r\'ev\'elation d'une telle information, il d\'ecide de ne plus mentionner en public l'int\'er\^et soutenu des Irakiens pour les calutrons.  Il quitte la Suisse en 1987 et retourne \`a ses premi\`eres amours: la recherche th\'eorique en physique fondamentale.

\subsection*{La confirmation des calutrons}
\label{Sec.A2.2.3}

   Mais en 1991, apr\`es la guerre du Golfe, il sursaute en voyant les images en provenance d'Irak: pr\'ecis\'ement des calutrons utilis\'es dans le programme de la bombe irakienne!  Andre Gsponer collecte alors toute l'information disponible sur les calutrons dans la litt\'erature scientifique et dans les rapports des missions de l'ONU en Irak.  Il termine actuellement une \'etude technique sur les calutrons, dans laquelle il d\'ecrit en d\'etail la mani\`ere dont l'Irak entendait fabriquer l'uranium enrichi n\'ecessaire \`a la bombe.  [Andre Gsponer et Jean-Pierre Hurni, ``Calutrons: 1945-1995'', ISRI, Case postale 30, 1211 Gen\`eve-12.]  Les \'ev\'enements lui ont donn\'e raison: l'Irak, comme d'autres pays, a bel et bien emprunt\'e la voie des calutrons, confirmant ainsi le potentiel de prolif\'eration de cette technique. ``Si on avait pris cette affaire au s\'erieux \`a l'\'epoque, le cours des \'ev\'enements aurait peut-\^etre \'et\'e diff\'erent'', estime Andre Gsponer.

   Mais il reste une \'enigme de taille.  Les Services de renseignements occidentaux ont pr\'etendu avoir \'et\'e surpris de d\'ecouvrir un important programme de calutrons en Irak: ``Je n'arrive pas \`a croire que j'ai \'et\'e seul, avec quelques autres, \`a savoir que les Irakiens s'int\'eressaient de pr\`es aux calutrons d\`es la fin des ann\'ees 70, s'interroge Andre Gsponer.  Les Services secrets occidentaux devaient \^etre au courant.  Pourquoi avoir alors tellement attendu pour r\'eagir, et pourquoi avoir feint de d\'ecouvrir ce programme apr\`es la guerre du Golfe?''

\begin{figure}
\begin{center}
\resizebox{8cm}{!}{ \includegraphics{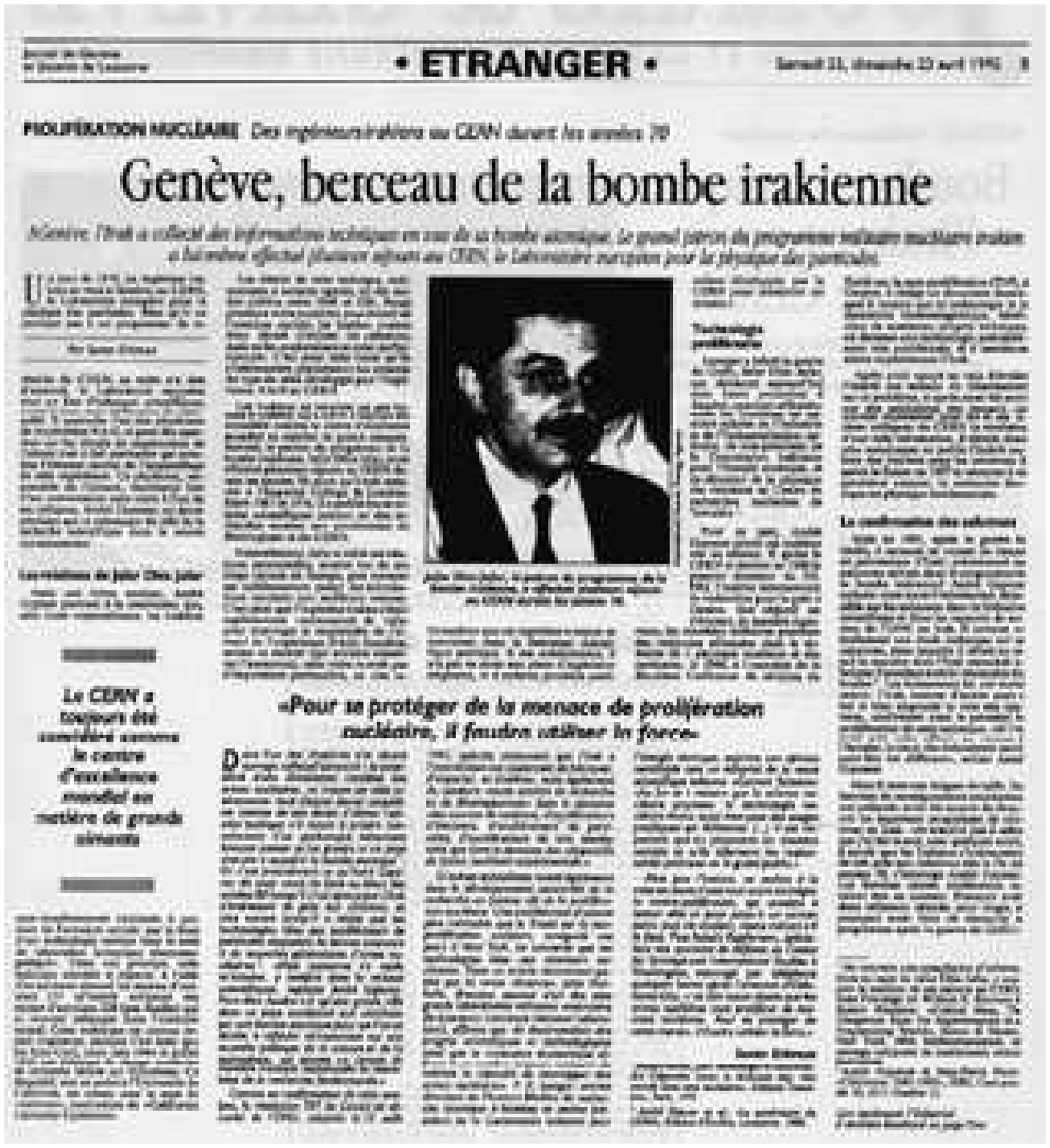}}
\end{center}
\caption[``Gen\`eve, berceau de la bombe irakienne.'']{\emph{Journal de Gen\`eve, page 5: ``Gen\`eve, berceau de la bombe irakienne.''}}
\label{figA:Page 5}
\end{figure}

\section*{``Pour se prot\'eger de la menace de prolif\'eration nucl\'eaire,
           il faudra utiliser la force''}
\label{Sec.A2.3}

    Dans l'un des chapitres d'un r\'ecent ouvrage collectif consacr\'e \`a la possibilit\'e d'une \'elimination compl\`ete des armes nucl\'eaires, on trouve une id\'ee int\'eressante: tout citoyen devrait consid\'erer comme son devoir d'alerter l'opinion publique s'il venait \`a prendre connaissance d'un quelconque \'ev\'enement laissant \`a penser qu'un groupe ou un pays cherche \`a acquerir la bombe atomique. [Joseph Rotblat, Jack Steinberger et Bhalchandra Udgaonkar (sous la direction de): ``Un monde sans armes nucl\'eaires'' Editions Transition, Paris, 1995.]  Or c'est pr\'ecis\'ement ce qu'Andre Gsponer dit avoir tent\'e de faire au d\'ebut des ann\'ees 80 lorsqu'il s'aper\,cu que l'Irak s'int\'eressait de pr\`es aux calutrons, et plus encore lorsqu'il a r\'ealis\'e que les technologies li\'ees aux acc\'el\'erateurs de particules risquaient de donner naissance \`a de nouvelles g\'en\'erations d'armes nucl\'eaires. [Andre Gsponer et al.: ``La Quadrature du CERN'', Editions d'En-Bas, Lausanne, 1984.]  ``Mais personne n'a voulu m'\'ecouter, y compris dans les milieux scientifiques, regrette Andre Gsponer.  Peut-\^etre faudra-t-il qu'une grande ville dans un pays occidental soit volatilis\'ee par une bombe atomique pour que l'on se d\'ecide \`a r\'efl\'echir s\'erieusement sur une nouvelle politique de la science et de la technologie, qui prenne en compte de mani\`ere vraiment responsable les retomb\'ees de la recherche fondamental.''

   Comme en confirmation de cette analyse, la r\'esolution 707 du Conseil de s\'ecurit\'e de l'ONU, adopt\'ee le 15 ao\^ut 1991, sp\'ecifie clairement que l'Irak a l'interdiction non seulement de fabriquer, d'importer, et d'utiliser, mais \'egalement de conduire ``toute activit\'e de recherche et de d\'eveloppement'' dans le domaine ``des sources de neutrons, d'acc\'el\'erateurs d'\'electrons, d'acc\'el\'erateurs de particules, d'acc\'el\'erateurs de ions lourds'' ainsi que dans le domaine des ``dispositifs de fusion nucl\'eaire exp\'erimentale.''

   D'autres sp\'ecialistes voient \'egalement dans le d\'evelopement incontr\^ol\'e de la recherche un facteur cl\'e de la prolif\'eration nucl\'eaire.  Une prolif\'eration d'autant plus in\'evitable que le Trait\'e sur la non-prolif\'eration nucl\'eaire, ren\'egoci\'e ces jours \`a New York, ne concerne que les technologies li\'ees aux r\'eacteurs nucl\'eaires.  Dans un article r\'ecemment publi\'e par la revue ``Science'', John Nuckolls, directeur associ\'e d'un des plus grands laboratoires militaires am\'ericains (le Lawrence Livermore National Laboratory), affirme que ``la diss\'emination des progr\`es scientifiques et technologiques ainsi que la croissance \'economique offrent \`a un nombre toujours croissant de nations la capacit\'e de d\'evelopper des armes nucl\'eaires''.  P.K. Iyengar, ancien directeur de l'Institut Bhabha de recherches atomiques \`a Bombay et ancien pr\'esident de la Commission indienne pour l'\'energie atomique, exprime un opinion semblable dans un \'editorial de la revue scientifique indienne ``Current Science'': ``Au fur et \`a mesure que la science nucl\'eaire progresse, la technologie nucl\'eaire \'evolue, aussi bien pour des usages pacifiques que militaires. (...)  Il est important que les physiciens en tiennent compte et qu'ils informent les responsables politiques et le grand public.''

   Mais pour l'instant, on assiste \`a la mise en oeuvre d'une toute autre strat\'egie: la contre-prolif\'eration, qui consiste \`a laisser aller un pays jusqu'\`a un certain point, puis \`a le stopper, manu militari s'il le faut.  Pour Robert Kupferman, sp\'ecialiste des questions de d\'efense au Center for Strategic and International Studies \`a Washington, interrog\'e par t\'el\'ephone quelques heures apr\`es l'attentat d'Oklahoma City, ``il ne fait aucun doute que les armes nucl\'eaires vont prolif\'erer de mani\`ere terrifiante.  Pour se prot\'eger de cette menace, il faudra utiliser la force.''


\end{document}